\documentclass[12pt]{article}

\usepackage{graphicx}
\usepackage{amsmath}
\usepackage{psfrag}
\usepackage{latexsym}

\begin{document}
\makeatletter
\@addtoreset{equation}{section}
\makeatother

\baselineskip=0.7cm

\newcommand{\Nt}{\widetilde{N}}
\newcommand{\bra}[1]{\langle #1\vert}
\newcommand{\ket}[1]{\vert #1\rangle}
\newcommand{\braket}[2]{\langle #1\vert #2\rangle}
\newcommand{\bbbk}[4]{{}_1\langle #1|{}_2\langle #2|
                      {}_3\langle #3|#4\rangle_{123}}
\newcommand{\C}{C_{\rm vac}}
\newcommand{\HI}{H_{\rm 3}}
\newcommand{\E}{E}
\renewcommand{\P}{{\bf P}}
\newcommand{\PSV}{P_{SV}}

\newcommand{\HSV}{\ket{\HI}_{SV}}
\newcommand{\Hh}{\ket{\HI}_{h}}
\newcommand{\HD}{\ket{\HI}_{D}}
\newcommand{\HCK}{\ket{\HI}_{CK}}

\newcommand{\vac}{{\rm vac}}
\renewcommand{\v}{{\rm v}}

\newcommand{\EQ}{\begin{equation}}
\newcommand{\EN}{\end{equation}}
\newcommand{\EQA}{\begin{eqnarray}}
\newcommand{\EQN}{\end{eqnarray}}
\newcommand{\EQAN}{\begin{eqnarray*}}
\newcommand{\EQNN}{\end{eqnarray*}}
\newcommand{\nn}{\nonumber}
\newcommand{\e}{{\rm e}}
\renewcommand{\l}{l}
\renewcommand{\a}{\alpha}
\renewcommand{\b}{\beta}
\renewcommand{\theequation}{\arabic{section}.\arabic{equation}}
\newcommand{\Tr}{{\rm Tr}}
\newcommand{\lpartial}{\buildrel \leftarrow \over \partial}
\newcommand{\rpartial}{\buildrel \rightarrow \over 
\partial}
\renewcommand{\thesection}{\arabic{section}.}
\renewcommand{\thesubsection}{\arabic{section}.\arabic{subsection}}
\renewcommand{\thefootnote}{\arabic{footnote}}

\newcommand{\inI}{2}
\newcommand{\inII}{3}
\newcommand{\out}{1}
\newcommand{\ainI}{\a_{(\inI)}}
\newcommand{\ainII}{\a_{(\inII)}}
\newcommand{\aout}{\a_{(\out)}}
\newcommand{\ar}{\a_{(r)}}
\newcommand{\arp}{\a_{(r')}}
\newcommand{\as}{\a_{(s)}}
\newcommand{\asp}{\a_{(s')}}
\newcommand{\at}{\a_{(t)}}
\newcommand{\atp}{\a_{(t')}}

\newcommand{\XI}{X_{\rm I}}
\newcommand{\XII}{X_{\rm II}}
\newcommand{\YI}{Y}
\newcommand{\yI}{Y_{\rm I}^{(1)}}
\newcommand{\YII}{Z}
\newcommand{\aDim}{\Delta^{\langle1\rangle}}
\newcommand{\Ctree}{C^{\langle0\rangle}}
\renewcommand{\r}{r}
\newcommand{\s}{s}
\newcommand{\ds}{\dot{s}}
\newcommand{\dr}{\dot{r}}
\newcommand{\gym}{g_{\rm YM}}
\newcommand{\lrp}{\overset{\leftrightarrow}{\partial}}
\newcommand{\h}{1cm}
\newcommand{\w}{1cm}

\newcommand{\hs}{\hspace}
\newcommand{\vs}{\vspace}

\def\thefootnote{\fnsymbol{footnote}}

\begin{flushright}
EPHOU-06-002\\
April, 2006
\end{flushright}

\vspace{1cm}
\begin{center}
\Large Impurity Non-Preserving 3-Point Correlators of BMN Operators \\
from PP-Wave Holography II : Fermionic Excitations

\vspace{0.7cm}

\normalsize
 \vspace{0.4cm}
Suguru {\sc Dobashi}
\footnote{e-mail address:\ \ {\tt doba@particle.sci.hokudai.ac.jp}}

\vspace{0.3cm}

\textit{Department of Physics, Faculty of Science, Hokkaido University\\
Sapporo, 060-0810, Japan}

\vspace{1cm}
Abstract
\end{center}

The holographic principle in the pp-wave limit proposed in our previous works
is further confirmed 
by studying  impurity non-preserving processes which contain 
a fermionic BMN operator with one scalar and one fermion impurities.
We show that the previously proposed duality relation 
between the matrix elements 
of the three point interaction Hamiltonian in
the holographic string field theory 
and the OPE coefficients in
super Yang-Mills theory holds
to the leading order in the large $\mu$ limit.
Operator mixing is required
to obtain the BMN operator  of definite conformal dimension 
which corresponds to 
the string state with one scalar and one fermion excitations.
The mixing term plays a crucial role 
for our duality relation to be valid.
Our results, combined with those in the previous papers,
provide a positive support that our duality relation holds 
for the general process regardless of the kind of impurities
and of whether impurities conserve or not.

\vspace{0.4cm}

\newpage
\section{Introduction}
Holography is the key concept to the duality between
string theory and gauge theory in the AdS/CFT correspondence.
The holographic
relation proposed by GKP/W \cite{gkpw} relates the 
degrees of freedom in the string theory defined in the
bulk AdS space and those in the gauge theory 
defined on the conformal boundary of the AdS spacetime.
Based on this proposal, a large amount of work has been done
giving many positive supports for this relation \cite{AdS/CFTreviews}.
In particular,
the non-renormalization property
of chiral primary operators has been confirmed \cite{lmrs}.
However in the BMN limit of the AdS/CFT correspondence \cite{bmn},
how this holographic relation is realized 
has not been clear because of the lack of the clear space-time picture.

In the previous papers \cite{dsy,yone,dy1,dy2}, 
we gave an answer to this problem.
We showed that 
the PP-wave background emerges
as the geometry around a \textit{tunneling null geodesic},
which starts from a point on the AdS boundary and 
returns to another point on the boundary.
This trajectory can be obtained 
as a dominant path 
in the large angular momentum limit 
for the path integral 
which is expected to give the two point function of gauge theory operators 
in the GKP/W prescription\footnote{See also \cite{yone,asy} for more general discussions.}.
The string/gauge duality in the PP-wave limit can be understood as the
correspondence between the oscillation modes of
strings around this trajectory and gauge theory operators sitting 
around its endpoints on the boundary.
This picture allows a natural correspondence between
an S-matrix calculation on the string theory side
and an operator product expansion on the gauge theory side.
As for three point functions,
with the assumption
that the free string basis corresponds
to the BMN operators of definite conformal dimension,
we proposed a duality relation between
the matrix elements of the string interaction Hamiltonian 
and the OPE coefficients.

Our proposal in \cite{dy1} has an advantage over the previous two types of
duality relations, presented in \cite{GMR,GMP1} and \cite{DPPRT},
in that it is applicable to impurity non-preserving cases.
Indeed, we checked that our duality relation holds 
for impurity non-preserving processes
which consist of BMN operators with only bosonic impurities 
in the previous paper \cite{dy2},
as well as for the impurity preserving processes \cite{dy1}.
To establish the duality relation for
impurity non-preserving processes is important
for studying the unsolved problem of 
string/gauge correspondence at higher genus level,
since impurity non-preserving processes are known to appear as intermediate states
of a string loop amplitude \cite{Roiban et. al.,GP,Grignani et. al.}.

The aim of this paper is to show that our proposal also holds 
for the impurity non-preserving cases involving fermionic impurities.
Checking our duality relation for such processes is
important because our proposal 
was originally obtained as an extension 
of the duality relation for chiral primary operators
to the general string excitations.
It is not trivial this relation is satisfied
for the non-BPS sector in general.
Since we have already checked our proposal
for impurity preserving sector with several different kinds of impurities,
including bosons and fermions,
and for impurity non-preserving sector with only bosonic impurities,
what remains to be studied is impurity non-preserving processes
which contain fermionic impurities.
Though it is desirable to study the general processes, 
we restrict ourself in this paper
to simple impurity non-conserving processes,
where the fermionic BMN operators involved 
contain only one scalar and one fermionic impurities.
Even for these simple examples,
our duality relation holds in a very non-trivial way.

When studying the duality relation,
it is important how to relate the basis between the two theories.
In our prescription, the correspondent on the gauge theory side
to the free string basis is a set of BMN operators 
of definite conformal dimension.
As in the case of the singlet sector of 
several two impurity states \cite{GMP2,GKT,GKT2,GT}, 
the operator mixing is required for the fermionic BMN operator 
we are now considering so that it obtains a definite conformal dimension.
We will see that this operator mixing plays a crucial role
for our duality relation to hold.
In addition, 
the phase factor of the BMN operator 
must be properly chosen up to ${\cal O}(1/J)$
in order to give the correct world sheet momentum dependence.

The result of this paper,
in addition to those in the previous papers,
strengthens
the expectation that our duality relation is valid for general processes
regardless of the kind of impurities 
and of whether impurities conserve or not.
We believe that the successful clarification of the duality relation
at the first order of the string interaction
lays the foundation
for the study of the duality relation at higher genus level.

The paper is organized as follows.
In section 2 we briefly
review our proposal for the duality relation.
In section 3
we calculate two examples of 
the string amplitude 
for fermionic impurity non-preserving processes.
In section 4 
we first identify the gauge theory operator
which corresponds to the string states 
with one scalar and one fermion excitations.
We show by explicit perturbation calculations that 
for the corresponding BMN operator to obtain a definite conformal 
dimension, in addition to the operator with
one fermion and one scalar impurities,
another operator with the same classical dimension 
and R-charge is required.
With this operator, we compute three point functions
which correspond to the string amplitudes calculated
in the previous section.
We will see that our duality relation holds 
to the leading order of the effective gauge coupling
and the operator required for the diagonalization 
plays a crucial role.
We conclude in section 5.
Appendix \ref{AppFey} is devoted to the details of
the perturbative calculations required 
for the two point function of the BMN operators 
with one scalar and one fermion impurities.
We list the explicit forms of the prefactors in
Appendix \ref{AppPre} and the asymptotic behavior of the 
Neumann coefficients in the large $\mu$ limit
in Appendix \ref{AppLaM}.

\section{Review of the holographic duality relation}

Our duality relation at the first order of the string interaction 
relates the 
basic ingredients which characterize the three point interaction 
on each theory side, the matrix element of the 3-point
interaction Hamiltonian $H_{123}$ on string theory side
and the OPE coefficient $C_{123}$ on gauge theory side.
Identifying the free string basis
with the set of BMN operators of definite conformal dimension,
we propose that the duality relation
is given by
\begin{align}
 C_{123}=
\frac{1}{\mu(\Delta_2+\Delta_3-\Delta_1)}
\left(f\frac{J_2J_3}{J_1}\right)^{-\frac{\Delta_2+\Delta_3-\Delta_1}{2}}
\Gamma\left(\frac{\Delta_2+\Delta_3-\Delta_1}{2}+1\right)H_{123},
\label{genuine duality}
\end{align}
where $\Delta_r$ and $J_r$ are the conformal
dimension and the U(1) R-charge, respectively, 
of the gauge theory operator $O_r$ ($r=1,2,3$)
which corresponds to the string state $\ket{r,J_r}$
satisfying the relation $H_2^{(r)}/\mu=\Delta_r-J_r$.
Here, $H_2^{(r)}$ and $\mu$ are the light-cone free Hamiltonian
and the mass parameter of the string theory, respectively,
and 
$J$ denotes the angular momentum around the large circle of $S^5$
on the string theory side.
The effective gauge theory coupling is related as
$\lambda'=\gym^2N/J^2=1/(\mu p^+\a')^2$ 
in the PP-wave limit of AdS/CFT correspondence.
The definition of $f$ in (\ref{genuine duality}) 
is given by some combination of 
Neumann coefficients (see the reference \cite{dy1}).
In what follows, 
we need only its asymptotic behavior in the large $\mu$ limit 
which is given by
\begin{align}
 f\frac{J_2J_3}{J_1}\to\frac{J_1}{4\pi\mu|\a_{(1)}|},
\end{align}
where $\a_{(r)}$ denotes $\a_{(r)}\equiv \a'p_r^+$ and 
we take $\a_{(1)}(=\a_{(2)}-\a_{(3)})<0, \a_{(2)}, \a_{(3)}>0$.
The relation (\ref{genuine duality}) is expected to hold
to all order of $\lambda'=1/(\mu \a_{(1)})^2$, 
but we restrict ourself to checking the relation 
to the leading order in the large $\mu$ limit in this article.

The OPE coefficient $C_{123}$ 
is defined as, for scalar operators, 
\begin{align}
 \langle O_1(x_1)O_{2}(x_2)O_3(x_3)\rangle=\frac{C_{123}}
{|x_{12}|^{\Delta_1+\Delta_2-\Delta_3}
|x_{23}|^{\Delta_2+\Delta_3-\Delta_1}
|x_{31}|^{\Delta_3+\Delta_1-\Delta_2}},
\end{align}
under the normalization of
$\langle O_{1}O_{2}\rangle={\delta_{12}}/{|x_{12}|^{2\Delta}}$.
It is known that for the impurity preserving sector, 
we should take into account the operator mixing of single and double 
trace operators \cite{Beisertetal,Constableetal2}. 
However, for the impurity non-preserving process,
the effect of such mixing is suppressed by $1/N$, 
so we need not consider this type of 
operator mixing as long as we are interested in the planar limit.
The operator mixing we should consider later to obtain the 
BMN operator of definite conformal dimension 
is the one among  single-trace operators 
with the same conformal dimension and R-charge.

The $H_{123}$ is the matrix element of the 3-string interaction
Hamiltonian defined as
\begin{align}
 H_{123}= ~_{(1)}\bra{1}~_{(2)}\bra{2}~_{(3)}\bra{3}
\frac{\sqrt{J_1J_2J_3}}{N}\ket{H_3}_{h},
\end{align}
under the canonical normalization of the string field theory action
such as
\begin{align}
&S_2=\int d\tau\left[
\frac{1}{2}\bra{\bar\psi}\partial_{\tau}\ket{\psi}
-\frac{1}{2}(\partial_\tau\bra{\bar\psi})\ket{\psi}
+\bra{\bar\psi}H_{2}\ket{\psi}
\right],\\
&S_3=\frac{1}{2}\int d\tau~_{(1)}\bra{\bar \psi}~_{(2)}\bra{\psi}~_{(3)}\bra{\psi}\frac{\sqrt{J_1J_2J_3}}{N}\ket{H_3}_{h}+h.c.,\\
&H_2^{(r)}=\frac{1}{|\a_{(r)}|}
\sum_{n=-\infty}^{\infty}
\omega_{n}^{(r)}
(\a_{n}^{(r)\dagger}\a_{n}^{(r)}
+\b_{n}^{(r)\dagger}\b_{n}^{(r)}),\quad \text{with}\quad
\omega_{n}^{(r)}=\sqrt{n^2+(\mu\a_{(r)})^2}.
\end{align}

We proposed that the specific form of the string interaction Hamiltonian $\Hh$
is the equal weight sum of the two interaction vertexes presented so far,
\begin{align}
\label{Hh}
 \Hh = \frac{1}{2} \ket{\HI}_{SV}+\frac{1}{2}\ket{\HI}_{D},
\end{align}
where $\HSV$ is constructed in \cite{SV,SV2}
and further developed in \cite{Pan1,PS1,Pan2,PS2}
as a generalization of the well-known interaction Hamiltonian in the flat
space to the PP-wave background,
while $\HD$ is proposed in \cite{DPPRT}
and its prefactor takes the form of energy difference.
These two vertexes both satisfy the SUSY algebra, 
as well as, the continuity condition and momentum conservation.
The interaction Hamiltonian of the latter type is
not adopted in the flat space case  because 
it does not affect S-matrix calculations 
due to the energy conservation, 
but in our formalism where world-sheet time is naturally Wick-rotated,
it can give indispensable contribution.
The combination (\ref{Hh}) is required
so that the bosonic zero-mode sector should properly reduce 
to the form which is obtained
by taking the BMN limit of the effective action 
for the fluctuation fields around AdS background 
which corresponds to the chiral primary operators
\footnote{In the scalar impurity sector this vertex reduces to 
the phenomenological interaction vertex proposed in \cite{CK},
which is successfully related with the genuine OPE coefficient
through the relation $\mu(\Delta_2+\Delta_3-\Delta_1)C_{123}=H_{123}$
in the impurity preserving process.}.
Though in \cite{LeeRusso} it is claimed that 
the $\HD$ part has to be modified in order to respect 
the U(1)$_{\text{Y}}$ symmetry in the supergravity sector,
the modification gives no effect for the process we are now considering,
as for all the cases we have confirmed so far.
We will discuss the processes in the non-BPS sector
for which such modification cannot be ignored in the future study \cite{dobashi}.

For the impurity preserving processes,
where $\Delta_2+\Delta_3-\Delta_1=0+{\cal O}(\lambda')$,
the duality relation (\ref{genuine duality}) reduces to a simple form 
$\mu(\Delta_2+\Delta_3-\Delta_1)C_{123}=H_{123}$,
the one first conjectured in \cite{Constable et. al. 1}
from a different viewpoint.
We checked that this relation holds
to the leading order in the large $\mu$ limit
for several non-trivial processes including bosonic and fermionic
impurities in the previous paper \cite{dy1}.
For the impurity non-preserving cases, the factor $f$
in the duality relation plays a crucial role:
due to the $\mu$ dependence of Neumann coefficients,
the asymptotic behaviour of $H_{123}$ in the large $\mu$ limit
is suppressed by $\lambda'=1/(\mu p^+\a')^2$ 
as the difference in the number of the impurities 
between in- and out-states increases,
while 
the factors $(fJ_2J_3/J_1)^{(\Delta_1-\Delta_2-\Delta_3)/2}$
give a compensating contribution.
Indeed, in \cite{dy2}, we showed that 
the contributions from each factor on the right hand
side of (\ref{genuine duality}) 
nicely combined to give the OPE coefficients $C_{123}$
which has an appropriate $\mu$ dependence.
In what follows, we confirm that this is also the case for
the impurity non-preserving processes which contain 
fermionic BMN operators.

\section{String theory calculation}
We consider fermionic impurity non-preserving processes
on the string theory side.
In this paper we restrict ourself to 
simple but nontrivial processes which consist 
only of one boson and one fermion impurities such as
\begin{align}
\vac &\to 
\alpha^{(2)i}_{m}\beta_{\a_1\a_2,-m}^{(2)} 
+ \alpha^{(3)j}_{n}\beta^{(3)}_{\b_1\b_2,-n},
\label{ex1}\\
\alpha^{(1)i}_{n_1}\alpha^{(1)j}_{-n_1} &\to 
\alpha^{(2)i}_{n_2}\beta_{\a_1\a_2,-n_2}^{(2)} 
+ \alpha^{(3)j}_{n_3}\beta^{(3)}_{\b_1\b_2,-n_3}.
\label{ex2}
\end{align}
Even for these simple processes, 
we will see that
the characteristic property that the interaction vertex is 
given by the equal weight sum of $\HSV$ and $\HD$, and
the non-trivial $\mu$ dependence on the right hand side of 
(\ref{genuine duality})	play important roles for 
our duality relation to hold.
In (\ref{ex1}) and (\ref{ex2}), 
$\a^{(r)i}_n (i=1,2,3,4)$ and $\b^{(r)}_{\a_1\a_2,n}$ denote
the $n$-th bosonic excitation mode in a scalar direction
and the $n$-th fermionic excitation mode, respectively,
in the exponential basis of $r$-th string
which are defined in terms of sin/cos oscillators, $a_{-n}/a_{n}$ 
( or $b_{-n}/b_{n}$ for a fermion), such as
\begin{eqnarray}
& \a_0=a_0,\quad \a_n=\frac{1}{\sqrt{2}}(a_n-i a_{-n}),
\quad \a_{-n}=\frac{1}{\sqrt{2}}(a_n+ia_{-n}),\\
& \b_0=b_0,\quad \b_n=\frac{1}{\sqrt{2}}(b_n-i b_{-n}),
\quad \b_{-n}=\frac{1}{\sqrt{2}}(b_n+ib_{-n}).\label{exp-sin/cos}
\end{eqnarray}
The SO(8) spinor indexes is decomposed as
SO(8)$=$(SU(2)$\times$SU(2))$\times$(SU(2)$\times$SU(2)),
and the subscripts of $\b_{\a_1\a_2,n}$ denotes
$(\mathbf{2},\mathbf{1})\times(\mathbf{2},\mathbf{1})$ sector of the decomposition
$\mathbf{8_s}=(\mathbf{2},\mathbf{1})\times(\mathbf{2},\mathbf{1})
\oplus(\mathbf{1},\mathbf{2})\times(\mathbf{1},\mathbf{2})$,
while we denote the $(\mathbf{1},\mathbf{2})\times(\mathbf{1},\mathbf{2})$ sector
as $\b_{\dot \a_1\dot \a_2,n}$.

\subsection{$\vac \to \alpha^{(2)i}_{m}\beta^{(2)}_{\a_1\a_2,-m} 
+ \alpha^{(3)j}_{n}\beta^{(3)}_{\b_1\b_2,-n}$}
The matrix element of the interaction vertex for the process (\ref{ex1})
is given by
\begin{eqnarray}
\bra{v}
\b_{\b_1\b_2,-n}^{(3)}\a_{n}^{(3)j}
\b_{\a_1\a_2,-m}^{(2)}\a_{m}^{(2)i}
\ket{\HI}_h,\label{StringAmp1}
\end{eqnarray}
where $\Hh$ is the equal weight sum of the $\HSV$ and $\HD$
such as in (\ref{Hh}).

Let us first calculate the contribution from $\HD$
of the interaction Hamiltonian $\Hh$.
Noting that the prefactor is given by 
$2\sum_{r=1}^{3}\omega_{n}^{(r)}/\a_{(r)}\sim4\mu$
due to the form of the prefactor,
$\ket{\HI}_D=(H^{(2)}_{2}+H^{(3)}_{2}-H^{(1)}_{2})\ket{E}$,
we can easily see that the contribution from the vertex $\HD$ is given by
\begin{align}
 \bra{v}
\b_{\b_1\b_2,-n}^{(3)}\a_{n}^{(3)j}
\b_{\a_1\a_2,-m}^{(2)}\a_{m}^{(2)i}
\HD
&=4\mu\times
\frac{i}{2}
(Q_{nm}^{32}-Q_{mn}^{23})\widetilde N_{-m-n}^{23}
\epsilon_{\a_1\b_1}\epsilon_{\a_2\b_2}\delta^{ij}.\label{Dcontribution0}
\end{align}
Here we used the 
the form of the overlap vertex state such as
\begin{align}
&\ket{E}=\ket{E_a}\ket{E_b},\\
&\ket{E_b}=\exp\left[\sum_{r,s=1}^{3}\sum_{m,n=1}^{\infty}
\left(-\frac{i}{2}Q_{mn}^{rs}
(\b_{m}^{(r)\dagger\a_1\a_2}
\b_{n,\a_1\a_2}^{(s)\dagger}
+\b_{m}^{(r)\dagger\a_1\a_2}
\b_{-n,\a_1\a_2}^{(s)\dagger}\right.\right.\notag\\
&\hspace{6cm}\left.\left.-\b_{-m}^{(r)\dagger\a_1\a_2}
\b_{n,\a_1\a_2}^{(s)\dagger}
-\b_{-m}^{(r)\dagger\a_1\a_2}
\b_{-n,\a_1\a_2}^{(s)\dagger}
\frac{}{})\right)\right]\ket{E_b^0},\\
&\ket{E_a}
=\exp \left[
-\sum_{r,s=1}^{3}\sum_{m,n=-\infty}^{\infty}
\a_{m}^{(r)\dagger}\widetilde N_{mn}^{rs}\a_{n}^{(s)\dagger}
\right]\ket{v},
\end{align}
where $\ket{E_b^0}$ is the overlap vertex including the fermionic zero-mode
sector whose detailed form does not concern the present calculation,
and the definition of $Q_{mn}^{rs}$ is 
\begin{eqnarray}
 Q_{mn}^{rs}=e(\ar)
\sqrt{\frac{|\as|}{|\ar|}}
\left(U_{(r)}^{1/2}C^{1/2}N^{rs}C^{-1/2}C_{(s)}^{1/2}\right)_{mn},
\end{eqnarray}
with $U_{(r)}=C^{-1}(C_{(r)}-\mu\a_{(r)})$, and
$C_{mn}=m\delta_{mn}$, $C_{(r)mn}=\omega_{m(r)}\delta_{mn}$.
The details are referred to the reference \cite{Pan2}.
In the large $\mu$ limit,
the form of $Q_{nm}^{32}$ and $Q_{mn}^{23}$ are given 
in terms of Neumann coefficients such as
\begin{eqnarray}
 Q_{mn}^{23}=\frac{m}{\mu y |\aout|}\tilde N_{mn}^{\inI\inII},\qquad
 Q_{nm}^{32}=\frac{n}{\mu(1-y)|\aout|}\tilde N_{nm}^{\inII\inI}.
\end{eqnarray}
With this relation and the asymptotic behaviour in the large $\mu$ limit
of the bosonic Neumann coefficient $\tilde N_{mn}^{23}=\tilde N_{nm}^{32}$,
which is listed in Appendix \ref{AppLaM},
the contribution from the interaction vertex $\HD$ is given by
\begin{align}
 \bra{v}
\b_{\b_1\b_2,-n}^{(3)}\a_{n}^{(3)j}
\b_{\a_1\a_2,-m}^{(2)}\a_{m}^{(2)i}
\HD
=-i\left(\frac{m}{y}-\frac{n}{1-y}\right)
\frac{1}{8\pi^2\mu^2|\a_{(1)}|^3y(1-y)}
\epsilon_{\a_1\b_1}\epsilon_{\a_2\b_2}\delta^{ij}\nn.
\label{Dcontribution}
\end{align}
Here we have used the notation $y=-\a_{(2)}/\a_{(1)}$,
($1-y=-\a_{(3)}/\a_{(1)}$).

As for the interaction Hamiltonian $\HSV$, there are two types of contraction,
\begin{eqnarray}
[\a\a](\b\b) \quad \text{or}\quad
[\a\a][\b\b],
\end{eqnarray}
where $[XY]$ and $(XY)$  denote the contractions 
of X and Y through the prefactor and overlap, respectively.
For example, $[\a\a](\b\b)$ means 
$\bra{v} \a \a \PSV \ket{v}\cdot\bra{v}\b\b\ket{\E_b}$.
There are no combination from $(\b\b)(\a\a)$, $[\b\b](\a\a)$, etc.
because of the form of the prefactor $\PSV$ presented in Appendix \ref{AppPre}.

Using the property that,
for purely bosonic external states,  
the interaction vertex $\HSV$ can be rewritten as
\begin{align}
&\HSV=\HD-\XII^2\ket{\E},\notag\\
&\XII^2\ket{\E}=
-\sum_{r,s=1}^{3}\sum_{m,n=1}^{\infty}
\frac{\omega_m^{(r)}}{\ar}
\left(\widetilde N^{rs}_{mn}-\widetilde N^{rs}_{m-n}\right)
\left(\a_{m}^{(r)\dagger} \a_{n}^{(s)\dagger}
+\a_{-m}^{(r)\dagger} \a_{-m}^{(s)\dagger}\right)\ket{\E}\\
&\hspace{1.7cm}+
\sum_{r,s=1}^{3}\sum_{m,n=1}^{\infty}
\left(\frac{\omega_m^{(r)}}{\ar}+\frac{\omega_n^{(s)}}{\as}\right)
\left(\widetilde N^{rs}_{mn}-\widetilde N^{rs}_{m-n}\right)
\a_{m}^{(r)\dagger} \a_{-n}^{(s)\dagger}\ket{\E}\notag,
\end{align}
where $\XII$ is a bosonic constituent of the prefactor
whose explicit form is given in Appendix \ref{AppPre},
we can see that the contribution from {$[\a\a](\b\b)$} 
is given by the product of two contributions,
\begin{align}
\bra{v}\a_{m}^{(2)i}\a_{n}^{(3)j}\PSV\ket{E}
&=\bra{v}\a_{-m}^{(2)i}\a_{-n}^{(3)j}
\left(
2\mu-\XII^2\right)\ket{E}
\notag\\
&=-
2\mu\widetilde N_{-m-n}^{23}\delta^{ij}
+\frac{\delta^{ij}}{2}
\left(
\frac{\omega_{m}^{(2)}}{\a_{(2)}}
+\frac{\omega_{m}^{(3)}}{\a_{(3)}}
\right)
\left(\widetilde N_{mn}^{23}-\widetilde N_{m-n}^{23}\right)\notag\\
&=-2\mu \widetilde N_{mn}^{23}\delta^{ij},\\
\bra{v}\b_{\b_1\b_2,-n}^{(3)}\b_{\a_1\a_2,-m}^{(2)}\ket{E}
&=\epsilon_{\a_1\b_1}\epsilon_{\a_2\b_2}
\times\left(-\frac{i}{2}\right)
\left(Q_{nm}^{32}-Q_{mn}^{23}\right).
\end{align}
Then, the net result of $[\a\a](\b\b)$ is half of the contribution from $\HD$,
\begin{align}
\label{SVcontribution1}
[\a\a](\b\b)&=i \mu \left(Q_{nm}^{32}-Q_{mn}^{23}\right) \widetilde N_{mn}^{23}
\epsilon_{\a_1\b_1}\epsilon_{\a_2\b_2}\delta^{ij}\notag\\
&=
-i\left(\frac{m}{y}-\frac{n}{1-y}\right)
\frac{1}{8\pi^2\mu^2|\a_{(1)}|^3y(1-y)}
\epsilon_{\a_1\b_1}\epsilon_{\a_2\b_2}\delta^{ij}.
\end{align}

As for the  $[\b\b][\a\a]$ contribution, the only relevant term in 
the prefactor $\PSV$ is 
\begin{eqnarray}
-\frac{i}{4}\widetilde K^i\widetilde K^j Y^2_{ij}
=\frac{i}{2}\XI^i\XII^jY^2_{ij}.\label{relPSV}
\end{eqnarray}
For the definition of $\XI$, $\XII$, and $Y_{ij}^2$, 
being bosonic and fermionic constituents of the prefactor $\PSV$,
see Appendix \ref{AppPre}.
Noticing that 
$Y^2_{ij}=Y^{\a_1}_{~~~\gamma_2}Y^{\b_1\gamma_2}\sigma^{ij}_{\a_1\b_1}$
is defined in terms of cos mode basis such as 
$Y^{\a_1\a_2}=\sum_{n=0}^{\infty}\bar G^{(r)}_nb_n^{(r)\a_1\a_2\dagger}$
and that the relation between sin/cos and exponential basis
is given by (\ref{exp-sin/cos}), the contribution of $[\b\b]$
contraction is
\begin{align}
\bra{v} \b_{\b_1\b_2,-n}^{(3)}\b_{-m,\a_1\a_2}^{(2)} Y_{ij}^2\ket{v}
&=
\frac{1}{2}
\bra{v}
b_{\b_1\b_2,n}^{(3)}
b_{\a_1\a_2,m}^{(2)}
Y^{\rho_1}_{~~~\gamma_2}Y^{\sigma_1\gamma_2}\sigma^{ij}_{\rho_1\sigma_1}
\ket{v}\notag\\
&=
-\epsilon_{\a_2\b_2}(\sigma^{ij})_{\a_1\b_1}
\bar G_{m}^{(2)}\bar G_n^{(3)}.\label{bb}
\end{align}
Here, $(\sigma^{ij})_{\a_1\b_1}$ is defined as 
$(\sigma^{ij})_{\a_1\b_1}=\epsilon_{\b_1\gamma_1}
1/2(\sigma^i\bar\sigma^j-\sigma^j\bar\sigma^i)_{\a_1}^{~\gamma_1}$.
Similarly, the bosonic contraction $[\a\a]$ is given by
\begin{align}
\bra{v}\a_{m}^{(2)i}\a_{n}^{(3)j}\XI^{[i'}\XII^{j']}\ket{v}
&=
\bra{v} \a_{m}^{(2)i}\a_{n}^{(3)j}
\cdot 
-i \sum_{r,s=1}^{3}\sum_{m=0,n=1}^{\infty}
\bar F_{m}^{(r)}\bar F_{n}^{(s)}U^{(s)}_{n}
a_{m}^{(r)^\dagger[i'}a_{-n}^{(s)j']\dagger}\ket{v}\notag\\
&=
\frac{1}{2}\bar F^{(2)}_{m}\bar F^{(3)}_{n}(U_{m}^{(2)}-U_{n}^{(3)})
\delta^{i[i'}\delta^{j']j}.\label{aa}
\end{align}
Combining the two contributions, (\ref{bb}) and (\ref{aa}),
and using the asymptotic form for $\bar F$ and $\bar G$ listed in
Appendix \ref{AppLaM},
the contribution from $[\a\a][\b\b]$ is evaluated as
\begin{align}
\label{SVcontribution2}
[\b\b][\a\a]
=
-i\left(\frac{m}{y}-\frac{n}{1-y}\right)
\epsilon_{\a_2\b_2}
(\sigma^{ij})_{\a_1\b_1}
\frac{1}{16\pi^2\mu^2|\a_{(1)}|^3 y(1-y)}.
\end{align}

Combining all the contributions, 
(\ref{Dcontribution}), (\ref{SVcontribution1}), and (\ref{SVcontribution2}), 
we obtain the matrix element of the interaction vertex $\Hh$
for the process (\ref{StringAmp1}) in the large $\mu$ limit,
\begin{align}
&\bra{v}
\b_{\b_1\b_2,-n}^{(3)}\a_{n}^{(3)i}
\b_{\a_1\a_2,-m}^{(2)}\a_{m}^{(2)i}
\ket{\HI}_h\notag\\
&=-i
\left(\frac{m}{y}-\frac{n}{1-y}\right)
\frac{1}{16\pi^2\mu^2|\a_{(1)}|^3 y(1-y)}
\epsilon_{\a_2\b_2}\left(\frac{3}{2}\epsilon_{\a_1\b_1}\delta^{ij}
+\frac{1}{2}(\sigma^{ij})_{\a_1\b_1}
\right)\label{matrix-element1}.
\end{align}

In order to obtain the OPE coefficient according to 
the duality relation (\ref{genuine duality}),
the $G$ factor 
\begin{align}
\label{Gfactor}
 G&\equiv
\frac{1}{\mu(\Delta_2+\Delta_3-\Delta_1)}
\left(f\frac{J_2J_3}{J_1}\right)^{-\frac{\Delta_2+\Delta_3-\Delta_1}{2}}
\Gamma\left(\frac{\Delta_2+\Delta_3-\Delta_1}{2}+1\right)
\frac{\sqrt{J_1J_2J_3}}{N}
\notag\\
&=\frac{1}{4\mu}\left(\frac{J_1}{4\pi\mu|\a_{(1)}|}\right)^{-2}
\Gamma(3)\frac{\sqrt{J_1J_2J_3}}{N}\notag\\
&=8\pi^2\mu|\a_{(1)}|^2
J_{1}^{-1/2}N^{-1}\sqrt{y(1-y)},
\end{align}
must be multiplied with (\ref{matrix-element1}).
Therefore, the CFT coefficient $C_{123}$ which should be obtained
from the corresponding three point correlation function
on the gauge theory side is expected to be
\begin{align}
 C_{123}=
-\frac{i}{2}
\left(\frac{m}{y}-\frac{n}{1-y}\right)
\frac{\gym J_1^{-3/2}N^{-1/2}}{\sqrt{y(1-y)}}
\epsilon_{\a_2\b_2}
\left(\frac{3}{2}\epsilon_{\a_1\b_1}
+\frac{1}{2}\left(\sigma^{ij}\right)_{\a_1\b_1}\right).
\label{C1}
\end{align}
We should note that the interaction vertexes $\HSV$ and $\HD$ 
are both necessary to obtain this results
and that the $G$ factor properly adjust
the $\mu$-dependence to give ${\cal O}(\gym)={\cal O}(1/\mu)$ result.

\subsection{$\alpha^{(1)i}_{n_1}\alpha^{(1)j}_{-n_1} \to 
\alpha^{(2)i}_{n_2}\beta^{(2)}_{\a_1\a_2,-n_2} 
+ \alpha^{(3)j}_{n_3}\beta^{(3)}_{\b_1\b_2,-n_3}$}

Next, we proceed to calculate 
the matrix element of the string interaction vertex
for the process (\ref{ex2}), which is given by
\begin{align}
\bra{v}
\b_{\b_1\b_2,-n_3}^{(3)}\a_{n_3}^{(3)j}
\b_{\a_1\a_2,-n_2}^{(2)}\a_{n_2}^{(2)i}
\a^{(1)i}_{n_1}\a^{(1)j}_{-n_1}
\ket{\HI}_h.\label{StringAmp2}
\end{align}
We restrict ourself to the case $i\ne j$ for simplicity.

In almost the same way as the previous case, 
the contribution from the $\HD$ is given by
\[
\bra{v}\b_{\b_1\b_2,-n_3}^{(3)}\a_{n_3}^{(3)j}
\b_{\a_1\a_2,-n_2}^{(2)}\a_{n_2}^{(2)i}
\a^{(1)i}_{n_1}\a^{(1)j}_{-n_1}
\HD \]
\[
\hs*{-3.4cm}=2\mu\times
\left(-\frac{i}{2}\right)
(Q_{nm}^{32}-Q_{mn}^{23})
(-\widetilde N_{n_3,-n1}^{31})
(-\widetilde N_{n_2n_1}^{21})
\epsilon_{\a_1\b_1}\epsilon_{\a_2\b_2}\delta^{ij}\nonumber 
\]\vs{-.5cm}
\begin{align}
\label{Amp2ofHD}
=
i
\frac{1}{4\pi^2\mu|\a_{(1)}|^2y(1-y)}
\epsilon_{\a_1\b_1}\epsilon_{\a_2\b_2}\delta^{ij}
\left(\frac{n_2}{y}-\frac{n_3}{1-y}\right)
\frac{\sin^2(\pi yn_1)}
{\left(n_1-\frac{n_2}{y}\right)\left(n_1+\frac{n_3}{1-y}\right)},
\end{align}
where we used the asymptotic form of the Neumann coefficients
in the large $\mu$ limit presented in Appendix \ref{AppLaM}.

On the other hand, 
because of the form of the prefactor $\PSV$,
for the case $i\ne j$,
the possible contribution from $\HSV$ is that 
only one pair of $\a$ is contracted through the prefactor
and others are 
through the overlap vertex:
\begin{align}
\big[\a_{n_3}^{(3)j}\a^{(1)j}_{-n_1}\big]
\big(\a_{n_2}^{(2)i}\a^{(1)i}_{n_1}\big)
\big(\b_{\b_1\b_2,-n_3}^{(3)}\b_{\a_1\a_2,-n_2}^{(2)}\big)
\; \text{or}\;
\big(\a_{n_3}^{(3)j}\a^{(1)j}_{-n_1}\big)
\big[\a_{n_2}^{(2)i}\a^{(1)i}_{n_1}\big]
\big(\b_{\b_1\b_2,-n_3}^{(3)}\b_{\a_1\a_2,-n_2}^{(2)}\big).\label{nonZero2}
\end{align}
In particular, when $i\ne j$, the contraction 
through the prefactor (\ref{relPSV}) is not possible
because of the anti-symmetric property of $i$ and $j$ in $Y_{ij}^2$
and the fact that only a pair of oscillators with the same scalar index
can be contracted through the overlap vertex.
This type of contraction is possible 
when we consider, for example, 
the case where the excitation of the incoming string is given by
$\a^{(1)i}_{n_1}\a^{(1)k}_{-n_1}\ket{v}\, (k\ne j)$ 
and will play an important role for our duality relation to hold.

The amplitude of the first term in (\ref{nonZero2}) is the product of 
three amplitude such as
\begin{align}
\bra{v}\a_{n_3}^{(3)j}\a^{(1)j}_{-n_1}\PSV\ket{v}\cdot
\bra{v}\a_{n_2}^{(2)i}\a^{(1)i}_{n_1}\ket{E}\cdot
\bra{v}\b_{\b_1\b_2,-n_3}^{(3)}\b_{\a_1\a_2,-n_2}^{(2)}\ket{E}.\label{a-1}
\end{align}
At the first factor, the prefactor $\PSV$ can be 
replaced with $\sum_{r}E_r-\XII^2$ in the the same way as in the previous 
subsection. But in contrast to the situation there, 
it gives the order ${\cal O}(1/\mu)$ contribution
as in the impurity preserving processes:
\begin{align}
\bra{v}\a_{n_3}^{(3)j}\a^{(1)j}_{-n_1}\sum_{r}E_r\ket{E}
&=
\left(\frac{\omega_{n}^{(3)}}{\a_{(r)}}
+\frac{\omega_{n}^{(1)}}{\a_{(r)}}\right)
\widetilde N^{31}_{n_3,-n_1}
={\cal O}(1/\mu).
\end{align}
Since the contribution from $\XII$ is of the same
order and 
the orders of the remaining contributions are
\begin{align}
\bra{v}\a_{n_2}^{(2)i}\a^{(1)i}_{n_1}\ket{E}={\cal O}(1),\quad
\bra{v}\b_{\b_1\b_2,-n_3}^{(3)}\b_{\a_1\a_2,-n_2}^{(2)}\ket{E}
={\cal O}(1/\mu^2),
\end{align}
the net result of (\ref{a-1}) gives a sub-leading 
contribution and negligible compared to the result 
(\ref{Amp2ofHD}) in the large $\mu$ limit. 
The same is true for the second case of (\ref{nonZero2}).

Therefore only the interaction vertex $\HD$ 
gives the leading contribution such as
\begin{align}
 &\bra{v}\b_{\b_1\b_2,-n_3}^{(3)}\a_{n_3}^{(3)j}
\b_{\a_1\a_2,-n_2}^{(2)}\a_{n_2}^{(2)i}
\a^{(1)i}_{n_1}\a^{(1)j}_{-n_1}
\Hh\\
&=i
\frac{1}{8\pi^2\mu|\a_{(1)}|^2y(1-y)}
\epsilon_{\a_1\b_1}\epsilon_{\a_2\b_2}
\left(\frac{n_2}{y}-\frac{n_3}{1-y}\right)
\frac{\sin^2(\pi yn_1)}
{\left(n_1-\frac{n_2}{y}\right)\left(n_1+\frac{n_3}{1-y}\right)}\nn.
\end{align}

Taking into account the $G$ factor for this process,
\begin{align}
 G=\left(\frac{J_1}{4\pi\mu|\a_{(1)}|}\right)^{-1}
\Gamma(2)\frac{\sqrt{J_1J_2J_3}}{N}\frac{1}{2\mu}
=2\pi|\a_{(1)}|J_1^{1/2}\sqrt{y(1-y)}N^{-1},
\end{align}
the OPE coefficient predicted from the string theory side is
\begin{align}
C_{123}=i
\frac{\gym J_1^{-1/2}N^{-1/2}}{4\pi^2\sqrt{y(1-y)}}
\epsilon_{\a_1\b_1}\epsilon_{\a_2\b_2}
\left(\frac{n_2}{y}-\frac{n_3}{1-y}\right)
\frac{\sin^2(\pi yn_1)}
{\left(n_1-\frac{n_2}{y}\right)\left(n_1+\frac{n_3}{1-y}\right)}.
\label{C2}
\end{align}

\section{Gauge theory calculation}
Now that we have calculated two examples of the three point amplitude
for the impurity non-preserving process which contain
fermionic impurity operators, (\ref{ex1}) and (\ref{ex2}), rewritten as
\begin{align}
\bra{v}
\b_{\b_1\b_2,-n}^{(3)} \a_{n}^{(3)j}
\b_{\a_1\a_2,-m}^{(2)} \a_{m}^{(2)i}
\Hh&,\label{stringAmp1}\\
\bra{v}
\b_{\b_1\b_2,-n_3}^{(3)}\a_{n_3}^{(3)j}
\b_{\a_1\a_2,-n_2}^{(2)}\a_{n_2}^{(2)i}
\a^{(1)i}_{n_1}\a^{(1)j}_{-n_1}
\Hh,&\label{stringAmp2}
\end{align}
let us proceed to calculate
the corresponding gauge theory three point correlators 
and derive the OPE coefficients.

For this purpose, it is important to identify properly 
the operator which corresponds to the string state
$\a_{n}^{i\dagger}\b_{\a_1\a_2,-n}^{\dagger}\ket{v}$ with
the angular momentum $J$ around $S^5$.
According to the prescription of inserting an impurity into the 
vacuum state $\Tr(Z^J)$ with an appropriate phase factor, 
the operator with the proper U(1) R-charge and 
the conformal dimension satisfying the relation $H/\mu=\Delta-J$ 
might be given by
\begin{align}
\label{O}
\sum_{l=0}^{J}\exp\Big(\frac{2\pi i n}{J+2} (l+1)\Big)
\Tr(\phi^{i}Z^l\lambda_{r\a}Z^{J-l}),
\end{align}
where the subscript $r$ of $\lambda_{r\a}$ is the spinor index of
the $(\mathbf{2},\mathbf{1})_{+1/2}$ sector in the decomposition 
of SU(4) R-symmetry vector index $A$
of Weyl fermion $\lambda^A_\a$ in the $d=4$, ${\cal N}=4$ SYM theory,
such as 
\begin{align}
 \mathbf{4} = (\mathbf{2},\mathbf{1})_{+1/2}\oplus(\mathbf{1},\mathbf{2})_{-1/2}.\label{decomposition}
\end{align}
Here the subscript $\pm1/2$ denotes 
the charge of U(1)$_J$ R-symmetry subgroup we have picked up
\footnote{The correspondence of the spinor indexes between string and gauge
theory side is $(\a_1,\a_2)\leftrightarrow(r,\a)$.}.
However, as we will show in what follows,
in order to obtain the operator of definite conformal dimension, 
we must consider a linear combination of (\ref{O}) and the operator
\begin{align}
(\tau^{i})^{\dot r}_r \Tr(\theta_{\dot r\a}Z^{J+1}),\label{o}
\end{align}
which is the unique possibility of the operator with the 
same classical dimension and U(1) R-charge as the operator (\ref{O}).
Here $\theta_{\dot r \a}$ is the $(\mathbf{1},\mathbf{2})_{-1/2}$ sector in the
decomposition of (\ref{decomposition})
and $(\tau^i)^{\dot r}_{r}$ is the Clebsch-Gordan coefficient for 
a SO(4) vector under SO(4)=SU(2)$\times$SU(2) decomposition.

The operator (\ref{o}) is expected to be obtained when the SUSY transformation
and the rotation in the $iZ$-plane 
of R-symmetry act on 
one and the same constituent $Z$ of the vacuum state $\Tr(Z^{J+2})$,
while (\ref{O}) is obtained when these two generators act 
on different $Z$'s with an appropriate phase dependence 
\cite{PR, Gursoy, beisert}.
Though the order of the number of the latter case is ${\cal O}(J)$ 
while  the former is of ${\cal O}(1)$, we cannot omit this term
when we identify the BMN operator which corresponds to the free string
basis
as in the case of the singlet sector of the operator with 
two bosonic or fermionic impurities studied so far \cite{GMP2,GT,GKT,GKT2}.

In what follows, we first determine the weight of the linear combination of
(\ref{O}) and (\ref{o}) so that the two point function of
the BMN operator takes the canonical form.
Next, with the proper linear combination thus determined,
we calculate the gauge theory three point functions which
correspond to (\ref{stringAmp1}) and (\ref{stringAmp2}),
respectively.
We will see that the OPE coefficients which can be
read from the three point functions  exactly agree with
(\ref{C1}) and (\ref{C2}).

\subsection{Diagonalization and the necessity of the operator mixing}

The operator on the gauge theory side which corresponds
to a free string state should have a definite conformal dimension
and the form of the two point function of these operators
takes the canonical form 
determined by the conformal symmetry.
The desired form of the two point function 
in the present case is
\begin{align}
\label{canonical}
&\langle O_{r\a,m}^{J,i}(x) \bar O_{s\dot\a,n}^{J,j}(y)\rangle\\
&=
-\delta^{ij}\delta_{mn}\epsilon_{rs}
\frac{1}{(x-y)^{2(J+2)}}
(\sigma^{\mu})_{\a\dot\a}
\frac{(x-y)^{\mu}}{(x-y)^2}
(1-\lambda'n^2\ln(x-y)^2\Lambda^2).\notag
\end{align}

In order to obtain an operator of definite conformal dimension, 
we consider the linear combination such as 
\begin{align}
O_{r\a,m}^{J,i}\equiv\sum_{l=0}^{J}q_l
\Tr(\phi^iZ^l\lambda_{r\a}Z^{J-l})
+b(\tau^{i})^{\dot r}_{r}\Tr(\theta_{\dot r\a} Z^{J+1}),\label{linear_comb}
\end{align}
and chose the parameter $b$ so that the two point function,
\begin{align}
\langle O_{r\a}^{J,i}\bar O_{s\dot\a}^{J,j} \rangle
&= \sum_{l,l'=0}^{J}q_l\bar r_{l'}
\langle 
\Tr(\phi^iZ^l\lambda_{r\a}Z^{J-l})
\Tr(\bar Z^{J-l'}\bar\lambda_{s\dot\a}\bar Z^{l'}\phi^j)
\rangle\notag\\
&+b
\langle
(\tau^{i})^{\dot r}_r\Tr(\theta_{\dot r\a}Z^{J+1})
\sum_{l=0}^{J}\bar r_l 
\Tr(\bar Z^{J-l}\bar \lambda_{s\dot \a}\bar Z^{l}\phi^j)
\rangle\label{proper2point}\\
&+\bar b
\langle
\sum_{l=0}^{J}
q_l \Tr(\bar Z^{J-l}\bar \lambda_{r\dot \a}\bar Z^{l}\phi^i)
(\tau^{j})^{\dot s}_s\Tr(\bar Z^{J+1}\bar\theta_{\dot s\dot \a})
\rangle\notag\\
&+|b|^2
\langle
(\tau^{i})^{\dot r}_r\Tr(\theta_{\dot r\a}Z^{J+1})
(\tau^{j})^{\dot s}_s\Tr(\bar Z^{J+1}\bar\theta_{\dot s\dot \a})
\rangle,\notag
\end{align}
takes the canonical form of (\ref{canonical}). 
Here we have defined the phase factor $q_l$ and $r_l$ such as
\begin{align}
q_l\equiv\exp\left(\frac{2\pi im}{J+2}(l+1)\right),
\quad  
r_l\equiv\exp\left(\frac{2\pi in}{J+2}(l+1)\right).
\end{align}
The complex conjugate with lower indexes is defined as 
$\bar O^{i}_{r\dot\a}\equiv\epsilon_{rs}
\bar O^{is}_{\dot\a}\equiv\epsilon_{rs}(O^{i}_{s\a})^{*}$.
Note that $\tau$ and $\bar\tau$ are defined as
\begin{align}
 (\tau^i)_{r\dot s}\equiv(i\sigma^{i},-1),\quad
(\bar \tau^i)^{\dot s r}\equiv(-i\sigma^i,-1),
\end{align}
where $\sigma^{i}$'s are Pauli matrices, and the complex conjugate of the 
matrix $\tau$ is $((\tau^i)_{r\dot s})^*=(\bar \tau^i)^{\dot s r}$.
We will call the contributions in the first, second, third and fourth
line in (\ref{proper2point})
as  $(++)$, $(-+)$, $(+-)$, and $(--)$ sector, respectively.

In order to determine $b$, it is enough to focus on the 
log divergent part in (\ref{canonical}).
Let us first compute the log divergences
which appear in the ($++$) sector.
The divergence comes form self-energy, $4Z$-interaction,
$2\lambda Z$-interaction,
$2\phi Z$-interaction, and $2\phi\lambda$-interaction, 
which will be explained bellow.
The calculation is partially same as
in the reference \cite{Kristjansen et. al.}
where the anomalous dimension of
the operator with two scalar impurities is calculated,
but here the calculation becomes
a little bit complicated because 
we must consider additional interactions which involve fermions.
We calculate here $4Z$-interaction, etc.\,
which have already been calculated in \cite{Kristjansen et. al.} for self-containedness.
The details of the calculations are referred to Appendix \ref{AppFey}.

The two point functions 
of $Z^a$'s and $\lambda_{r\a}^a$'s
at the one-loop level are given by
\begin{align}
&\langle Z^{a}(x)\bar Z^{b}(y)\rangle=
-\frac{1}{2}\left(\frac{\gym^2}{4\pi^2}\right)^2N\delta_{ab}
\frac{\ln(x-y)^2\Lambda^2}{(x-y)^2},\label{bosonSE}\\
&\langle \lambda^{a}_{r\a}(x)\bar\lambda_{s\dot\a}^{b}(y)\rangle
=-\left(\frac{\gym^2}{4\pi^2}\right)^2 N\delta_{ab}\epsilon_{rs}
(\sigma^{\mu})_{\a\dot\a}
\partial_{x^\mu}\left(\frac{1}{(x-y)^2}\right)\ln(x-y)^2\Lambda^2.
\label{fermionSE}
\end{align}
Hereafter, we focus only on the log divergent part and
omit the other terms.
The net log divergence from the self-energy 
in the planer limit
is obtained by replacing one of the free bosonic or fermionic propagator 
with (\ref{bosonSE}) or (\ref{fermionSE}), respectively,
for each possible free contraction.
Then we obtain
\begin{align}
&C_{SE}=\sum_{l=0}^{J}q_{l}\bar r_{l}
\big(-\frac{1}{2}\times(J+1)+(-1)\times1\big)\delta_{ij}\epsilon_{rs}
\Sigma,
\end{align}
where, $q$, $r$, and $\Sigma$ are defined as
\begin{align}
&\Sigma\equiv\frac{\gym^2 N}{4\pi^2}\left(\frac{N}{2}
\frac{\gym^2}{4\pi^2}\right)^{N+2}
\frac{1}{(x-y)^{2(J+1)}}
(\sigma^{\mu})_{\a\dot\a}
\partial_{x^\mu}\left(\frac{1}{(x-y)^2}\right)\ln(x-y)^2\Lambda^2.
\end{align}

The contribution from the $4Z$-interaction 
is obtained by replacing the two adjacent
free propagators of $Z$ and $\bar Z$ with the 4-point correlator
calculated at one-loop order such as
\begin{align}
\langle Z^{a_1}(x)Z^{a_2}(x)\bar Z^{b_1}(y)\bar Z^{b_2}(y)\rangle
&=\left(\frac{\gym^2}{4\pi^2}\right)^3
(f_{pa_1b_1}f_{pa_2b_2}+f_{pa_1b_2}f_{pa_2b_1})
\frac{\ln(x-y)^2\Lambda^2}{(x-y)^4}.
\end{align}
Apart from the phase factor,
the log divergence from one possible contraction
through the $4Z$-interaction is given by
\renewcommand{\h}{3cm}
\begin{align}
&\frac{1}{2}\left(\frac{N}{2}\frac{\gym^2}{4\pi^2}\right)^{J-1}
\frac{1}{(x-y)^{2(J-1)}}
\delta_{ij}\epsilon_{rs}\sigma^{\mu}_{\a\dot\a}
\partial_{x^\mu}\left(\frac{\gym^2}{4\pi^2(x-y)^2}\right)
\Tr(T^{a_1}T^{a_2}T^{b_2}T^{b_1})\notag\\
&\times \left(\frac{\gym^2}{4\pi^2}\right)^3
(f_{pa_1b_1}f_{pa_2b_2}+f_{pa_1b_2}f_{pa_2b_1})
\frac{\ln(x-y)^2\Lambda^2}{(x-y)^4}\notag\\
&=\frac{1}{2}\delta_{ij}\epsilon_{rs}\Sigma,
\end{align}
where we have used the formula
\begin{align}
f_{pa_1b_2}f_{pa_2b_1}\Tr(T^{a_1}T^{a_2}T^{b_1}T^{b_2})
&=N\left(\frac{N}{2}\right)^3\big(1-\frac{1}{N^2}\big)\label{ffTTTT1}\\
f_{pa_1b_1}f_{pa_2b_2}\Tr(T^{a_1}T^{a_2}T^{b_1}T^{b_2})&=0.\label{ffTTTT2}
\end{align}
When $\phi$ and $\lambda$ sit adjacent to each other, there are
$J-1$ ways of insertion of the $4Z$-interaction, otherwise
$J-2$ ways. Therefore,
with the phase factor appropriately taken into account,
the one-loop order contribution from $4Z$-interaction is
\begin{align}
&C_{4Z}=\left((q_0\bar r_0+q_{J}\bar r_{J})(J-1)
+\sum_{l=1}^{J-1}q_l\bar r_l(J-2)\right)
\times\frac{1}{2}\delta_{ij}\epsilon_{rs}\Sigma.
\end{align}

\begin{figure}
\begin{center}
\begin{minipage}{5cm}\vs{-.2cm}
\begin{align}
&\psfrag{a}{$\phi^{i}$}
\psfrag{b}{\hs{-.1cm}$Z$}
\psfrag{d}{\hs{-.1cm}$Z$}\psfrag{e}{}\psfrag{f}{}
\psfrag{g}{\hs{-.1cm}$Z$}\psfrag{h}{$Z$}
\psfrag{i}{\hs{-.1cm}$\phi^{i}$}
\psfrag{j}{\hs{-.1cm}$\bar Z$}
\psfrag{l}{\hs{-.2cm}$\bar Z$}
\psfrag{m}{}\psfrag{n}{}
\psfrag{o}{\hs{-.1cm}$\bar Z$}\psfrag{p}{\hs{-.1cm}$\bar Z$}
\psfrag{z}{$l$}
\psfrag{w}{\hs{-.5cm}$J-1-l$}
\includegraphics[height=\h]{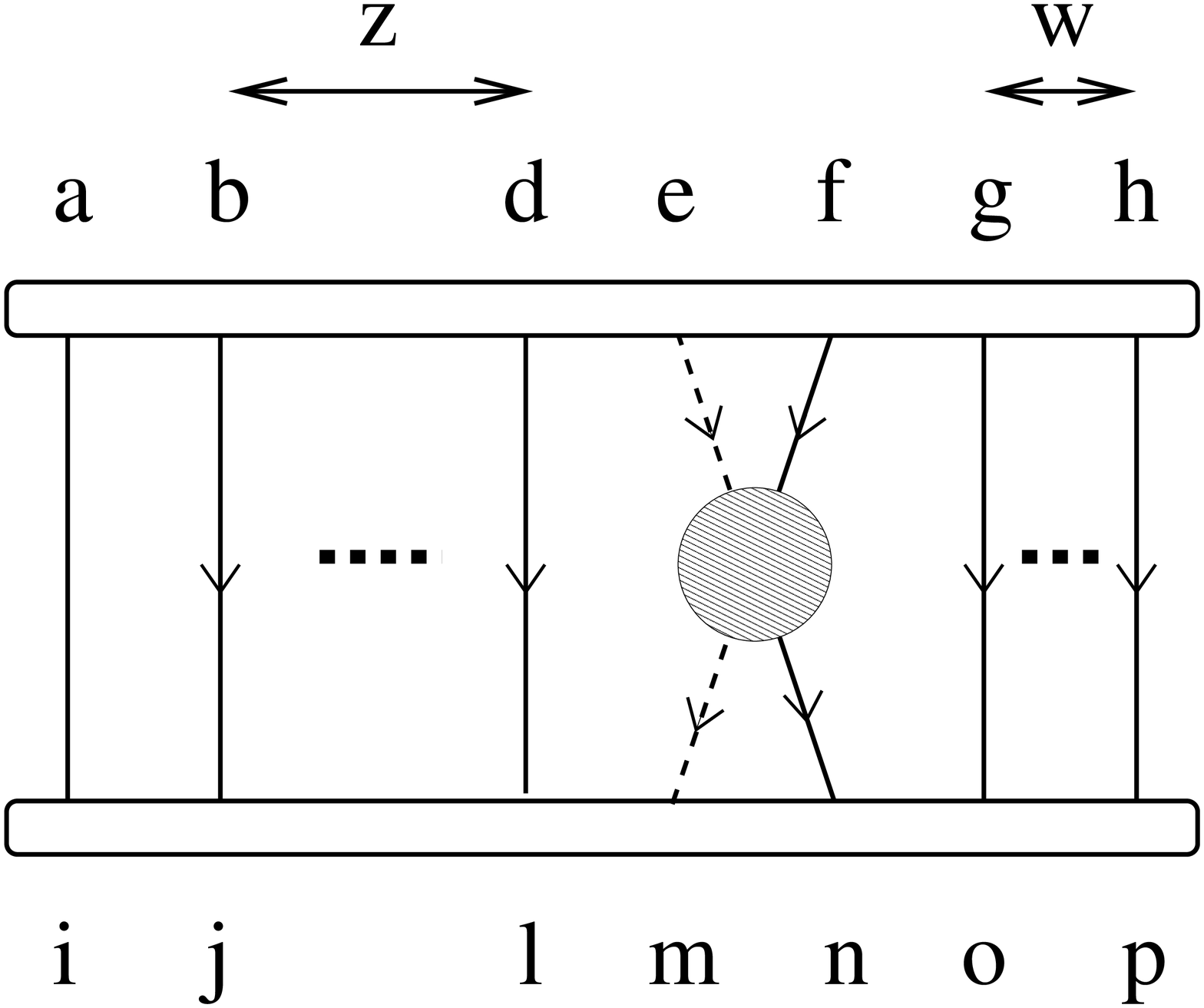}\notag\\[-.2cm]
&\hs{1.7cm}q_l\bar r_l\notag
\end{align}
\end{minipage}
\begin{minipage}{1cm}
\vs{1cm}
\end{minipage}
$\times\hs{.6cm}\Big(\hs{.2cm}$
\begin{minipage}{7cm}\vs{.5cm}
\renewcommand{\h}{2cm}
\renewcommand{\w}{1.2cm}
\begin{align}
&\begin{minipage}{\w}
\psfrag{a}{\hs{-.1cm}$\lambda$}
\psfrag{b}{\hs{-.1cm}$Z$}
\psfrag{c}{\begin{minipage}{.5cm}\vs{.2cm}\hs{-.1cm}$\bar \lambda$\end{minipage}}
\psfrag{d}{\begin{minipage}{.5cm}\vs{.2cm}\hs{-.1cm}$\bar Z$\end{minipage}}
\includegraphics[height=\h]{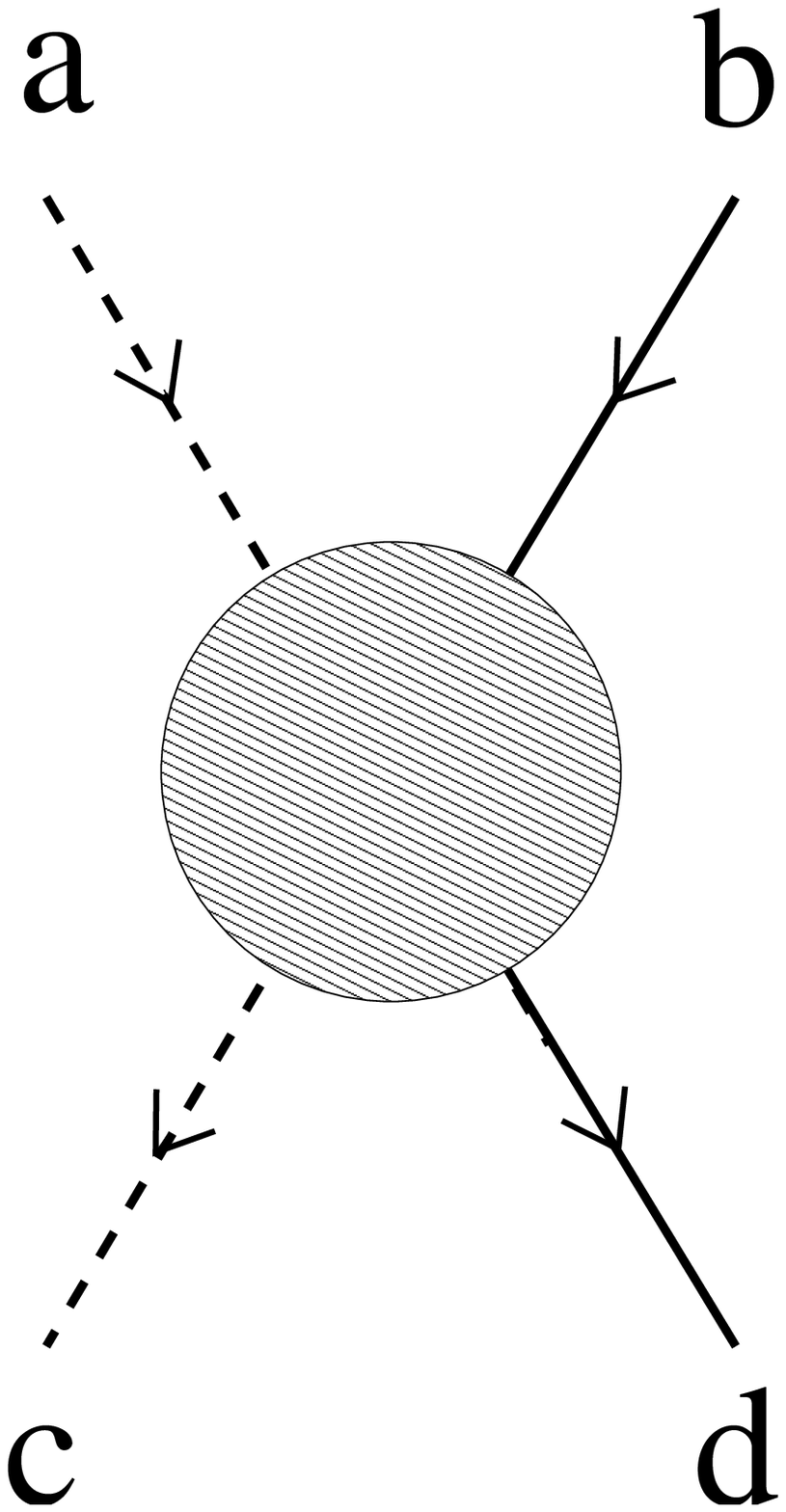}
\end{minipage}
+\begin{minipage}{\w}
\psfrag{a}{\hs{-.1cm}$\lambda$}
\psfrag{b}{\hs{-.1cm}$Z$}
\psfrag{d}{\begin{minipage}{.5cm}\vs{.2cm}\hs{-.1cm}$\bar \lambda$\end{minipage}}
\psfrag{c}{\begin{minipage}{.5cm}\vs{.2cm}\hs{-.1cm}$\bar Z$\end{minipage}}
\includegraphics[height=\h]{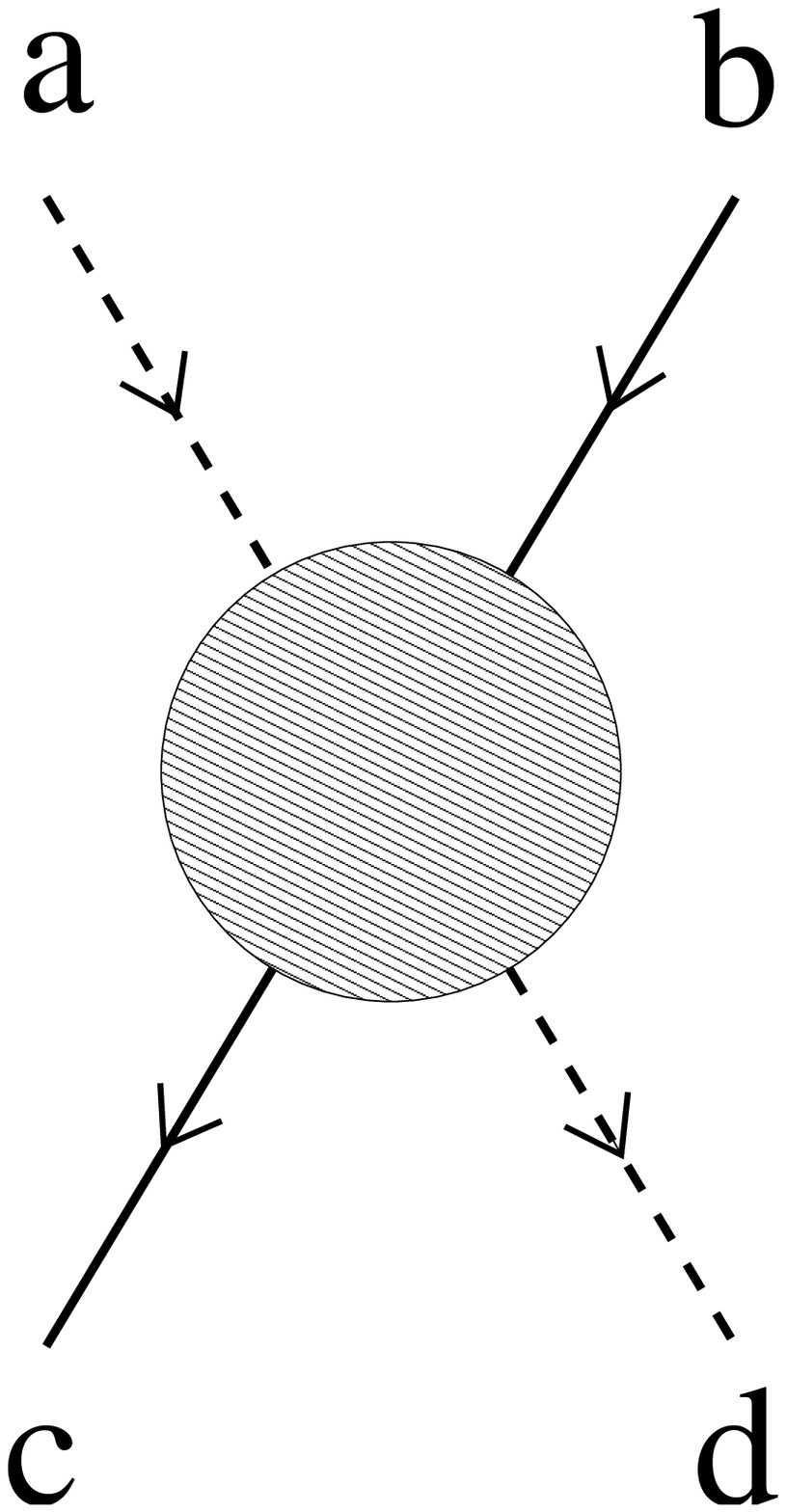}
\end{minipage}
\psfrag{b}{\hs{-.1cm}$\lambda$}
\psfrag{a}{\hs{-.1cm}$Z$}
\psfrag{c}{\begin{minipage}{.5cm}\vs{.2cm}\hs{-.1cm}$\bar \lambda$\end{minipage}}
\psfrag{d}{\begin{minipage}{.5cm}\vs{.2cm}\hs{-.1cm}$\bar Z$\end{minipage}}
+\begin{minipage}{\w}
\includegraphics[height=\h]{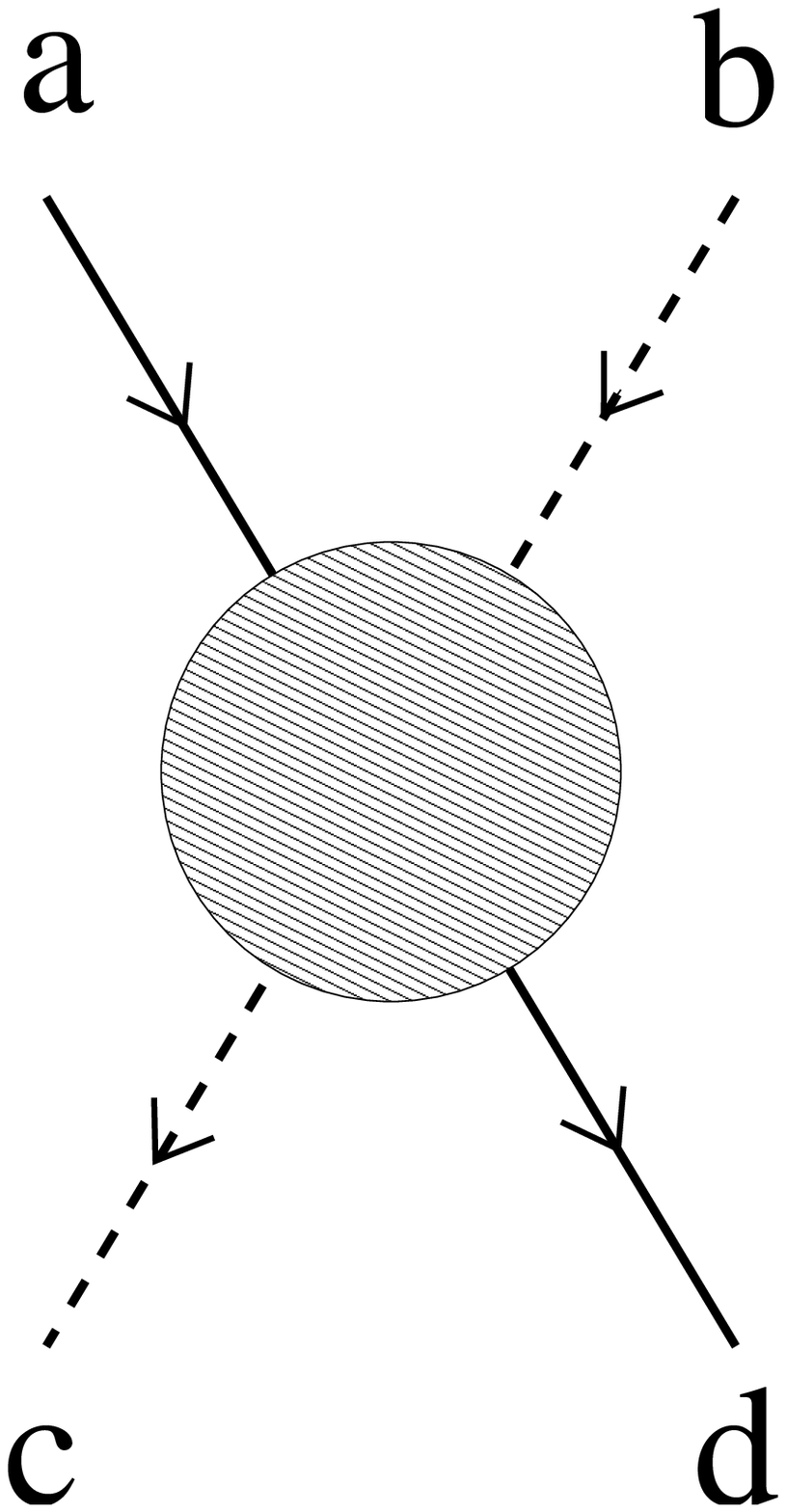}
\end{minipage}
\psfrag{d}{\begin{minipage}{.5cm}\vs{.2cm}\hs{-.1cm}$\bar \lambda$\end{minipage}}
\psfrag{c}{\begin{minipage}{.5cm}\vs{.2cm}\hs{-.1cm}$\bar Z$\end{minipage}}
+\begin{minipage}{1.8cm}
\includegraphics[height=\h]{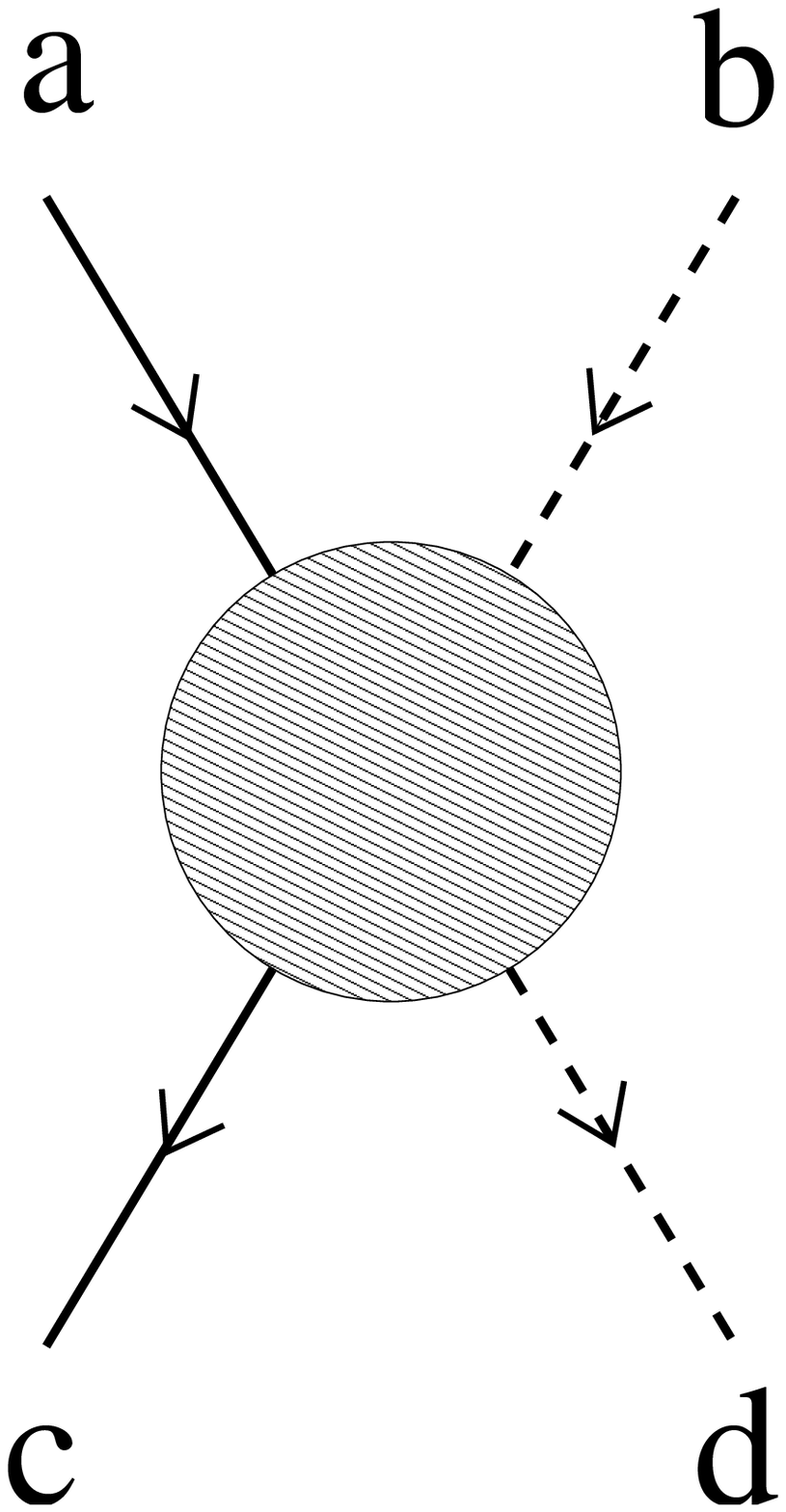}
\end{minipage}\notag\\[.2cm]
&\hs{.4cm}1\hs{1.5cm}\bar r\hs{1.5cm}q\hs{1.5cm}q\bar r\notag
\end{align}
\end{minipage}$\hs{-.3cm}\Big)$
\end{center}
\caption{$2\lambda Z$-interaction}\label{2lambdaZ}
\end{figure}

In the case of $2\lambda Z$-interaction, 
the weight of the log divergence is different
according to the relative position of $\lambda$ and $Z$,
since 
the correlator of $2\lambda Z$-interaction is 
given by
\begin{align}
\label{2LZ}
&\langle
\lambda_{r\a}^{a_1}(x)
Z^{a_2}(x)
\bar\lambda_{s\dot\a}^{b_1}(y)
\bar Z^{b_2}(y)
\rangle
=\left(\frac{\gym^2}{4\pi^2}\right)^3
\epsilon_{rs}
\left(
f_{pa_1b_2}f_{pa_2b_1}+
\frac{1}{2}f_{pa_1b_1}f_{pa_2b_2}
\right)\notag\\
&\hs{5.5cm}\times\frac{1}{(x-y)^2}\sigma^\mu_{\a\dot\a}
\partial_{x^\mu}
\left(\frac{1}{(x-y)^2}\right)\ln(x-y)^2\Lambda^2.
\end{align}
In this case, there are 4 types of configuration of 
$\lambda$ and $Z$
for each possible replacement as depicted in Figure \ref{2lambdaZ}.
The log divergence from the 
crossing configurations of $\lambda$ and $Z$,
which are depicted in Figure \ref{2lambdaZ} 
with the relative phase dependence $q$ or $\bar r$,
is obtained by contracting (\ref{2LZ}) with 
$\Tr(T^{a_1}T^{a_2}T^{b_1}T^{b_2})$, giving
$1/2\delta_{ij}\epsilon_{rs}\Sigma$, while the others 
obtained by contracting (\ref{2LZ}) with
$\Tr(T^{a_1}T^{a_2}T^{b_2}T^{b_1})$,
giving $1/4\delta_{ij}\epsilon_{rs}\Sigma$.
Here, we have taken into account the free contractions 
of remaining operators.
Then, the total contribution 
from $2\lambda Z$-interaction is
\begin{align}
C_{2\lambda Z}=
\sum_{l=0}^{J-1}q_{l}\bar r_l
\times\Big(\frac{1}{4}(1+q\bar r)+\frac{1}{2}(\bar
 r+q)\Big)\delta_{ij}\epsilon_{rs}\Sigma,
\end{align}
where $q=\exp(2\pi im/(J+2))$ and $r=\exp(2\pi in/(J+2))$,
respectively.

The correlator of the $2\phi Z$-interaction is
\begin{align}
&\langle 
\phi^{ia_1}(x)Z^{a_2}(x)\phi^{jb_1}(y)\bar Z^{b_2}(y)
\rangle
=\left(\frac{\gym^2}{4\pi^2}\right)^3\delta_{ij}
f_{pa_1b_2}f_{pa_2b_1}
\frac{\ln(x-y)^2\Lambda^2}{(x-y)^4},
\end{align}
From this form of the 4-point correlator
and the formula (\ref{ffTTTT2}),
we can easily see that only the crossing configuration 
of $\phi$ and $Z$ gives non-zero contribution.
So, we obtain, including the free contractions of the other operators,
the log divergence from the $2\phi Z$-interaction,
\begin{align}
&C_{2\phi Z}=\sum_{l=0}^{J-1}q_l\bar r_{l}
\times(r+\bar q)\frac{1}{2}
\delta_{ij}\epsilon_{rs}\Sigma.
\end{align}

The last contribution we should consider in the $(++)$ sector
is the $2\phi\lambda$-interaction, 
which occurs only when $\phi$ and $\lambda$ impurities 
sit adjacent to each other.
Since the four point correlation function is given by
\begin{align}
&\langle
\lambda_{r\a}^{a_1}(x)
\phi^{ia_2}(x)
\bar\lambda^{b_1}_{s\dot\a}(y)
\phi^{ib_2}(y)
\rangle\\
&=\frac{1}{2}
\left(\frac{\gym^2}{4\pi^2}\right)^3
\left(
(-\tau^j\bar\tau^i)_{rs}f_{pa_1b_2}f_{pa_2b_1}
+\frac{1}{2}(\tau^i\bar\tau^j)_{rs}f_{pa_1a_2}f_{pb_1b_2}
+\delta_{ij}\epsilon_{rs}f_{pa_1b_1}f_{pa_2b_2}
\right)\notag\\
&\hspace{8cm}\times\frac{1}{(x-y)^2}\sigma^\mu_{\a\dot\a}
\partial_{x^\mu}
\left(\frac{1}{(x-y)^2}\right)\ln(x-y)^2\Lambda^2,\notag
\end{align} 
we can easily see that the contribution from $2\phi\lambda$-interaction is
\begin{align}
&C_{2\phi\lambda}=
(q_0\bar r_0+q_{J}\bar r_{J})\Big(
\frac{1}{4}\delta_{ij}\epsilon_{rs}+\frac{1}{8}(\tau_i\tau_j)_{rs}
\Big)\Sigma\notag\\
&\qquad+(q_0\bar r_{J}+q_{J}\bar r_0)\Big(
\frac{1}{2}\delta_{ij}\epsilon_{rs}+\frac{1}{8}(\tau_i\tau_j)_{rs}
\Big)\Sigma.
\end{align}

Combining all the results so far,
the net contribution from one-loop interactions
in the $(++)$ sector of the two point function 
of the operators $O_{r\a}^{J,i}$ and $\bar O_{s\dot\a}^{J,i}$  is
\begin{align}
& C_{SE}+C_{4Z}+C_{2\lambda Z}+C_{2\phi Z}+C_{2\phi\lambda}\notag\\
&=\left\{
\begin{matrix}
J\left(\frac{2\pi in}{J}\right)^2\delta_{ij}\epsilon_{rs}\Sigma
+\frac{1}{2}(\tau^{i}\bar\tau^{j})_{rs}\Sigma
+{\cal O}(1/J^2),
&(m=n)\\
+\frac{1}{2}(\tau^{i}\bar\tau^{j})_{rs}\Sigma+{\cal O}(1/J^2),
&(m\ne n)
\end{matrix}
\right.\label{result0}
\end{align}
which shows that the two point function does not 
take the canonical form,
therefore the operator (\ref{O}) cannot be the correspondent
of the $\a^{i}_{n}\b_{\r\a-n}\ket{v}$.
Since the ${\cal O}(1)$ factor $(\tau^i\bar\tau^j)_{rs}$ 
in (\ref{result0})
breaks the canonical form,
the basic requirement for the BMN conjecture
is not satisfied 
that the anomalous dimension 
to the first order of $\lambda'\equiv{\gym^2N}/{J^2}$
should match the first order perturbation expansion 
of the free string energy spectrum in terms of $1/\mu^2$,
under the identification of $\lambda'=1/(\mu p^+\a')^2$.

The extra term in (\ref{result0}) 
is eliminated by choosing $b$ properly.
In the $(-+)$  sector,
there appears the contribution from the 
4-point correlator 
\begin{align}
\langle
\lambda_{r\a}^{a_1}(x)\phi^{ia_2}(x)
\bar\theta_{\dot r\dot\a}^{b_1}(y)\bar Z^{b_2}(y)
\rangle
&=\frac{1}{\sqrt{2}}
\left(\frac{\gym^2}{4\pi^2}\right)^{3}
(\tau^{i})_{r\dot r}
(f_{pa_1b_2}f_{pa_2b_1}+\frac{1}{2}f_{pa_1a_2}f_{pb_1b_2})\notag\\
&\times\frac{1}{(x-y)^2}
(\sigma^{\mu})_{\a\dot\a}\partial_{x^\mu}
\left(\frac{1}{(x-y)^2}\right)\ln(x-y^2)\Lambda^2.\label{LPTZ}
\end{align}
With the formula (\ref{ffTTTT1}) and (\ref{ffTTTT2}), 
we can easily see that
each of 4 possible configurations
of $\lambda(x)$, $\phi(x)$, $\bar\theta(y)$ and $\bar Z(y)$
gives the same contribution, $\sqrt{2}/8(\tau^i)_{r\dot r}\Sigma$,
except for a phase factor. The same is true for the $(-+)$ sector.
Then, the log divergences from the $(+-)$ and $(-+)$ sectors 
with appropriate phase dependences, are given by
\begin{align}
b(q_0+q_{J})\frac{1}{2\sqrt{2}}(\tau^i\bar \tau^j)_{rs}\Sigma
+\bar b(\bar r_{0}+\bar r_{J})\frac{1}{2\sqrt{2}}(\tau^i\bar \tau^j)_{rs}\Sigma.
\end{align}

As for the $(--)$ sector, in addition to the 
self-energy contribution, 
\begin{align}
-|b|^2
\left(-\frac{1}{2}\times(J+1)-1\times1\right)(\tau^i\bar\tau^j)_{rs}\Sigma,
\label{--1}
\end{align}
where $-1/2$ comes from the self-energy of a boson propagator and $-1$ from
fermion,
and the $4Z$-interaction contribution,
\begin{align}
-|b|^2\frac{1}{2}J (\tau^i\bar\tau^j)_{rs}\Sigma,\label{--2}
\end{align}
we have to consider the $2\theta Z$-interaction
which gives the log divergence,
\begin{align}
\langle
\theta^{a_1}_{\dot r\a}(x)Z^{a_2}(x)
\bar\theta^{b_1}_{\dot s\dot\a}(y)\bar Z^{b_2}(y)
\rangle
&=-\frac{1}{2}\left(\frac{\gym^2}{4\pi^2}\right)^3
(f_{pa_1a_2}f_{pb_1b_2}-f_{pa_1b_1}f_{pa_2b_2})\notag\\
&\times\frac{1}{(x-y)^2}\epsilon_{\dot r\dot s}
(\sigma^{\mu})_{\a\dot\a}\partial_{x^\mu}
\left(\frac{1}{(x-y)^2}\right)\ln(x-y)^2\Lambda^2.
\end{align}
We have non-vanishing contribution only when 
$\theta$ and $Z$ sit in the crossing position,
and therefore the $\theta Z$-interaction gives the log divergence such as
\begin{align}
 -|b|^22\times\frac{1}{4}(\tau^i\bar\tau^j)_{rs}\Sigma.\label{--3}
\end{align}
From (\ref{--1}), (\ref{--2}) and (\ref{--3}),
the total contribution in the $(--)$ sector is
\begin{align}
|b|^2(\tau^i\bar\tau^j)_{rs}\Sigma.
\end{align}

Combining all the contributions 
form the $(++)$, $(-+)$, $(+-)$ and $(--)$ sectors,
the total log divergence is given by
\begin{align}
\left\{\begin{matrix}
J\left(\frac{2\pi in}{J}\right)^2\delta_{ij}\epsilon_{rs}\Sigma
+\frac{1}{2}(1+\sqrt{2}(b+\bar b)+2|b|^2)
(\tau^{i}\bar\tau^{j})_{rs}\Sigma
+{\cal O}(1/J^2)
&(m=n)\\
+\frac{1}{2}(1+\sqrt{2}(b+\bar b)+2|b|^2)
(\tau^{i}\bar\tau^{j})_{rs}\Sigma+{\cal O}(1/J^2)
&(m\ne n)
\end{matrix}\right.
\end{align}
from which we conclude that we should choose 
\begin{align}
 b=-\frac{1}{\sqrt{2}},
\end{align}
in order that the two point function of the operator $O_{rs}^{J,i}$ 
takes the canonical form.

Here we should comment on the choice of the phase factor of 
the operator $O^{J,i}_{r\a,m}$ in (\ref{linear_comb}).
In fact, we can obtain the operators of definite conformal dimension
which are ``diagonalized'' up to ${\cal O}(1/J^2)$,
even if we adopt the other phase dependence,
for example, $\exp(2\pi iml/(J+1))$ or $\exp(2\pi iml/J)$.
However, according to \cite{PR, Gursoy},
we have chosen the phase factor so that the level matching 
condition for the world sheet momentum of the closed string 
is naturally realized when we regard the BMN operator
$O^{J,i}_{rm}$ is obtained by acting the rotation in the R-symmetry
space and the SUSY transformation on the vacuum state $\Tr(Z^{J+2})$
accompanied by the phase factor $q$ such as
\begin{align}
 \delta\left(O_1(x)O_2(x)\cdots O_n(x)\right)
=(\delta O_1(x))O_2(x)\cdots O_n(x)+qO_1(x)(\delta O_2(x))\cdots O_n(x)\notag\\
+\cdots+q^{n-1}O_{1}(x)O_{2}(x)\cdots (\delta O_{n}(x)).
\end{align}
With this prescription, we have
\begin{align}
 \delta_2\delta_1 \Tr(Z^2)
=\sum_{l=0}^{J+1}(q_1q_2)^l
\left[
\sum_{l=0}^{J}q_2^{l+1}\Tr((\delta_1Z)Z^l(\delta_2Z)Z^{J-l})
+\Tr((\delta_2\delta_1Z)Z^{J+1})
\right].
\end{align}
Then, the level matching condition is realized 
when we choose $q_i$ such as $q_i=\exp(-2\pi im_i/(J+2))\,(i=1,2)$.
As explained at the end of the next subsection, 
this choice of the phase factor is necessary for our 
duality condition to hold.

\subsection{Three point function}

Having properly identified the operator which 
corresponds to the string state with one scalar and one fermion
excitation, 
$\a^i_{n}\b_{r\a,-n}\ket{v}$,
we proceed to calculate the three point functions
which correspond to (\ref{stringAmp1}) and (\ref{stringAmp2}).

\subsubsection{$\vac \to 
\alpha^{(2)i}_{n_2}\beta^{(2)}{\a_1\a_2,-n_2} 
+ \alpha^{(3)j}_{n_3}\beta^{(3)}_{\b_1\b_2,-n_3}$}

\begin{figure}[t]
\begin{align}
\renewcommand{\h}{5cm}\notag
\psfrag{a}{\hs{-.3cm}$\lambda_{r\a}$}
\psfrag{c}{$\lambda_{s\b}$}
\psfrag{b}{\hs{-.1cm}$\phi^i$}
\psfrag{d}{\hs{-.1cm}$\phi^j$}
\psfrag{f}{$Z$}\psfrag{h}{$Z$}
\psfrag{g}{$\bar Z$}\psfrag{i}{$\bar Z$}
\psfrag{e}{$\bar Z$}
\includegraphics[width=\h]{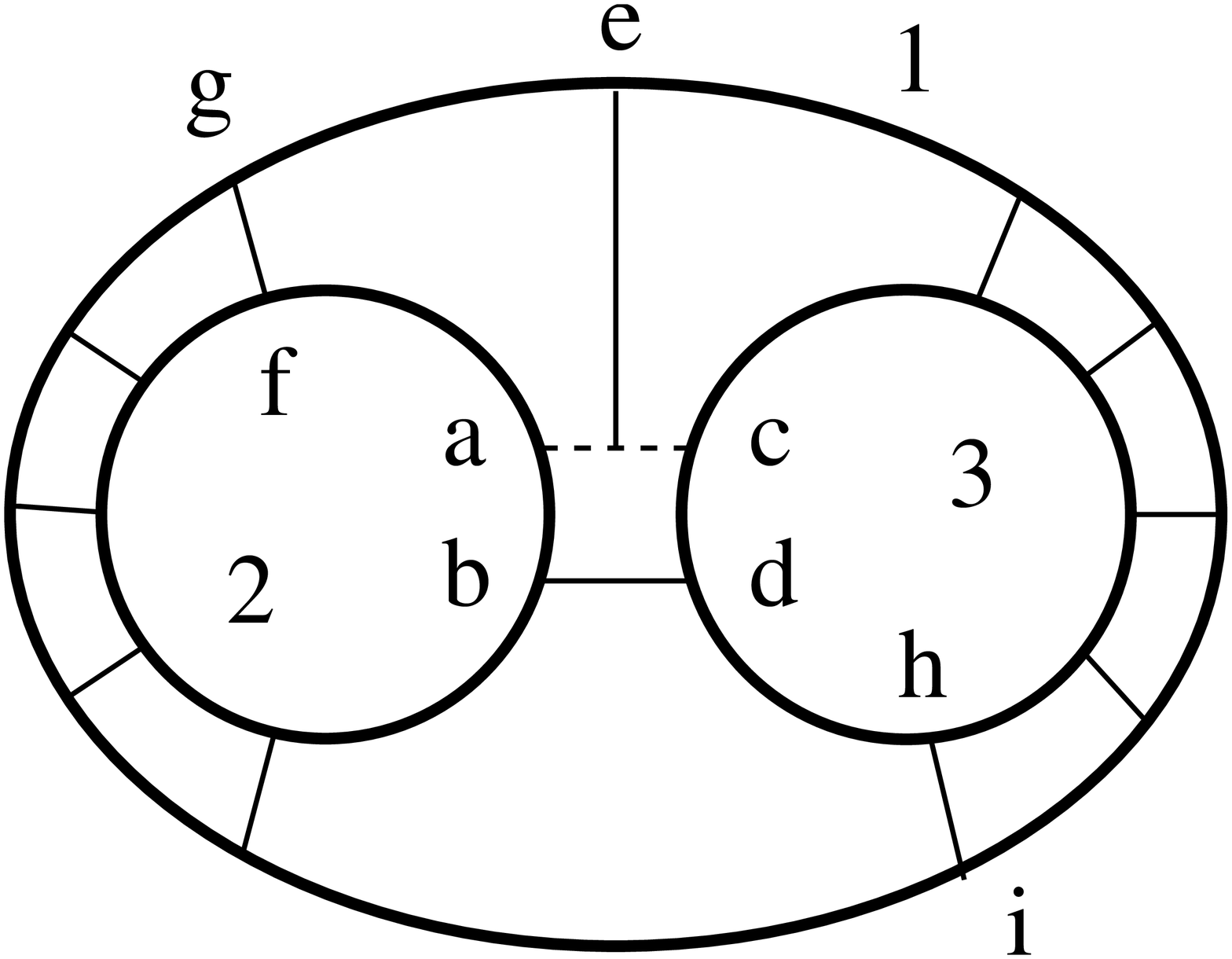}\hs{2cm}
\begin{minipage}[c]{\h}\vs{-3.2cm}\includegraphics[width=\h]{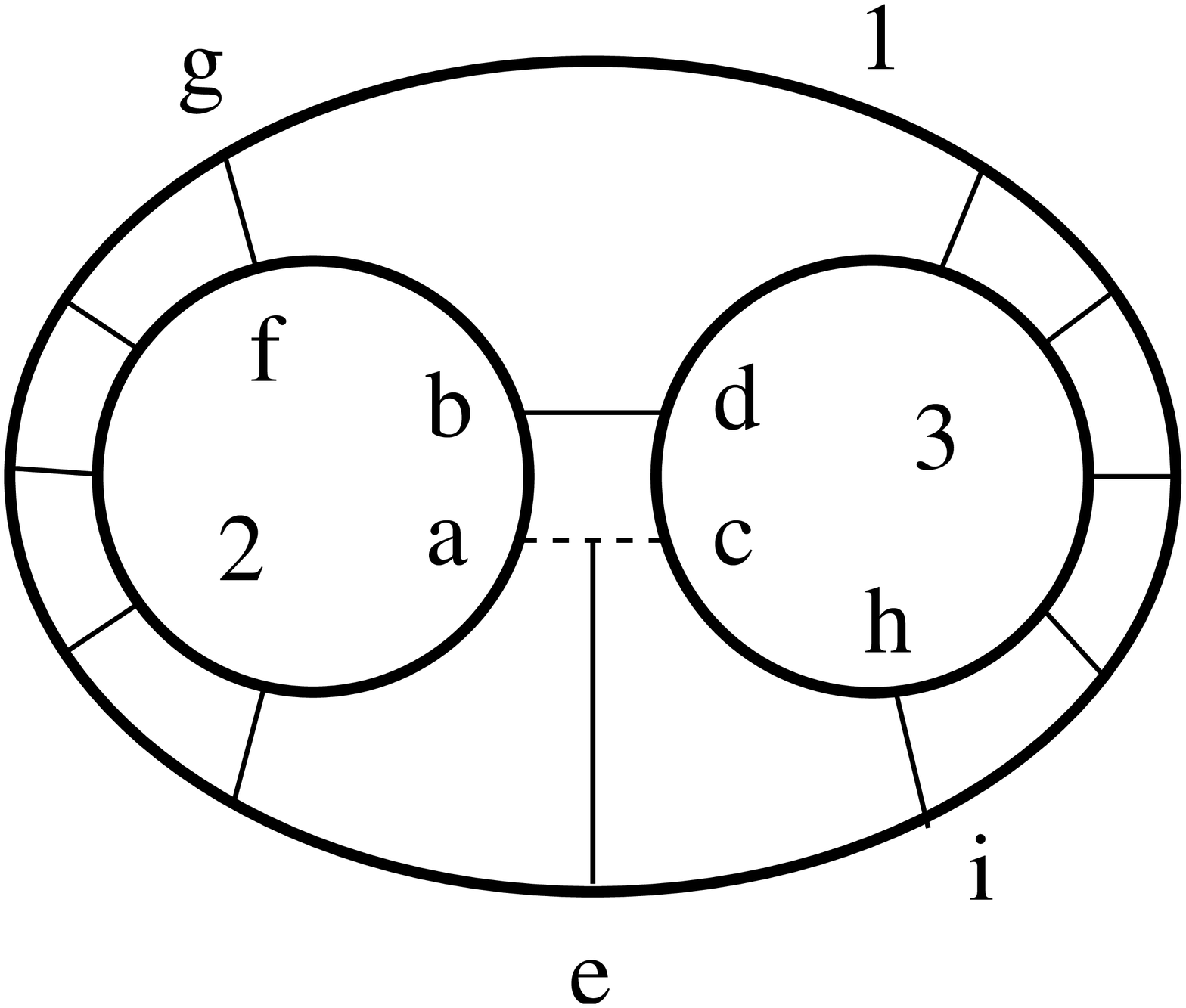}\end{minipage}
\end{align}
\caption{ $(++)$ sector of three point function}\label{pp}
\end{figure}

\begin{figure}[t]
\begin{align}
\renewcommand{\h}{4cm}\notag
\psfrag{a}{\hs{-.3cm}$\lambda_{r\a}$}
\psfrag{c}{$\theta_{\dot s\b}$}
\psfrag{b}{\hs{-.2cm}$\phi^i$}
\psfrag{d}{\hs{-.1cm}$\phi^j$}
\psfrag{f}{$Z$}\psfrag{h}{$Z$}
\psfrag{g}{$\bar Z$}\psfrag{i}{$\bar Z$}\psfrag{e}{}
\includegraphics[height=\h]{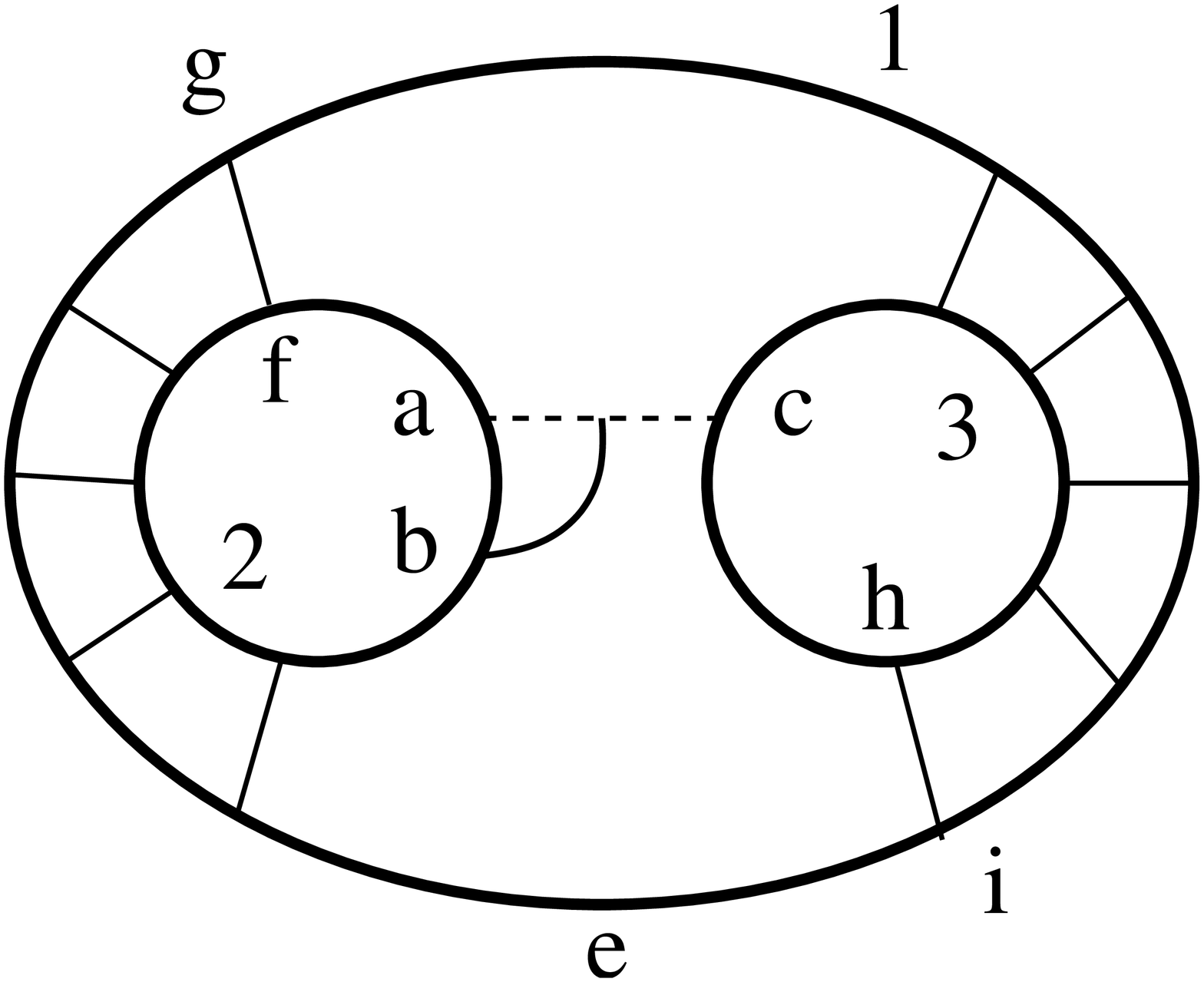}\hs{2cm}
\includegraphics[height=\h]{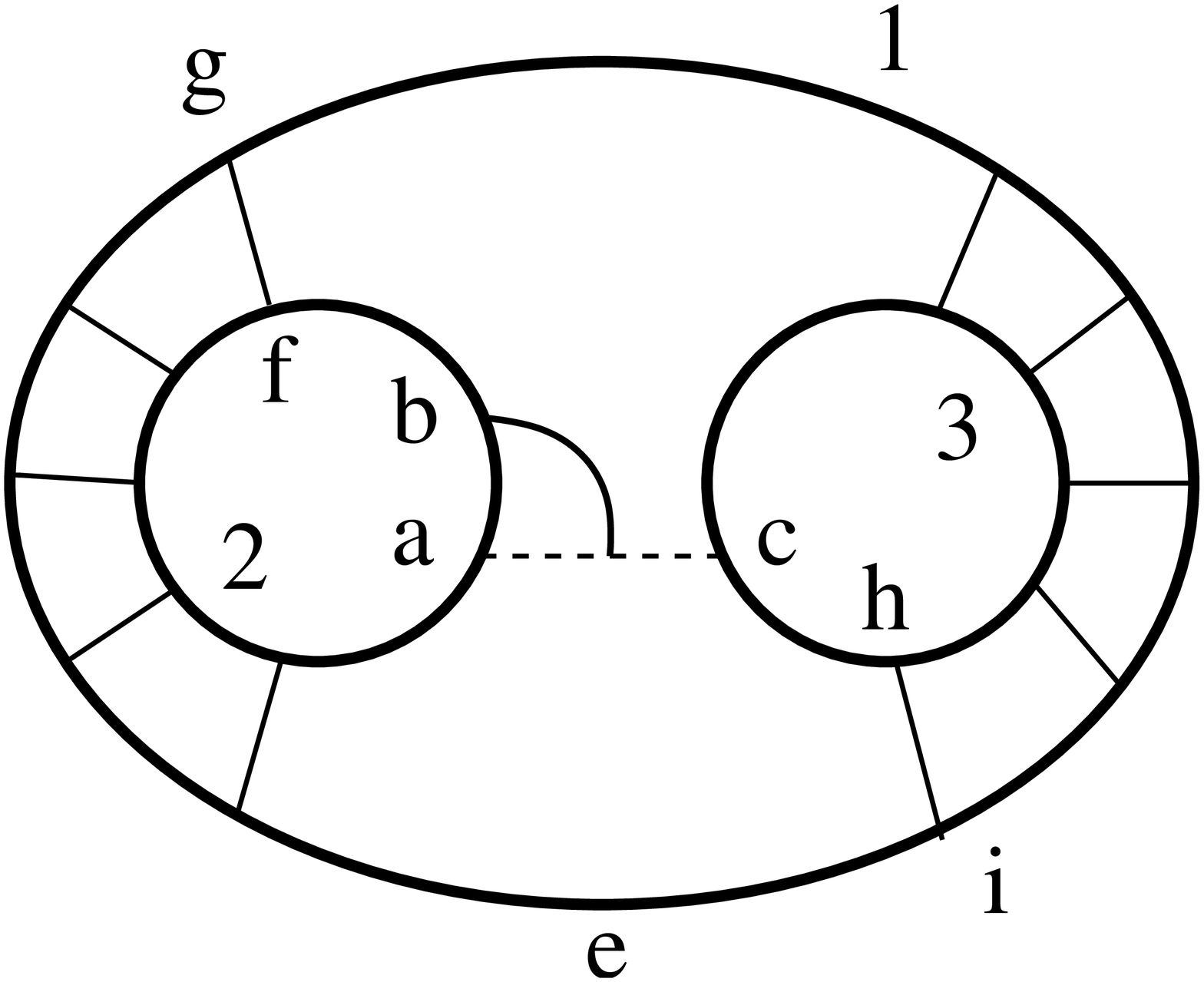}
\end{align}
\caption{ $(+-)$ sector of three point function}\label{pm}
\end{figure}

We first calculate the correspondent on the gauge theory side to 
the string  three point function (\ref{stringAmp1}), such as
\begin{align}
\langle\bar O^{J_1}_{vac}(x_1)O^{J_2,i}_{r\a,m}(x_2)O^{J_3,j}_{s\b,n}(x_3)\rangle.
\label{gaugeCol1}
\end{align}
The normalization factor of each operator is chosen as
\begin{align}
&O_{vac}^{J_1}=\frac{1}{\sqrt{(J_1+1)N_{0}^{J_1+1}}}\Tr(Z^{J_1+1}),
\quad N_0\equiv\frac{N}{2}\frac{\gym^2}{4\pi^2},\label{Ovac}\\
&O_{s\a,m}^{J_r,i}=
\frac{1}{\sqrt{2(J_r+1)N_0^{J_r+2}}}
\left(\sum_{l=0}^{J_r}
\e^{\frac{2\pi im(l+1)}{J_r+2}}
\Tr(\phi^iZ^l\lambda_{s\a}Z^{J_r-l})+
b(\tau^i)_s^{\dot s}
\Tr(\theta_{\dot s}Z^{J_r+1})
\right),\label{Ois}
\end{align}
so that
the two point function 
is normalized to $1/|x|^2$ for $O^{J}_{vac}$ and 
takes the canonical form (\ref{canonical}) for $O_{r\a,m}^{J,i}$.
The coefficient $b$ has been determined
in the previous subsection as $b=-1/{\sqrt{2}}$.
Similarly to the two point function, there are four 
sectors of
the three point function to be calculated:
three of them are depicted in the Figure
\ref{pp}, \ref{pm} and \ref{mm}, which we also call
$(++)$, $(+-)$ and $(--)$ sector, respectively,
and the remaining one is the $(-+)$ sector, 
for which we have omitted the figure.

Apart from the normalization factors in (\ref{Ovac}) and (\ref{Ois}),
the $(++)$ sector of the three point function (\ref{gaugeCol1})
at the leading order in terms of 
the genus and the gauge interaction expansions is given by
\begin{align}
\frac{1}{4}\left(\frac{N}{2}\right)^{J_1-1}
\left(\frac{\gym^2}{4\pi^2}\right)^{J_1+1}
\frac{(J_1+1)\delta_{ij}}{|x_{23}|^2|x_{12}|^{2J_2}|x_{31}|^{2J_3}}
\left(
q_{0}r_{J_3}\langle\Tr(\lambda_{s\b}\lambda_{r\a}\bar Z)\rangle
+q_{J_2}r_{0} \langle\Tr(\lambda_{r\a}\lambda_{s\b}\bar Z)\rangle
\right),
\end{align}
where the three point correlator in this expression 
is given by
\begin{align}
\label{integral}
&\langle \Tr(
\bar Z^{a_1}(x_1)
\lambda_{r\a}^{b_1}(x_2)
\lambda_{s\b}^{c_1}(x_3)
)\rangle\\
&=i\sqrt{2}
\frac{1}{\gym^2}
\left(\frac{\gym^2}{4\pi^2}\right)^3
f_{a_1b_1c_1}\epsilon_{rs}(\sigma^{\mu}\bar\sigma^{\nu})_{\a\b}
\int\frac{1}{(x_1-u)^2}
\partial_{u^{\mu}}\frac{1}{(x_2-u)^2}\cdot
\partial_{u^{\nu}}\frac{1}{(x_3-u)^2}.\notag
\end{align}
Integrating by part and using  the formula (\ref{delta}),
the symmetric part in terms of the spacetime indexes 
$\mu$ and $\nu$ in the integral can be evaluated as
\begin{align}
&\int d^4u
\frac{1}{(x_1-u)^2}
\partial_{u^{\mu}}\frac{1}{(x_2-u)^2}\cdot
\partial_{u_{\mu}}\frac{1}{(x_3-u)^2}\notag\\
&=
\frac{1}{2}
\int d^4u
\frac{1}{(x_1-u)^2}
\left[
\Box\left(\frac{1}{(x_2-u)^2}\frac{1}{(x_3-u)^2}\right)
\right.\\
&\hspace{4cm}\left.
-\Box\frac{1}{(x_2-u)^2}\cdot\frac{1}{(x_3-u)^2}
-\frac{1}{(x_2-u)^2}\Box\frac{1}{(x_3-u)^2}
\right]\notag\\
&=
-2\pi^2
\left(
\frac{1}{|x_{12}|^2|x_{31}|^2}
-\frac{1}{|x_{12}|^2|x_{23}|^2}
-\frac{1}{|x_{23}|^2|x_{31}|^2}
\right),\notag
\end{align}
Therefore, with the relations, 
$(\sigma^{\mu}\sigma^{\nu})_{\a\b}=-\eta_{\mu\nu}\epsilon_{\a\b}+(\sigma^{\mu\nu})_{\a\b}$,
where $(\sigma^{\mu\nu})_{\a\b}\equiv
(\sigma^\mu\bar\sigma^{\nu}-\sigma^{\nu}\bar\sigma^{\mu})_{\a\b}/2$,
and 
$f_{a_1b_1c_1}\Tr(T^{a_1}T^{b_1}T^{c_1})=iN^3/4+{\cal O}(N^2)$,
the $(++)$ sector of the three point function is given by
\begin{align}
\label{result-of-pp}
&\sqrt{2}\pi i\left(\frac{m}{y}-\frac{n}{1-y}\right)
\left(\frac{N}{2}\frac{\gym^2}{4\pi^2}\right)^{J_1+2}\delta_{ij}\epsilon_{rs}
\frac{1}{|x_{12}|^{2J_2}|x_{31}|^{2J_3}|x_{23}|^2}\notag\\
&\times\left[\epsilon_{\a\b}
\left(
\frac{1}{|x_{12}|^2|x_{31}|^2}
-\frac{1}{|x_{12}|^2|x_{23}|^2}
-\frac{1}{|x_{23}|^2|x_{31}|^2}
\right)\right.
\\
&\qquad\left.+(\sigma^{\mu\nu})_{\a\b}
\int d^4u
\frac{1}{(x_1-u)^2}
\partial_{u^{\mu}}\frac{1}{(x_2-u)^2}\cdot
\partial_{u^{\nu}}\frac{1}{(x_3-u)^2}\right].\notag
\end{align}
The world sheet momentum dependence
comes from
\begin{align}
&q_{1}r_{J_3+1}-q_{J_2+1}r_{1}=\notag\\
&\exp\left(\frac{2\pi i m}{J_2+2}\right)
\exp\left(\frac{2\pi i n(J_3+1)}{J_3+2}\right)
-\exp\left(\frac{2\pi i m(J_2+1)}{J_2+2}\right)
\exp\left(\frac{2\pi i n}{J_3+2}\right)\notag\\
&\sim 4\pi iJ_1\left(\frac{m}{y}-\frac{n}{1-y}\right)\label{phase-dependence}
\end{align}
where $y=-\a_{(2)}/\a_{(1)}=J_2/J_1$ and $1-y=-\a_{(3)}/\a_{(1)}=J_3/J_1$ 
and we take the leading term in the large $J$ limit.
We should note that this momentum dependence cannot 
be obtained if we choose 
the phase factor such as $\sum_{l=0}^{J}\exp(2\pi i
n/J)\Tr(\phi^iZ^l\lambda Z^{J-l})$ often adopted in the literature.

\begin{figure}
\begin{align}
\renewcommand{\h}{4cm}\notag
\psfrag{a}{\hs{-.3cm}$\theta_{\dot r\a}$}
\psfrag{c}{$\theta_{\dot s\b}$}
\psfrag{b}{\hs{-.1cm}$Z$}
\psfrag{f}{}\psfrag{h}{}
\psfrag{g}{}\psfrag{i}{}\psfrag{e}{}
\begin{minipage}{6cm}
\begin{center}
\includegraphics[height=\h]{pm1.eps}
\end{center}
\end{minipage}
+
\begin{minipage}{6cm}
\begin{center}
\includegraphics[height=\h]{pm2.eps}
\end{center}
\end{minipage}
=0
\end{align}
\caption{ $(--)$ sector of three point function}\label{mm}
\end{figure}
As in the case of two point function, the $(+-)$ and $(-+)$ sectors
are indispensable for our duality relation,
while the $(--)$ sector depicted in Figure \ref{mm} gives a vanishing results
since it does not contain a phase factor.
The $(+-)$ sector of the three point function depicted 
in Figure \ref{pm} is 
\begin{align}
&\frac{1}{4}\left(\frac{N}{2}\right)^{J_1-1}
\left(\frac{\gym^2}{4\pi^2}\right)^{J_1+1}
\frac{(J_1+1)}{|x_{12}|^{2J_2}|x_{31}|^{2J_3+2}}(\tau^j)_{s}^{\dot s}
\notag\\
&\times\left(
\langle\Tr(\phi^i(x_2)\lambda_{r\a}(x_2)\theta_{\dot\s\b}(x_3))\rangle
+q_{J}\langle\Tr(\lambda_{r\a}(x_2)\phi^i(x_2)\theta_{\dot\s\b}(x_3))\rangle
\right).
\end{align}
The perturbation calculation to the one-loop order gives 
\begin{align}
\label{integral2}
&\langle\Tr(
\phi^{a_1i}(x_2)\lambda_{r\a}^{b_1}(x_2)\theta_{\dot\s\b}^{c_1}(x_3))\rangle\\
&=
\frac{1}{\gym^2}
\left(\frac{\gym^2}{4\pi^2}\right)^3
if_{a_1b_1c_1}(\tau^i)_{r\dot s}(\sigma^{\mu}\bar\sigma^{\nu})_{\a\b}
\int d^4u\frac{1}{(x_2-u)^2}
\partial_{u^{\mu}}\frac{1}{(x_2-u)^2}\cdot
\partial_{u^{\nu}}\frac{1}{(x_3-u)^2}.\notag
\end{align}
This time, since the antisymmetric pert 
in terms of the indexes $\mu$ and $\nu$
can be eliminated, the integral is transformed as
\begin{align}
\frac{1}{2}(\sigma^{\mu}\bar\sigma^{\nu})_{\a\b}
\int d^4u
\partial_{u^{\mu}}\frac{1}{(x_2-u)^4}\cdot
\partial_{u^{\nu}}\frac{1}{(x_3-u)^2}
&=\frac{1}{2}\epsilon_{\a\b}
\int d^4u
\frac{1}{(x_2-u)^4}
\Box\frac{1}{(x_3-u)^2}\notag\\
&=-2\pi^2\epsilon_{\a\b}
\frac{1}{(x_2-x_3)^4},
\end{align}
and then the $(+-)$ sector of the three point function becomes
\begin{align}
\label{result-of-pm}
&ib\pi\frac{m}{y}
\left(\frac{N}{2}\frac{\gym^2}{4\pi^2}\right)^{J_1+2}
(\tau^i\bar\tau^{j})_{rs}\epsilon_{\a\b}
\frac{1}{|x_{12}|^{2J_2}|x_{31}|^{2J_3+2}|x_{23}|^4}.
\end{align}
Similarly, the amplitude of the $(-+)$ sector is given by
\begin{align}
\label{result-of-mp}
&-ib\pi\frac{n}{1-y}
\left(\frac{N}{2}\frac{\gym^2}{4\pi^2}\right)^{J_1+2}
(\tau^i\bar\tau^{j})_{rs}\epsilon_{\a\b}
\frac{1}{|x_{12}|^{2J_2+2}|x_{31}|^{2J_3}|x_{23}|^4}.
\end{align}

In the situation we are now considering, 
where 
$\epsilon=|x_2-x_3|\ll|x_1-x_2|$,
the integral in (\ref{result-of-pp}) can be
neglected because
\begin{align}
(\sigma^{\mu\nu})_{\a\b}
\int d^4u
\frac{1}{(x_1-u)^2}
\partial_{u^{\mu}}\frac{1}{(x_2-u)^2}\cdot
\partial_{u^{\nu}}\frac{1}{(x_3-u)^2}\to 0, \quad (x_3\to x_2).
\end{align}
Therefore,  combining all the contributions,
(\ref{result-of-pp}), (\ref{result-of-pm}) and (\ref{result-of-mp}),
and taking into account the normalization factors 
in (\ref{Ovac}) and (\ref{Ois}),
the three point functions where $O^{j}_{s\b,n}(x_3)$ 
sits in the neighbourhood of
$O^{i}_{r\a,m}(x_2)$ is given by
\begin{align}
&\langle\bar O^{J_1}_{vac}(x_1)O^{J_2,i}_{r\a,m}(x_2)O^{J_3,j}_{s\b,n}(x_2+\epsilon)
\rangle\notag\\
&=-\frac{i}{2}
\left(\frac{m}{y}-\frac{n}{1-y}\right)
\frac{\gym J_1^{-3/2}N^{-1/2}}{\sqrt{y(1-y)}}
\frac{1}{|x_{12}|^{2J_1+2}\epsilon^4}
\epsilon_{\a\b}
\left(\big(2+\frac{b}{\sqrt{2}}\big)\delta_{ij}\epsilon_{rs}
-\frac{b}{\sqrt{2}}(\tau^{ij})_{rs}\right),
\label{gauge3point2}
\end{align}
where $(\tau^{ij})_{rs}$ is defined as
$(\tau^{ij})_{rs}=\epsilon_{st}1/2(\tau^i\bar\tau^j-\tau^j\bar\tau^i)_r^{~t}$.
The final result (\ref{gauge3point2}) 
shows that the OPE coefficient 
which can be obtained in the limit $x_3\to x_2$
exactly agrees with the expected result 
from the string theory calculation, (\ref{C1}),
just when $b=-1/\sqrt{2}$.
This non-trivial matching confirms that 
the ``diagonalization'' up to ${\cal O}(1/J)$
is crucial for our duality relation.

\subsubsection{$\alpha^{(1)i}_{n_1}\alpha^{(1)j}_{-n_1} \to 
\alpha^{(2)i}_{n_2}\beta^{(2)}_{\a_1\a_2,-n_2} 
+ \alpha^{(3)j}_{n_3}\beta^{(3)}_{\b_1\b_2,-n_3}$}

Next, we consider the gauge theory correlation function 
corresponding to the process (\ref{stringAmp2}),
\begin{align}
 \langle
\bar O^{J_1,ij}_{n_1}(x_1)O^{J_2,i}_{r\a,n_2}(x_2)O^{J_3,j}_{s\b,n_3}(x_3)
\rangle,\label{gaugeAmp2}
\end{align}
at the leading order in terms of 
the genus and the gauge interaction expansion.
Here the operator $O_{n_1}^{J_1,ij}(x_1)$ is the two-scalar impurity
state defined as
\begin{align}
 &O_{n_1}^{J_1,ij}=\frac{1}{\sqrt{(J_1+2)N_{0}^{J_1+3}}}
\sum_{l_1=0}^{J_1}e^{\frac{2\pi i n_1 (l_1+1)}{J_1+2}}
\Tr(\phi^{i}Z^{l_1}\phi^jZ^{J_1-l_1}).
\end{align}
We consider the case where $i\ne j$ for simplicity
as in the corresponding calculation on the string theory side.
Since in this case too the ($--$) sector does not contribute
to the lowest order calculation with respect to 
the gauge theory interaction, it is enough to
consider the $(++)$, $(+-)$, and $(-+)$ sectors.

\begin{figure}
\begin{align}
\renewcommand{\h}{5.5cm}\notag
\psfrag{a}{\hs{-.3cm}$\lambda_{r\a}$}
\psfrag{c}{$\lambda_{s\b}$}
\psfrag{f}{$\phi^i$}\psfrag{h}{\hs{-.2cm}$\phi^j$}
\psfrag{g}{$\phi^i$}\psfrag{i}{$\phi^j$}
\psfrag{e}{$\bar Z$}
\psfrag{j}{$\hs{-.6cm}J_2-\l_2$}
\psfrag{k}{$\l_2$}
\psfrag{l}{$J_3-\l_3$}
\psfrag{m}{$\l_3$}
\includegraphics[width=\h]{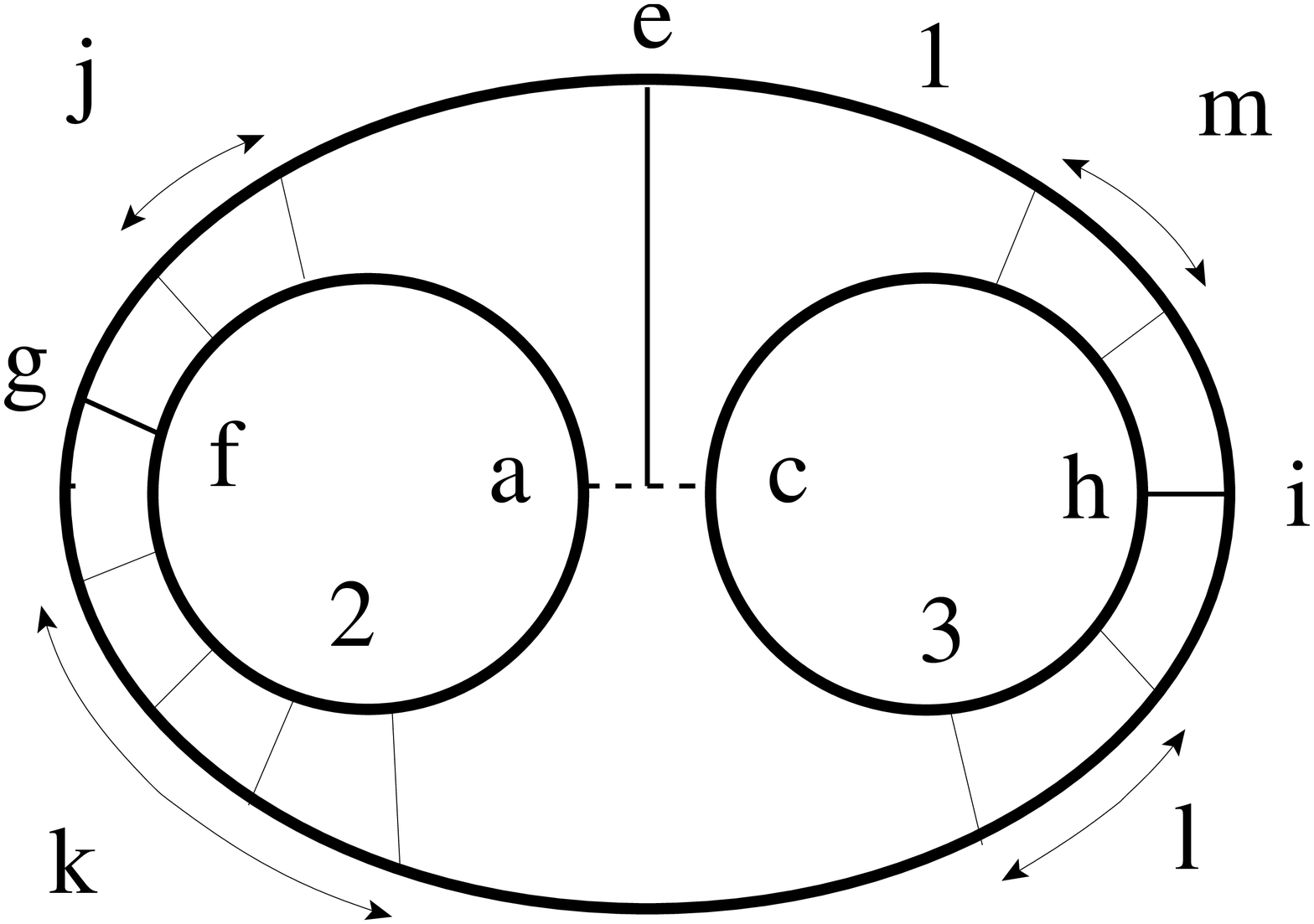}\hs{2cm}
\begin{minipage}[c]{\h}\vs{-3.2cm}\includegraphics[width=\h]{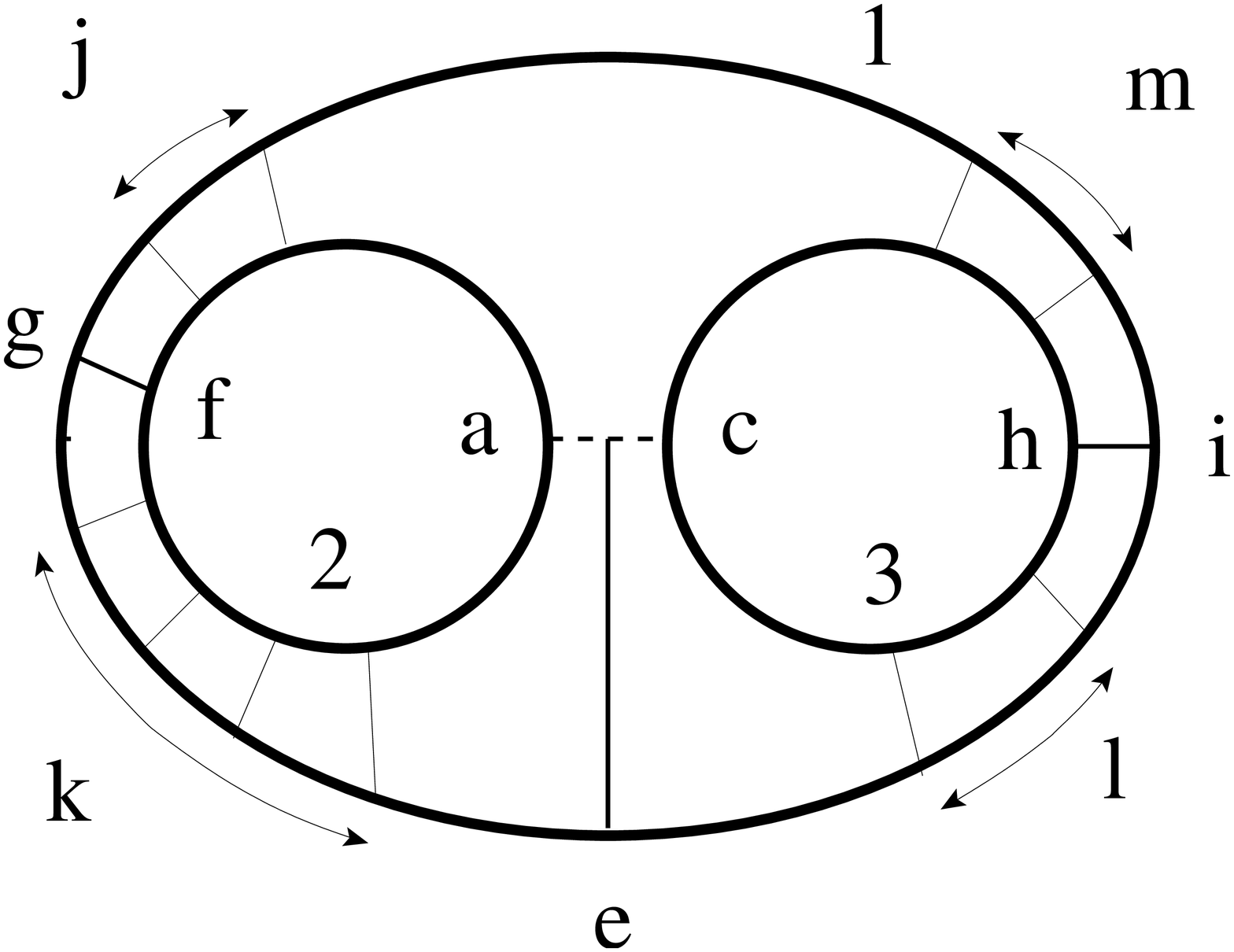}\end{minipage}
\end{align}
\caption{ $(++)$ sector of three point function}\label{PP}
\end{figure}
The $(++)$ sector of this correlator at the planer level
is depicted in Figure \ref{PP} and given by 
\begin{align}
&\frac{1}{4}\left(\frac{\gym^2}{4\pi^2}\right)^{J_1+2}
\left(\frac{N}{2}\right)^{J_1}
\frac{1}{|x_{12}|^{2(J_2+1)}|x_{13}|^{2(J_3+1)}}
\sum_{l_2,l_3=0}^{J_2,J_3}
e^{\frac{2\pi i (l_2+1)}{J_2+2}\left(n_2-\frac{J_2+2}{J_1+2}n_1\right)}
e^{\frac{2\pi i (l_3+1)}{J_3+2}\left(n_3-\frac{J_3+2}{J_1+2}n_1\right)}\notag\\
&\times\left(
e^{-\frac{2\pi i n_1 J_3}{J_1+2}}
\langle \Tr(\lambda_{r\a}(x_2)\bar Z(x_1) \lambda_{s\b}(x_3)\rangle)
+
e^{-\frac{2\pi i n_1 (J_3+1)}{J_1+2}}
 \langle\Tr(\bar Z(x_1) \lambda_{r\a}(x_2) \lambda_{s\b}(x_3)\rangle)
\right).\label{PP0}
\end{align}
The vacuum expectation value of 
$\Tr(\lambda_{r\a}(x_2)\bar Z(x_1) \lambda_{s\b}(x_3)\rangle)$
in (\ref{PP0}) can be evaluated in the same way as 
(\ref{integral}).
The sum of the phase factor at the leading order of 
the large $J$ limit is estimated as
\begin{align}
\sum_{l_2,l_3=0}^{J_2,J_3}
e^{\frac{2\pi i (l_2+1)}{J_2+2}\left(n_2-\frac{J_2+2}{J_1+2}n_1\right)}
e^{\frac{2\pi i (l_3+1)}{J_3+2}\left(n_3-\frac{J_3+2}{J_1+2}n_1\right)}
=-\frac{J_1} {\pi^2}
e^{-2\pi i\frac{J_2+2}{J_1+2}n_1}
\frac{\sin^2(\pi y n_1)}
{\left(n_1-\frac{n_2}{y}\right)\left(n_1+\frac{n_3}{1-y}\right)}.
\end{align}
Then, the $(++)$ sector of the 
three point function in the situation $|x_{23}|=\epsilon\ll|x_{12}|$ is
given by 
\begin{align}
&i{\sqrt{2}}
\frac{\gym^2}{4\pi^2}
\left(\frac{\gym^2}{4\pi^2}\frac{N}{2}\right)^{J_1+3}
\epsilon_{rs}
\frac{1}{|x_{12}|^{2(J_1+3)}\epsilon^2}
\frac{J_1}{\pi}
\frac{n_1\sin^2(\pi yn_1)}
{\left(n_1-\frac{n_2}{y}\right)\left(n_1+\frac{n_3}{1-y}\right)}.
\end{align}

\begin{figure}
\begin{align}
\renewcommand{\h}{5.5cm}\notag
\psfrag{a}{\hs{-.3cm}$\lambda_{r\a}$}
\psfrag{c}{$\theta_{\dot s\b}$}
\psfrag{f}{$\phi^i$}\psfrag{h}{\hs{-.2cm}$\phi^j$}
\psfrag{g}{$\phi^i$}\psfrag{i}{$\phi^j$}
\psfrag{e}{$\phi^j$}
\psfrag{j}{$\hs{-.6cm}J_2-\l_2$}
\psfrag{k}{$\l_2$}
\psfrag{l}{$J_3$}
\includegraphics[width=\h]{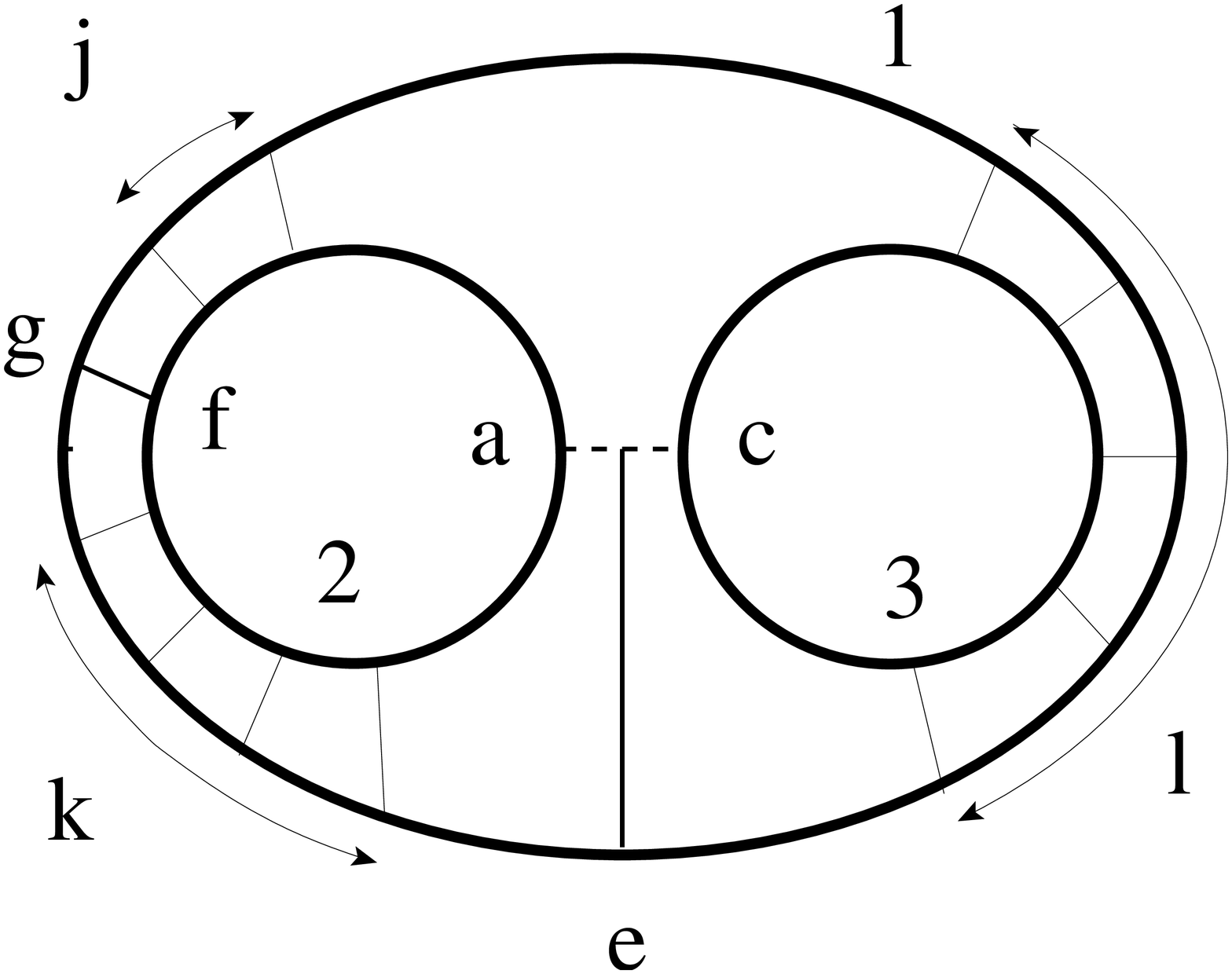}\hs{2cm}
\begin{minipage}[c]{\h}\vs{-4.5cm}\includegraphics[width=\h]{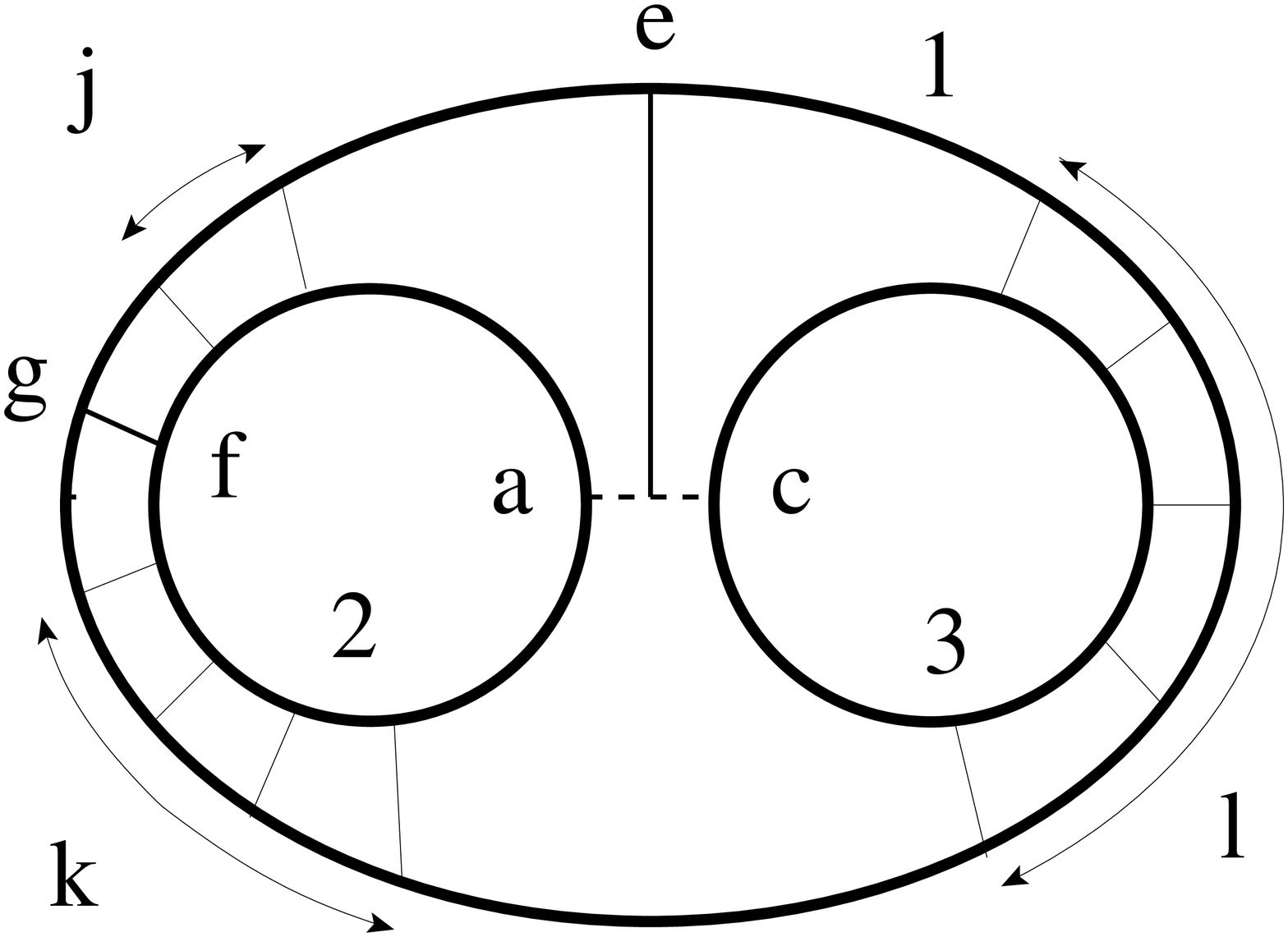}\end{minipage}
\end{align}
\caption{ $(+-)$ sector of three point function}\label{PM}
\end{figure}

On the other hand, the $(+-)$ sector in Figure \ref{PM} is
\begin{align}
&b\frac{1}{4}\left(\frac{\gym^2}{4\pi^2}\right)^{J_1+2}
\left(\frac{N}{2}\right)^{J_1}
\frac{1}{|x_{12}|^{2(J_2+1)}|x_{13}|^{2(J_3+1)}}
\sum_{l_2=0}^{J_2}
e^{\frac{2\pi i (l_2+1)}{J_2+2}\left(n_2-\frac{J_2+2}{J_1+2}n_1\right)}
(\tau^j)_s^{\dot s}\\
&\hspace{1cm}\times\left(
\langle \Tr(\phi^j(x_1)\lambda_{r\a}(x_2)\theta_{\dot s\b}(x_3)\rangle)
+
e^{-\frac{2\pi i n_1 (J_3+1)}{J_1+2}}
\langle \Tr(\theta_{\dot s\b}(x_3)\lambda_{r\a}(x_2)\phi^j(x_1)\rangle)
\right).\notag
\end{align}
The vacuum expectation value
$\langle\Tr(\phi^j(x_1)\lambda_{r\a}(x_2)\theta_{\dot s\b}(x_3)\rangle$
can be calculated in the same way as (\ref{integral2}),
and the sum of the phase factor  in the large $J$ limit is
\begin{align}
\sum_{l_2=0}^{J_2}
e^{\frac{2\pi i (l_2+1)}{J_2+2}\left(n_2-\frac{J_2+2}{J_1+2}n_1\right)}
=\frac{J_1}{2\pi i}\frac{1}{n_1-\frac{n_2}{y}}(1-\e^{-2\pi i y n_1}).
\end{align}
Therefore, the result for the $(+-)$ sector is
\begin{align}
 &
i b\frac{\gym^2}{4\pi^2}
\left(\frac{\gym^2}{4\pi^2}\frac{N}{2}\right)^{J_1+3}
\frac{1}{|x_{12}|^{2(J_1+3)}\epsilon^2}
(\tau^j\bar \tau^j)_{rs}
\frac{J_1}{\pi}
\frac{\sin^2(\pi yn_1)}
{n_1-\frac{n_2}{y}},
\end{align}
where we do not take the sum over the superscript $j$.
A parallel computation for the $(-+)$ sector gives
\begin{align}
&ib\frac{\gym^2}{4\pi^2}
\left(\frac{\gym^2}{4\pi^2}\frac{N}{2}\right)^{J_1+3}
\frac{1}{|x_{12}|^{2(J_1+3)}\epsilon^2}
(\tau^i\bar \tau^i)_{rs}
\frac{J_1}{\pi}
\frac{\sin^2(\pi yn_1)}
{n_1+\frac{n_3}{1-y}}.
\end{align}
With the relation $(\tau^i\bar\tau^i)_{rs}=-\epsilon_{rs}$
(with no sum over $i$),
and taking into account the normalization factors 
of BMN operators, 
we obtain
\begin{align}
 & \langle
\bar O_{n_1}^{J_1,ij}(x_1)O^{J_2,i}_{r\a,n_2}(x_2)
O^{J_3,j}_{s\b,n_3}(x_2+\epsilon)
\rangle\notag\\
&=
\frac{\sqrt{2}\gym J_1^{-1/2}N^{-1/2}}{4\pi^2\sqrt{y(1-y)}}
\frac{1}{|x_{12}|^{2(J_1+3)}\epsilon^2}
\epsilon_{rs}\epsilon_{\a\b}
\notag\\
&\quad\times\frac{\sin^2(\pi yn_1)}{\left(n_1-\frac{n_2}{y}\right)\left(n_1+\frac{n_3}{1-y}\right)}
\left[
2\left(i\frac{1}{\sqrt{2}}+ib\right)n_1
-i{b}
\left(\frac{n_2}{y}-\frac{n_3}{1-y}\right)
\right].
\end{align}
This shows that 
the contribution from all sectors
nicely combine to match exactly the result (\ref{C2}) when
$b=-1/\sqrt{2}$.
As in the previous example, 
the non-trivial matching is realized for the operators
of definite conformal dimension.

It is not difficult to consider a more general case where $O^{J_1,ij}_{n_1}$ 
is replaced by $O^{J_1,kl}_{n_1}$. In this case,
the interaction Hamiltonian $\HSV$ contributes 
to the calculation on the string theory side and 
gives the term proportional to 
$\delta^{ik}\tau^{jl}_{rs}$ or $\delta^{jl}\tau^{ik}_{rs}$
(or ($i\leftrightarrow j$) counterparts).
We can easily see that these terms also appear 
in the corresponding gauge theory calculation and 
will give the complete agreement.
This gives another verification of the necessity
for the both contributions from the vertexes $\HD$ and $\HSV$.
We should note that when $k=l$, the operator mixing
between a operator with scalar two impurities and 
the operator $\Tr(\bar Z Z^{J+1})$
will also play an important role for the duality relation.

Finally, we should comment on the normalization and the phase factor
of the BMN operator $O^{J,i}_{r\a,m}$.
The normalization has been chosen so that two point function
takes the form (\ref{canonical}). 
With this choice of the normalization, 
our duality relation holds for the case (\ref{gaugeAmp2})
regardless of the choice of the phase factor, that is,
whether we chose $\exp({2\pi im/J})$, $\exp({2\pi im/(J+1)})$,
or $\exp({2\pi im/(J+2)})$.
On the other hand, with the normalization thus fixed,
in order for our duality relation to be valid,
we must chose the phase factor $\exp({2\pi im/(J+2)})$;
otherwise we cannot obtain the result with
the appropriate world-sheet momentum dependence
and the normalization factor,
as is easily seen from the calculation (\ref{phase-dependence}).
Two simple examples we studied here is enough to fix
both the normalization and the phase factor of the BMN operator
$O^{J,i}_{r\s,m}$.

\section{Conclusion and Discussion}

In this paper we confirm that the duality relation
we proposed in \cite{dy1} holds for the impurity non-conserving
process which contains fermionic BMN operators.
In the process of the proof, it is crucial that
(a) the operator correspondent to the free string basis
has the definite conformal dimensions, 
(b) the string interaction Hamiltonian takes 
the form of the equal weight sum of 
the two interaction vertex $\HSV$ and $\HD$
and (c) the structure of $G$ factor in (\ref{Gfactor}), 
in particular, the factor $f$
balances the $\mu$ dependence
between $C_{123}$ and $H_{123}$ especially 
in the impurity non-preserving processes.

However, though our duality relation holds in a very non-trivial way 
with the choice of the equal weight sum of $\HSV$ and $\HD$, 
we cannot conclude that 
this form of the string 
interaction vertex has proved to be completely correct.
Since we cannot uniquely fix 
the form of the interaction vertex 
only from the requirement to respect the SUSY 
in the PP-wave background,
it is not enough to check only a particular sector.
In the previous paper \cite{dy1}
we have fixed the form of the interaction vertex 
so that it is consistent with the 
effective action for the chiral primary fields constructed
in \cite{lmrs}.
However, it has been claimed that the holographic string 
interaction vertex should be further modified 
in order to respect the U(1)$_{Y}$ symmetry 
existing in the supergravity sector \cite{LeeRusso}.
They claimed that the $\HD$ should be replaced as
\begin{align}
 \HD=\sum_{r=1}^{3}H^{(r)}_2\ket{E}\to\sum_{r=1}^{3}H^{(r)}_2
\left(1-\frac{1}{12}Y^4\right)\left(1-\frac{1}{12}Z^4\right)\ket{E}.
\end{align}
The new terms do not affect all the processes
we have studied in previous works.
The simplest example to test this modification is 
a impurity non-preserving process
where the vacuum state splits into two BMN operators with
two fermionic impurities.
For this process, in contrast to the impurity preserving process
where the energy difference gives ${\cal O}(1/\mu)$ contributions,
the new term $\sum_rH_2^{(r)}Y^4\ket{E}$ can play a significant role.
In fact, it seems that for the process
$\bra{v}b^{(2)}_{11,m}b^{(2)}_{12,-m}b^{(3)}_{21,n}b^{(3)}_{22,-n}\ket{H}_h$
the new term cancels the contribution from the prefactor in $\HSV$
and only $\HD$ contribution survives, which is of sub-leading order 
in the large $\mu$ limit compared with $\HSV$.
For our duality relation to be valid,
this cancellation seems to be necessary in order to obtain 
the correct $\mu$ dependence expected 
from the gauge theory calculation.
In this case too, the proper identification of the BMN operator
which corresponds to the free string states is crucial
to check our duality relation.
The details will be reported in the future work \cite{dobashi}.

Anyway, 
with the results obtained in the previous works
\cite{dy1, dy2} that our duality relation holds
for the impurity preserving process which consists of
the BMN operators with 
scalar, vector and fermionic impurities, as well as,
for the impurity non-preserving processes which consist of
those with scalar and vector impurities, 
our duality relation itself is strongly expected to 
hold for the general processes regardless of the kind of impurities
and of whether impurities conserve or not,
as long as we choose the appropriate form of the interaction Hamiltonian 
in the PP-wave background
and properly identify the dictionary between string and gauge theory basis.

\vspace{0.5cm}
\noindent
{\large Acknowledgements}

We would like to thank T. Yoneya for valuable discussions 
and the collaboration at an early stage of this work. 

\appendix 
\setcounter{equation}{0}
\renewcommand{\theequation}{\Alph{section}.\arabic{equation}}
\renewcommand{\thesubsection}{\Alph{section}.\arabic{subsection}}

\section{Feynman Graph}\label{AppFey}
In this Appendix, we give the details of the calculation
required to derive the log divergence of the two point function
(\ref{proper2point}).
As the regularization scheme, we adopt the differential regularization
method developed in the reference \cite{FJL}.
With this method the anomalous dimension of the 
operator with two vector impurities is calculated in \cite{Gursoy}.
The formulae necessary for our calculation are
\begin{align}
\frac{1}{(x-y)^4}=-\frac{1}{4}\Box
\frac{\ln((x-y)^2\Lambda^2)}{(x-y)^2},\quad
\Box\frac{1}{(x-y)^2}=-4\pi^2\delta^{(4)}(x-y).\label{delta}
\end{align}
With these formulae and partial integration, 
the two integrals needed in the calculations below 
to derive log divergences can be evaluated as
\begin{align}
\int d^4u\frac{1}{(x-u)^4}\frac{1}{(y-u)^4}
&=
2\pi^2\,\frac{\ln(x-y)^2\Lambda^2}{(x-y)^4},\label{x4y4}\\
\int d^4u\frac{1}{(x-u)^4}\frac{1}{(y-u)^2}
&=
\pi^2\,\frac{\ln(x-y)^2\Lambda^2}{(x-y)^2}.\label{x4y2}
\end{align}
This regularization method is used in the calculations of 
the three point correlators in (\ref{integral}) and (\ref{integral2}).
\subsection{${\cal N}=4$ Super Yang-Mills Action}
Here we present the ${\cal N}=4$ Super Yang-Mills action 
in the 4 dimensional space-time and summarize the notation.
The field content is 4 gauge fields $A^\mu (\mu=1,2,3,4)$,
6 real scalar fields $\phi^I (I=1,2,3,4,5,6)$ and 
4 Weyl spinor fields $\lambda^A_{\a} (A=1,2,3,4)$.
All these are in the adjoint representation such as
\begin{align}
 A^\mu=A^{a\mu}T^a, \quad
\phi^I=\phi^{aI}T^a, \quad \lambda^{aA}_{\a}T^a,
\end{align}
where $T^a$ satisfies 
\begin{align}
 [T^a, T^b]=if_{abc}T^c, \qquad \Tr[T^a,T^b]=\frac{1}{2}\delta^{ab},
\end{align}
with the structure constant $f_{abc}$ for SU(N) symmetry, which satisfies
\begin{align}
f_{apq}f_{bpq}=N\delta_{ab},\qquad
f_{pqr}f_{pqr}=N(N^2-1).
\end{align}

We pick up the U(1) subgroup of the R-symmetry such as
SO(6)=U(1)$\times$SO(4) for scalar field and
SU(4)=U(1)$\times$SU(2)$\times$SU(2) for spinor fields
and decompose the field such as
\begin{align}
\phi^I \to (\phi^i, Z, \bar Z),\qquad
\lambda^A_{\a}\to (\lambda_{r\a},\theta_{\dot r\a}),\label{decomposed field}
\end{align}
where $Z\equiv(\phi^5+i\phi^6)/\sqrt{2}$ which has the U(1) R-charge
1 under the rotation in the 56-plane.
The indexes $I$, $i$ and $A$ are vector indexes of SO(6), SO(4) and SU(4)
symmetry, respectively.
The decomposed fermionic fields $\lambda_{r\a}$ and $\theta_{\dot r\a}$
denote the $(\mathbf{2},\mathbf{1})_{+1/2}$ and $(\mathbf{1},\mathbf{2})_{-1/2}$ sectors 
in the decomposition above, where $\pm 1/2$ is the charge
under the U(1) R-symmetry we have picked up.
With the decomposed field in (\ref{decomposed field}) the
$d=4$, ${\cal N}=4$ super Yang-Mills action  is given by
\begin{align}
\label{SYMaction}
\gym^2 S=
&-\frac{1}{4}F^{a}_{\mu\nu}F^{a\mu\nu}
-\frac{1}{2}D_{\mu}\phi^{a}_{i}D^{\mu}\phi^{a}_{i}
-\overline{D_{\mu}Z^{a}}D^{\mu}Z^{a}
-\bar\theta^{a\r}D_{\mu}\bar\sigma^{\mu}\theta^{a}_{\r}
+\bar\lambda^{a}_{\dr}D_{\mu}\bar\sigma^{\mu}\lambda^{a\dr}\notag\\
&
-\frac{1}{4}f_{pab}f_{pcd}\phi^{a}_{i}\phi^{b}_{j}\phi^{c}_i\phi^{d}_j
-f_{pab}f_{pcd}\phi_i^aZ^b\phi^c_i\bar Z^d
+\frac{1}{2}f_{pab}f_{pcd}Z^{a}\bar Z^{b}Z^{c}\bar Z^{d}\\
&-if_{abc}\Big[
\theta^{a}_{\dr}(\bar\tau^{i})^{\dr\r}\lambda^{b}_{\r}\phi^{c}_{i}
+\bar\lambda^{a\r}(\tau^{i})_{\r\dr}\bar\theta^{b\dr}\phi^{c}_{i}\notag\\
&\hspace{3.5cm}
-\frac{1}{\sqrt{2}}
\big(\theta^{a}_{\dr}\theta^{b\dr}
-\bar\lambda^{a\r}\bar\lambda^{b}_{\r}\big) Z^{c}
+\frac{1}{\sqrt{2}}
\big(\bar\theta^{a}_{\dr}\bar\theta^{b\dr}
-\lambda^{a\r}\lambda^{b}_{\r}\big) \bar Z^{c}
\Big],\notag
\end{align}
where $F_{\mu\nu}=\partial_{x^\mu}A_{\nu}-\partial_{\nu}A_{\mu}
-i[A_{\mu},A_{\nu}]$, $D_\mu Z =\partial_\mu Z-i[A_\mu, Z]$.
The matrix $\sigma_{\mu}$ and $\tau_{i}$ are defined as
\begin{alignat}{2}
(\sigma^{\mu})_{\a\dot \a}&=(i\sigma^{i},-1),
\qquad
(\bar\sigma_{\mu})^{\dot \a\a}&=(-i\sigma^{i},-1),\\
(\tau^{i})_{\r\dr}&=(i\sigma^{i},-1),
\qquad
(\bar\tau_{i})^{\dr\r}&=(-i\sigma^{i},-1).
\end{alignat}
These satisfy the relation
\begin{alignat}{2}
(\sigma_\mu\bar\sigma_\nu)_{\a}^{~\b}
+(\sigma_\nu\bar\sigma_\mu)_{\a}^{~\b}&=2\eta_{\mu\nu}\delta_{\a}^{\b},
&\qquad
(\bar\sigma_\mu\sigma_\nu)^{\dot\a}_{~\dot\b}
+(\bar\sigma_\nu\sigma_\mu)^{\dot\a}_{~\dot\b}
&=2\eta_{\mu\nu}\delta_{\dot\a}^{\dot\b},\\
(\tau_i\bar\tau_j)_{r}^{~s}
+(\tau_j\bar\tau_i)_{r}^{~s}&=2\delta_{ij}\delta_r^{s},
&(\bar\tau_i\tau_j)^{\dot r}_{~\dot s}
+(\bar\tau_j\tau_i)^{\dot r}_{~\dot s}&
=2\delta_{ij}\delta_{\dot r}^{\dot s}.
\end{alignat}
Note that for the lowered indexes we have $(\tau_i\bar\tau_j)_{rs}
+(\tau_j\bar\tau_i)_{rs}=-2\delta_{ij}\epsilon_{rs}$.
The spinor indexes,
as for both Lorentz and R-symmetry,
are raised and lowered by the invariant tensor
$\epsilon^{rs}$ and $\epsilon_{rs}$ respectively such as
\begin{align}
\xi^{\r}\equiv\epsilon^{\r\s}\xi_{\s},\qquad
\xi_{\r}=\epsilon_{\r\s}\xi^{\s},\qquad
\epsilon^{rs}\equiv
\left(\begin{matrix}
0&1\\
-1&0\\
\end{matrix}\right)
,\quad
\epsilon_{\r\s}\equiv
\left(\begin{matrix}
0&-1\\
1&0\\
\end{matrix}\right)
\end{align}
and we use the usual convention as
\begin{align}
\xi\eta\equiv \xi^{r}\eta_{r}=\eta\xi,\qquad
\bar\xi\bar\eta\equiv\xi_{\dr}\eta^{\dr}=\bar\eta\bar\xi.
\end{align}
We have omitted the Lorentz spinor indexes in (\ref{SYMaction}) 
using this abbreviation.

From the SYM action (\ref{SYMaction}), 
the propagators are given, in the Euclidean space, by
\begin{align}
&\langle\phi^{ai}(x)\phi^{bj}(y)\rangle
=\delta^{ab}\delta^{ij}\frac{\gym^2}{4\pi^2}\frac{1}{(x-y)^2},
\quad
\langle Z^{a}(x)Z^{b}(y)\rangle
=\delta^{ab}\frac{\gym^2}{4\pi^2}\frac{1}{(x-y)^2},\\
&\langle\lambda^{a}_{r\alpha}(x)\bar\lambda^{bs}_{\dot\alpha}(y)\rangle
=-\delta^{ab}\delta^{\s}_{\r}\frac{\gym^2}{4\pi^2}
\sigma^{\mu}_{\alpha\dot\alpha}
\partial_{x^\mu}\Big(\frac{1}{(x-y)^2}\Big),\\
&\langle\theta^{a}_{\dr\alpha}(x)\bar\theta^{b\ds}_{\dot\alpha}(y)\rangle
=-\delta^{ab}\delta^{\ds}_{\dr}\frac{\gym^2}{4\pi^2}
\sigma^{\mu}_{\alpha\dot\alpha}
\partial_{x^\mu}\Big(\frac{1}{(x-y)^2}\Big).
\end{align}

\subsection{Feynman diagram}

We give the details of 
the perturbation calculations required to derive the log divergence
appearing in the two point function of the operator $O^i_{r\a,m}$.

\subsubsection{$\langle Z^a(x)\bar Z^b(y)\rangle$}
The diagram of fermion loop self-energy is 
\begin{align}
\begin{minipage}{2cm}
\psfrag{a}{\hs{-.1cm}$Z^a$}\psfrag{b}{\hs{-.1cm}$\bar Z^b$}
\psfrag{c}{$\lambda$}\psfrag{d}{$\lambda$}
\includegraphics[height=\h]{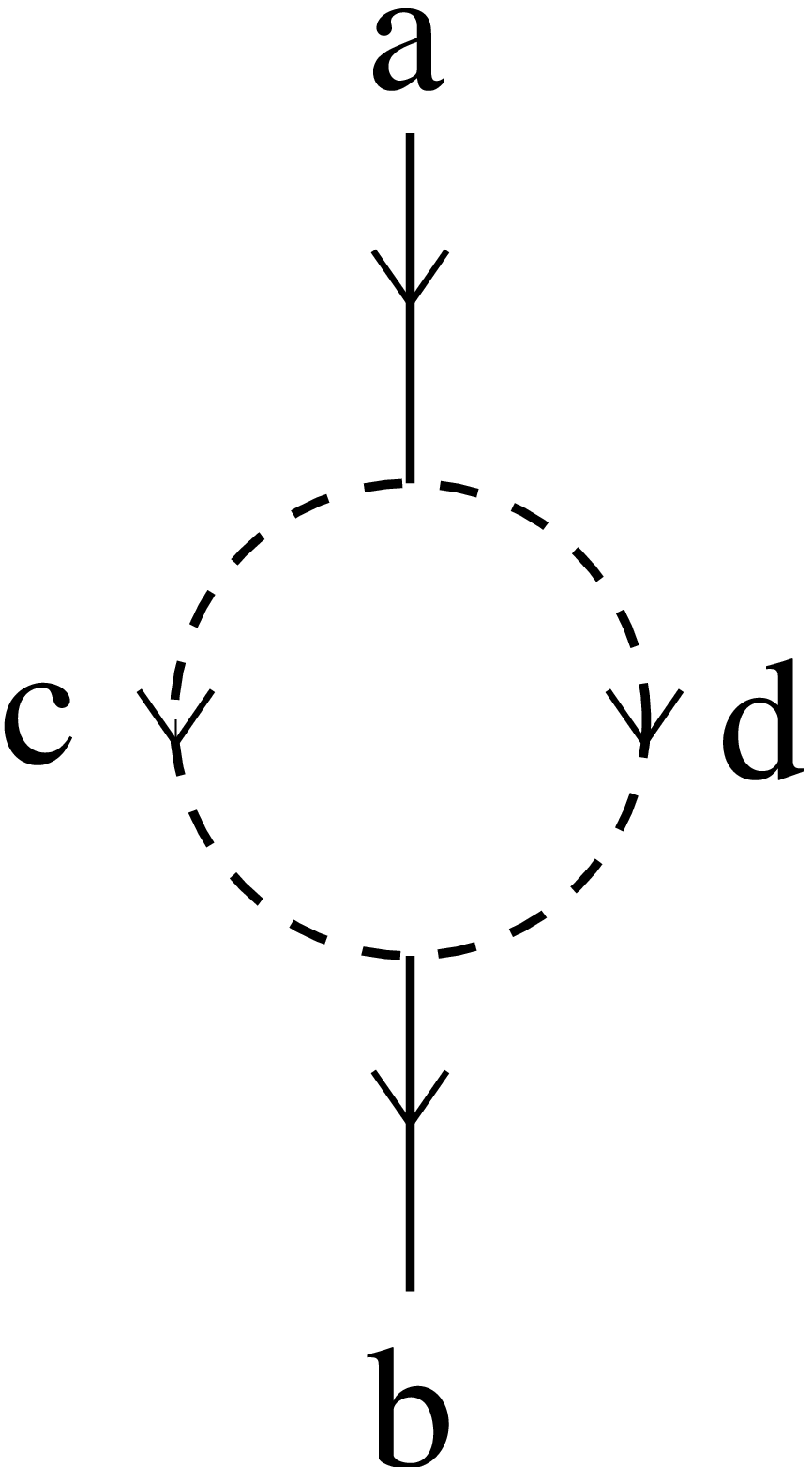}
\end{minipage}
&=\frac{1}{2}\times4\times
\frac{if_{apq}}{\sqrt{2}\gym^2}\frac{-if_{bpq}}{\sqrt{2}\gym^2}
\epsilon_{rs}\epsilon^{rs}
(\sigma^{\mu})^{\a}_{~\dot\b}(\sigma^{\nu})_{\a}^{~\dot\b}\notag\\[-.5cm]
&\hspace{1cm}\times
\left(\frac{\gym^2}{4\pi^2}\right)^4
\int d^4ud^4v
\frac{1}{(x-v)^2}
\frac{1}{(y-u)^2}
\partial_{u^{\mu}}\frac{1}{(u-v)^2}\cdot
\partial_{u^{\nu}}\frac{1}{(u-v)^2},\label{Zself}
\end{align}
where the factor $1/2\times4$ comes from the perturbation expansion 
and the symmetric factor.
With the basic formula, 
$(\sigma^\mu)^\a_{~\dot\b}(\sigma^\nu)_{\a}^{~\dot\b}=-2\eta^{\mu\nu}$,
the integrand can be transformed by partial integral such that
\begin{align}
&\int d^4ud^4v
\frac{1}{(x-v)^2}\frac{1}{(y-u)^2}
\partial_{u^{\mu}}\frac{1}{(u-v)^2}\cdot
\partial_{u_{\mu}}\frac{1}{(u-v)^2}\notag\\
&=
\int d^4ud^4v
\frac{1}{(x-v)^2}\frac{1}{(y-u)^2}
\left(
\frac{1}{2}\Box\frac{1}{(u-v)^4}
-\Box\frac{1}{(u-v)^2}\cdot\frac{1}{(u-v)^2}
\right).\label{byPart}
\end{align}
Since we focus only on the log divergence in this paper,
we neglect the second term, which also should be canceled by 
the contribution of the same degree of divergence from other diagrams. 
The log divergence of the first term in (\ref{byPart}) can 
be estimated using the formula (\ref{delta}) and (\ref{x4y2}).
Then, the log divergence from the diagram (\ref{Zself}) is given by
\begin{align}
 -\frac{1}{2}\left(\frac{\gym^2}{4\pi^2}\right)^2
N\delta_{ab}\frac{\ln(x-y)^2\Lambda}{(x-y)^2}.
\end{align}
The diagram whose intermediate states consist of $\theta$'s 
also gives the same contribution.

The contribution of the gluon emission-absorption process is
\begin{align}
\begin{minipage}{1.cm}
\psfrag{a}{\hs{-.1cm}$Z^a$}\psfrag{b}{\hs{-.1cm}$\bar Z^b$}
\includegraphics[height=\h]{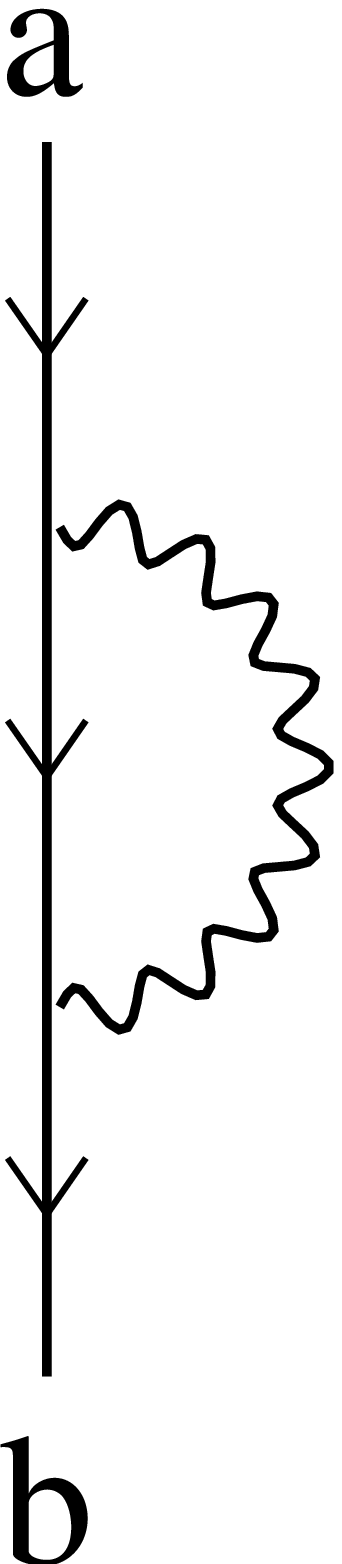}
\end{minipage}
=\frac{1}{2}\times2\times
\frac{f_{apq}}{\gym^2}\frac{f_{bqp}}{\gym^2}
\left(\frac{\gym^2}{4\pi^2}\right)^4
\int d^4ud^4v
\frac{1}{(u-v)^2}
\frac{1}{(x-u)^2}
\lrp_{u^{\mu}}\frac{1}{(u-v)^2}
\lrp_{v_{\mu}}\frac{1}{(y-v)^2}.
\end{align}
Integrating by part and using the formula (\ref{delta}) and
(\ref{x4y2}), the log divergence from gluon exchange 
is reduced to
\begin{align}
 \frac{1}{2}\left(\frac{\gym^2}{4\pi^2}\right)^2
N\delta_{ab}
\frac{\ln(x-y)^2\Lambda^2}{(x-y)^2}.
\end{align}
The log divergence which appears in the self-energy of $Z$ is
\renewcommand{\h}{3cm}
\begin{align}
\psfrag{a}{\hs{-0.1cm}$Z^a$}
\psfrag{b}{\begin{minipage}{.5cm}\vs{0cm}\hs{-.1cm}$\bar Z^b$\end{minipage}}
\begin{minipage}{1.5cm}
\includegraphics[height=\h]{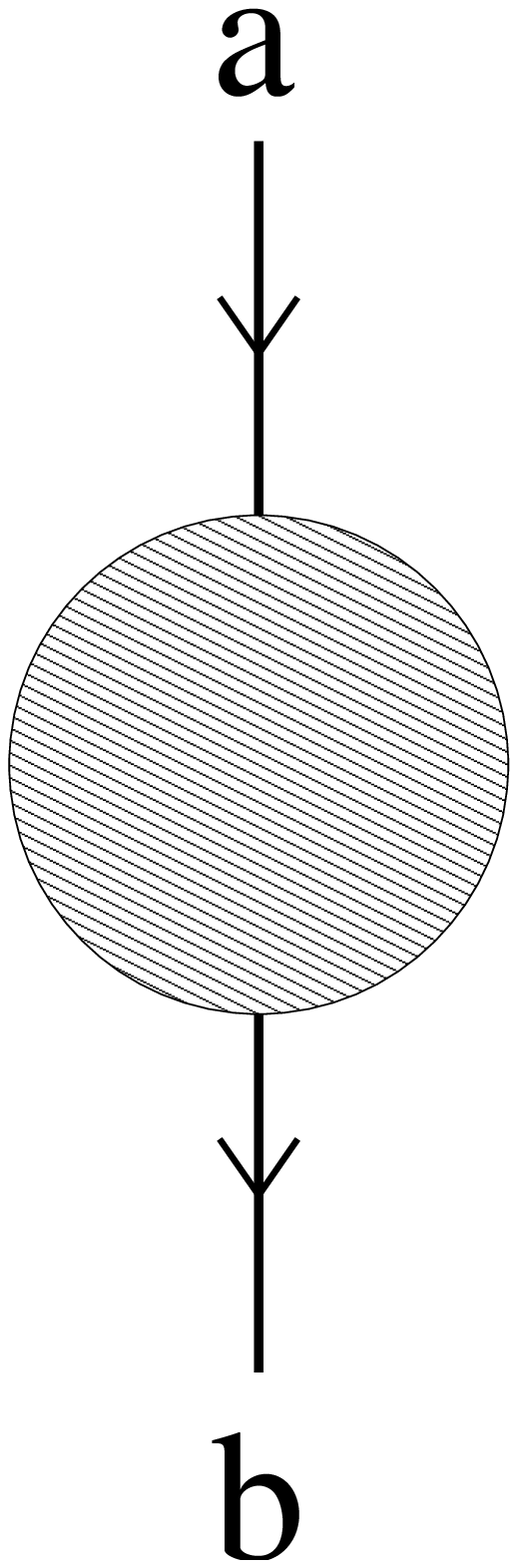}
\end{minipage}
=
\psfrag{a}{}
\psfrag{b}{}
\begin{minipage}{1.8cm}
\psfrag{c}{$\lambda$}\psfrag{d}{$\lambda$}
\includegraphics[height=\h]{ZLZ.eps}
\end{minipage}
+
\begin{minipage}{2cm}
\psfrag{c}{$\theta$}\psfrag{d}{$\theta$}
\includegraphics[height=\h]{ZLZ.eps}
\end{minipage}
+
\psfrag{a}{}\psfrag{b}{}
\begin{minipage}{1cm}
\includegraphics[height=\h]{ZGZ.eps}
\end{minipage}
=
-\frac{1}{2}\left(\frac{\gym^2}{4\pi^2}\right)^2N\delta_{ab}\frac{\ln(x-y)^2\Lambda^2}{(x-y)^2}.
\end{align}

\subsubsection{$\langle\lambda_{r\a}(x)\bar\lambda_{s\dot\a}(y)\rangle$}
The gluon emission-absorption diagram gives
\begin{align}
\begin{minipage}{1cm}
\psfrag{a}{\hs{-.1cm}$\lambda^{a}_{r\a}$}\psfrag{b}{\hs{-.1cm}$\bar\lambda_{s\dot\a}^{b}$}
\includegraphics[height=\h]{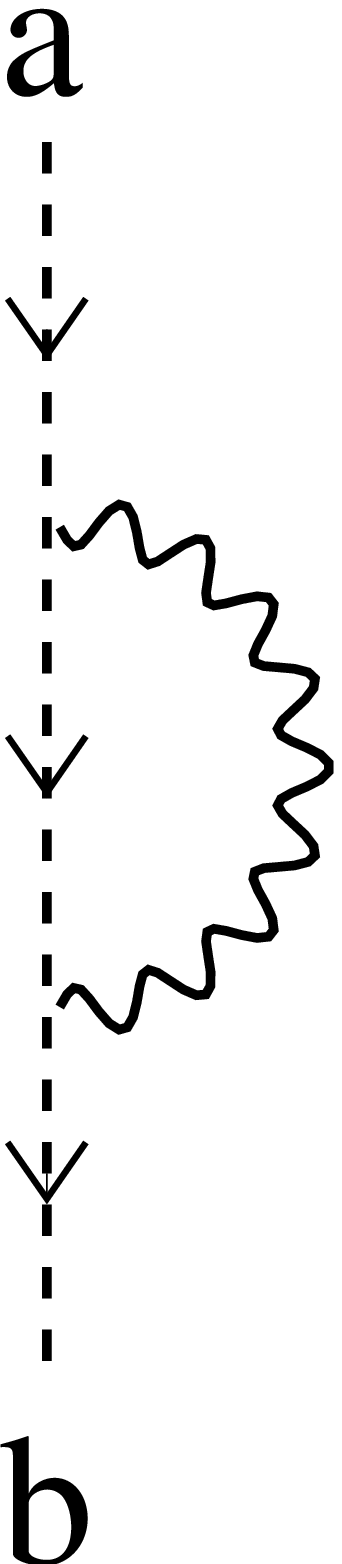}
\end{minipage}
&=\frac{1}{2}\times2\times
\epsilon_{rs}\frac{f_{apq}}{\gym^2}\frac{f_{bqp}}{\gym^2}
(\sigma^{\mu}\bar\sigma^{\lambda}\sigma^{\nu}
\bar\sigma_{\lambda}\sigma^{\rho})_{\a\dot\a}\notag\\[-.8cm]
&\hspace{1cm}\times
\left(\frac{\gym^2}{4\pi^2}\right)^4\int d^4u d^4v
\frac{1}{(u-v)^2}
\partial_{x^{\mu}}\frac{1}{(x-u)^2}
\partial_{u^{\nu}}\frac{1}{(u-v)^2}
\partial_{v^{\rho}}\frac{1}{(y-v)^2}.
\end{align}
With the identity 
\begin{align}
&(\sigma^{\mu}\bar\sigma^{\lambda}\sigma^{\nu}
\bar\sigma_{\lambda}\sigma^{\rho})_{\a\dot\a}
=-2(\sigma^\mu\bar\sigma^{\nu}\sigma^{\rho})_{\a\dot\a},\\
&\sigma^{\mu}\bar{\sigma^\nu}\sigma^{\rho}
=\eta^{\mu\nu}\sigma^{\rho}+\eta^{\nu\rho}\sigma^{\mu}
-\eta^{\mu\rho}\sigma^{\nu}-i\epsilon^{\mu\nu\rho\tau}\sigma_{\tau},
\label{sss1}
\end{align}
and by partial integral, this integral reduces to 
\begin{align}
&\left(\frac{\gym^2}{4\pi^2}\right)^4\epsilon_{rs}
\frac{f_{apq}}{\gym^2}\frac{f_{bpq}}{\gym^2}
\left[(\sigma^{\mu})_{\a\dot\a}\partial_{x^{\mu}}
\int d^4u\frac{1}{(x-u)^2}\frac{1}{(u-v)^4}\Box\frac{1}{(y-v)^2}
\right.\notag\\
&\hspace{3.5cm}\left.-i\epsilon^{\mu\nu\rho\tau}(\sigma_{\tau})_{\a\dot\a}
\partial_{x^{\mu}}
\int d^4u
\frac{1}{(x-u)^2}
\partial_{u^\nu}\partial_{u^{\rho}}\frac{1}{(u-v)^4}\cdot
\frac{1}{(y-v)^2}\right].
\end{align}
The second term vanish because of the antisymmetric property of 
$\epsilon^{\mu\nu\rho\tau}$, and with the formula (\ref{delta}) and
(\ref{x4y2}), we obtain the log divergence
\begin{align}
& -\frac{1}{4}
\left(\frac{\gym^2}{4\pi^2}\right)^2N\epsilon_{rs}\delta_{ab}
(\sigma^{\mu})_{\a\dot\a}\partial_{x^\mu}
\left(\frac{1}{(x-y)^2}\right)\ln(x-y)^2\Lambda^2.
\end{align}

There are two kinds of contribution from the Yukawa interaction,
where the intermediate states are $\lambda, Z$ and $\theta, \phi$,
respectively.
The former diagram is 
\begin{align}
\begin{minipage}{2cm}
\psfrag{a}{\hs{-.1cm}$\lambda_{r\a}^{a}$}\psfrag{b}{$\bar\lambda_{s\dot\a}^{b}$}
\psfrag{c}{\hs{-.1cm}$Z$}\psfrag{d}{$\lambda$}
\includegraphics[height=\h]{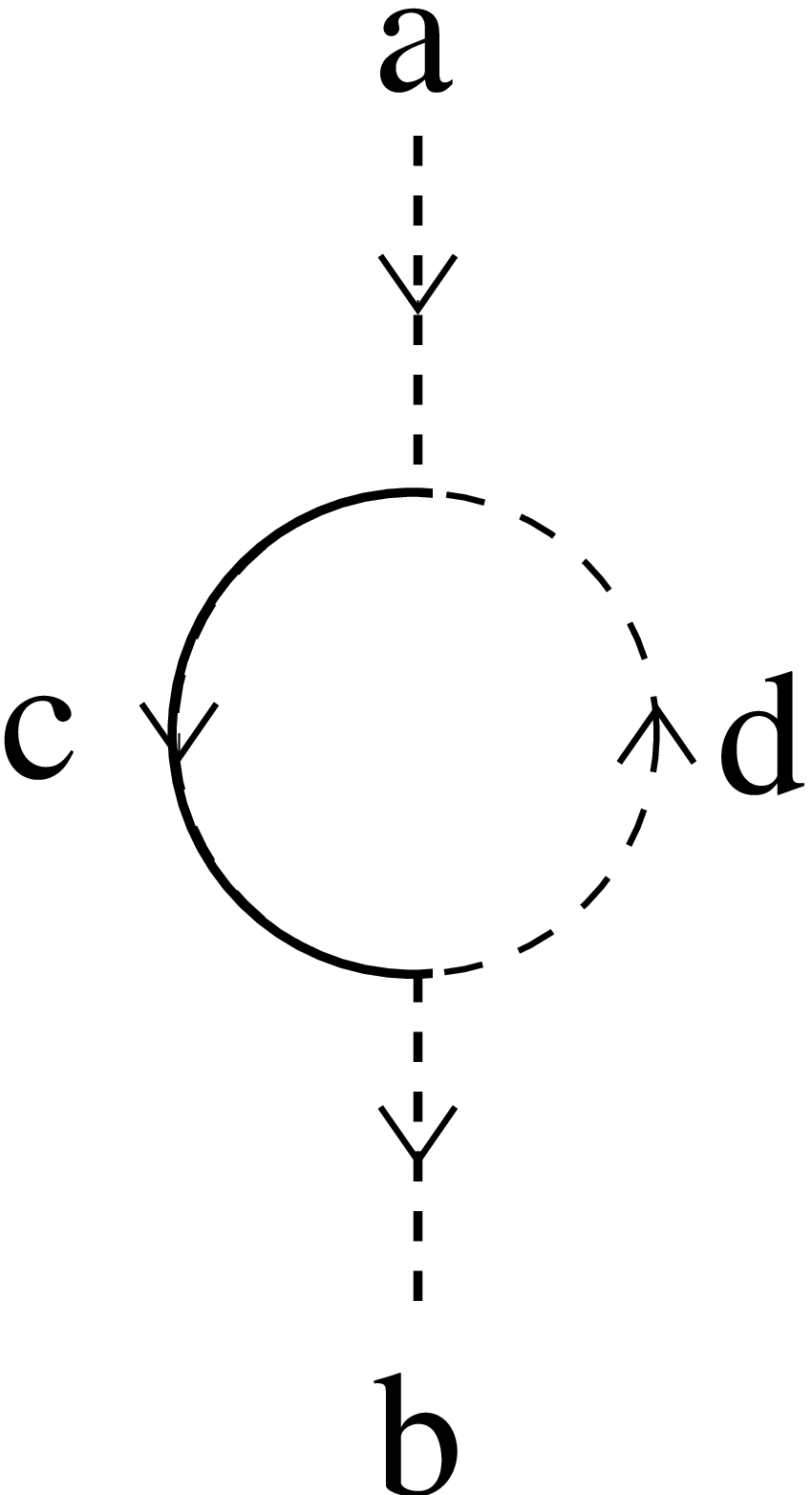}
\end{minipage}
&=\frac{1}{2}\times8\times
\frac{if_{apq}}{\sqrt{2}\gym^2}\frac{-if_{bpq}}{\sqrt{2}\gym^2}
\epsilon_{sr}(\sigma^\mu\bar\sigma^\nu\sigma^\rho)_{\a\dot\a}\notag\\[-1cm]
&\quad\times\left(\frac{\gym^2}{4\pi^2}\right)^4\int d^4ud^4v
\frac{1}{(u-v)^2}\partial_{x^\mu}\frac{1}{(x-u)^2}
\partial_{v^\nu}\frac{1}{(u-v)^2}
\partial_{v^\rho}\frac{1}{(y-v)^2}\notag\\
&=
-\frac{1}{4}
\left(\frac{\gym^2}{4\pi^2}\right)^2N
\epsilon_{rs}\delta_{ab}
(\sigma^\mu)_{\a\dot\a}\partial_{x^\mu}
\left(\frac{1}{(x-y)^2}\right)\ln(x-y)^2\Lambda^2,\label{YukawaSelf1}
\end{align}
where the integral has been performed in the same way as 
the gluon emission-absorption process.
The latter contribution is 
\begin{align}
\begin{minipage}{2cm}
\psfrag{a}{\hs{-.1cm}$\lambda_{r\a}^{a}$}
\psfrag{b}{$\bar\lambda_{s\dot\a}^{b}$}
\psfrag{c}{$\phi$}\psfrag{d}{$\theta$}
\includegraphics[height=\h]{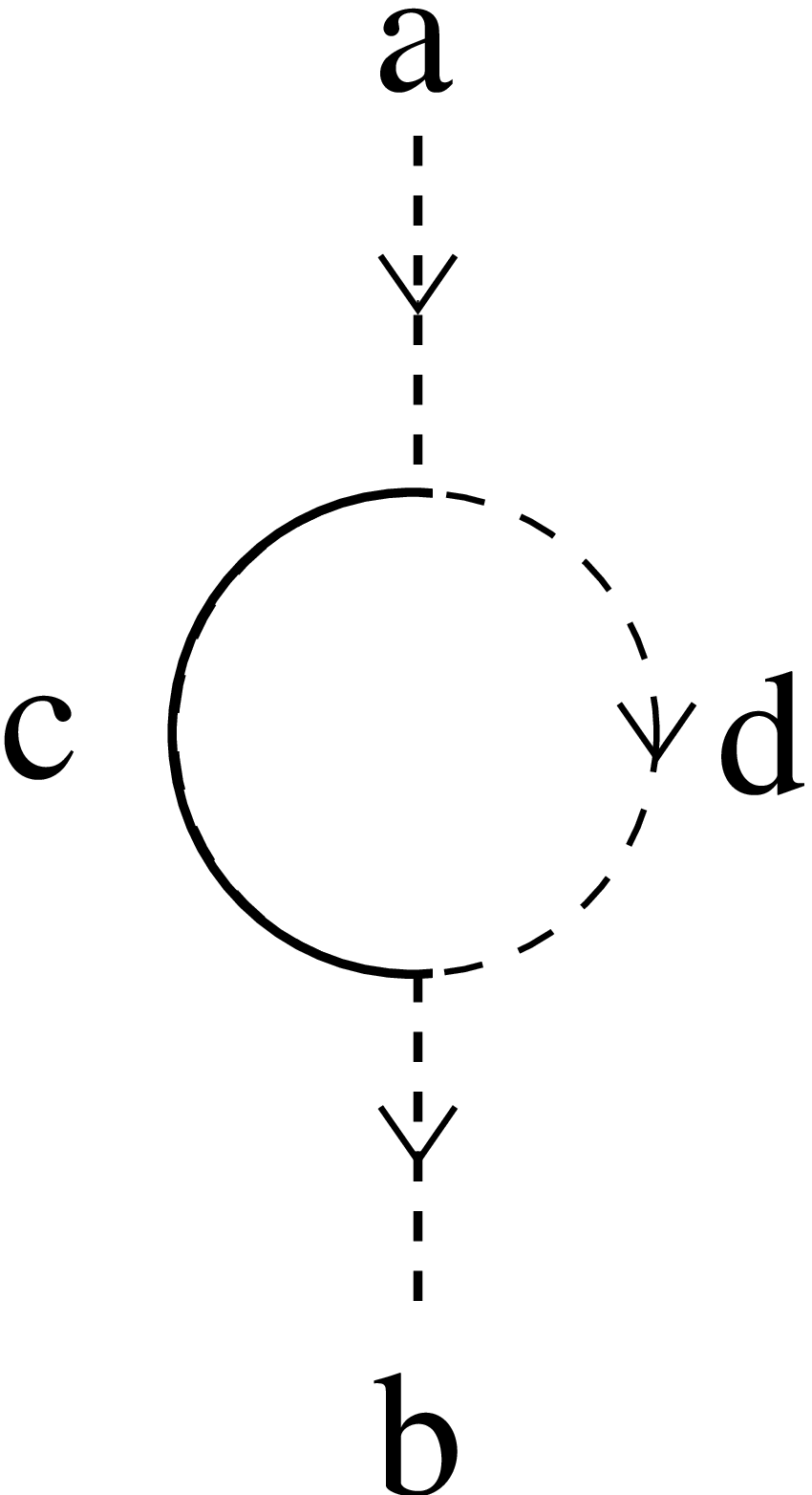}
\end{minipage}
&=\frac{1}{2}\times2\times \frac{if_{apq}}{\gym^2}\frac{if_{bpq}}{\gym^2}(\tau^i)_{r\dot u}(\bar \tau^i)^{\dot u t}\epsilon_{ts}
(\sigma^\mu\bar\sigma^\nu\sigma^\rho)_{\a\dot\a}\notag\\[-1cm]
&\quad\times\left(\frac{\gym^2}{4\pi^2}\right)^4\int d^4ud^4v
\frac{1}{(u-v)^2}
\partial_{x^\mu}\frac{1}{(x-u)^2}
\partial_{v^\nu}\frac{1}{(u-v)^2}
\partial_{v^\rho}\frac{1}{(y-v)^2}.
\end{align}
Since $(\tau^i)_{r\dot u}(\bar \tau^i)^{\dot u t}=+4\delta^t_{r}$, this
gives the result two times of (\ref{YukawaSelf1}), 
\begin{align}
-\frac{1}{2}
\left(\frac{\gym^2}{4\pi^2}\right)^2N
\epsilon_{rs}\delta_{ab}
(\sigma^\mu)_{\a\dot\a}\partial_{x^\mu}
\left(\frac{1}{(x-y)^2}\right)\ln(x-y)^2\Lambda^2.\label{YukawaSelf2}
\end{align}

The net contribution of logarithmic
divergence from the fermion self-energy is
\begin{align}
\psfrag{a}{\hs{-.05cm}$\lambda^a_{r\a}$}
\psfrag{b}{\begin{minipage}{.5cm}\vs{.1cm}\hs{-.1cm}$\bar \lambda^b_{s\b}$\end{minipage}}
\begin{minipage}{1cm}
\includegraphics[height=\h]{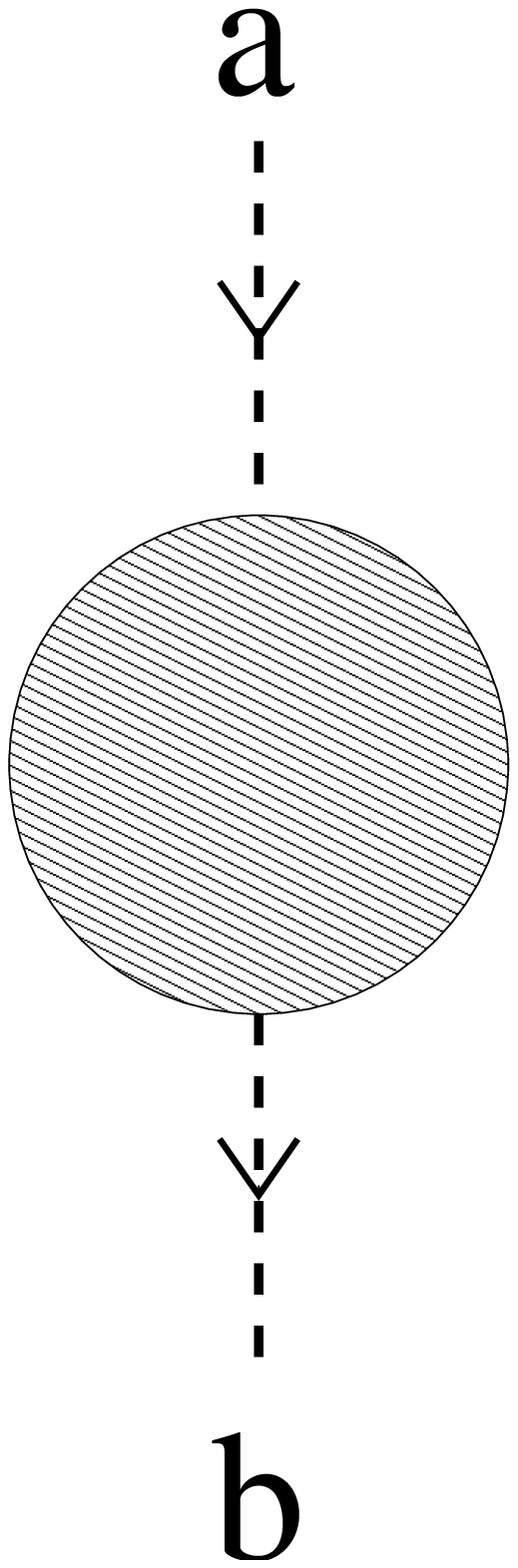}
\end{minipage}
=
\psfrag{a}{}\psfrag{b}{}
\begin{minipage}{.8cm}
\includegraphics[height=\h]{LGL.eps}
\end{minipage}
+
\begin{minipage}{1.8cm}
\psfrag{c}{\hs{-.05cm}$Z$}\psfrag{d}{$\lambda$}
\includegraphics[height=\h]{LZL.eps}
\end{minipage}
+
\begin{minipage}{1.8cm}
\psfrag{c}{$\phi$}\psfrag{d}{$\theta$}
\includegraphics[height=\h]{LPL.eps}
\end{minipage}
&=
-\left(\frac{\gym^2}{4\pi^2}\right)^2
N\epsilon_{rs}\delta_{ab}
(\sigma^{\mu})_{\a\dot\a}
\partial_{x^\mu}\left(\frac{1}{(x-y)^2}\right)\notag\\[-1cm]
&\hspace{3cm}\times\ln(x-y)^2\Lambda^2.
\end{align}

\subsubsection{
$\langle 
Z^{a_1}(x)Z^{a_2}(x)\bar Z^{b_1}(y)\bar Z^{b_2}(y)
\rangle$
}
The amplitude for the gluon exchange diagram 
is given by
\begin{align}
\begin{minipage}{1.8cm}
\renewcommand{\h}{2.5cm}
\psfrag{a}{$\hs{-.1cm}Z^{a_1}$}\psfrag{b}{$Z^{a_2}$}
\psfrag{c}{\begin{minipage}{1cm}\vs{.1cm}\hs{-.1cm}$\bar Z^{b_1}$\end{minipage}}
\psfrag{d}{\begin{minipage}{1cm}\vs{.1cm}\hs{-.1cm}$\bar Z^{b_2}$\end{minipage}}
\includegraphics[height=\h]{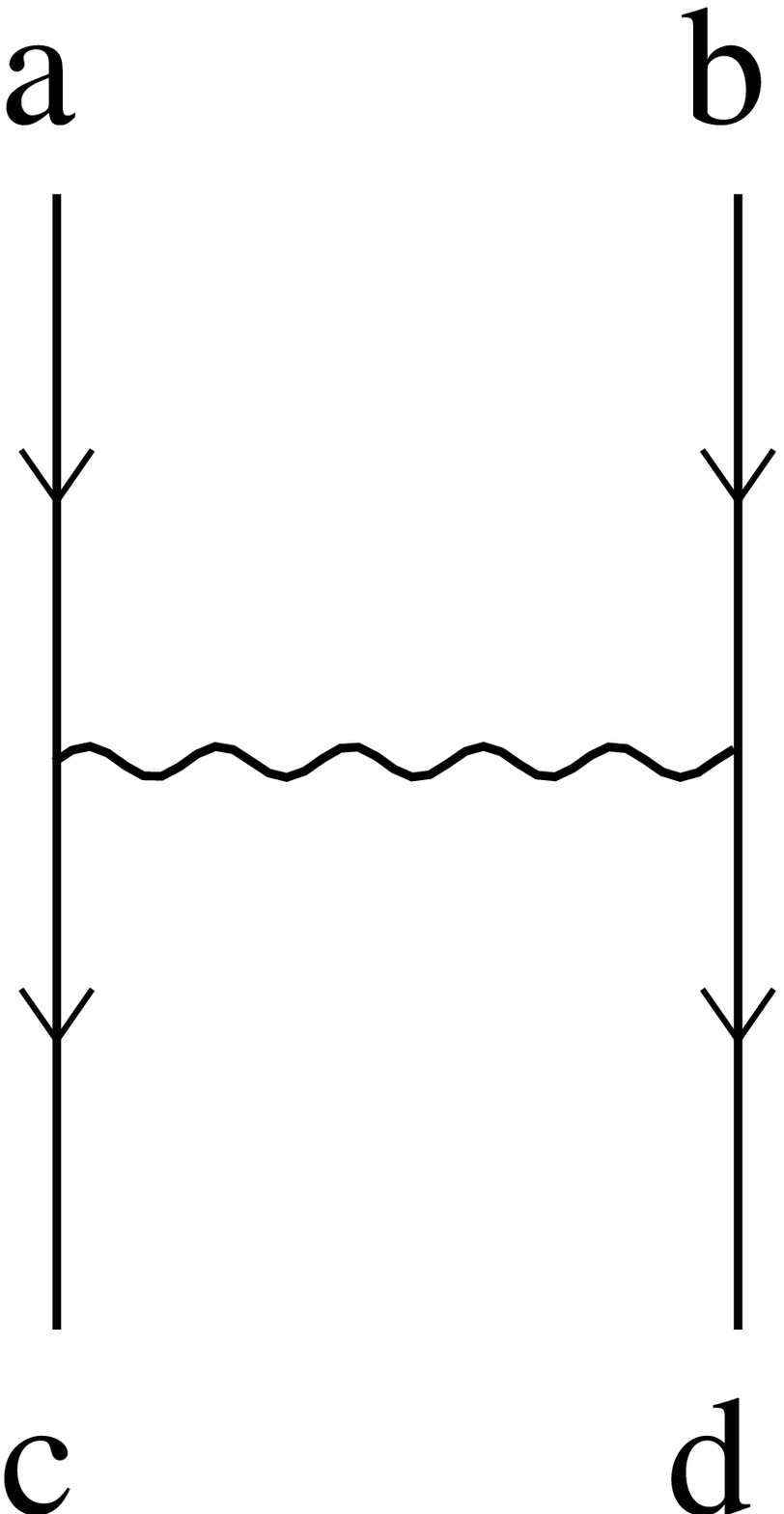}
\end{minipage}
+\hs{.2cm}
\begin{minipage}{1.8cm}
\renewcommand{\h}{2.5cm}
\psfrag{a}{$\hs{-.1cm}Z^{a_1}$}\psfrag{b}{$Z^{a_2}$}
\psfrag{c}{\begin{minipage}{1cm}\vs{.1cm}\hs{-.1cm}$\bar Z^{b_1}$\end{minipage}}
\psfrag{d}{\begin{minipage}{1cm}\vs{.1cm}\hs{-.1cm}$\bar Z^{b_2}$\end{minipage}}
\includegraphics[height=\h]{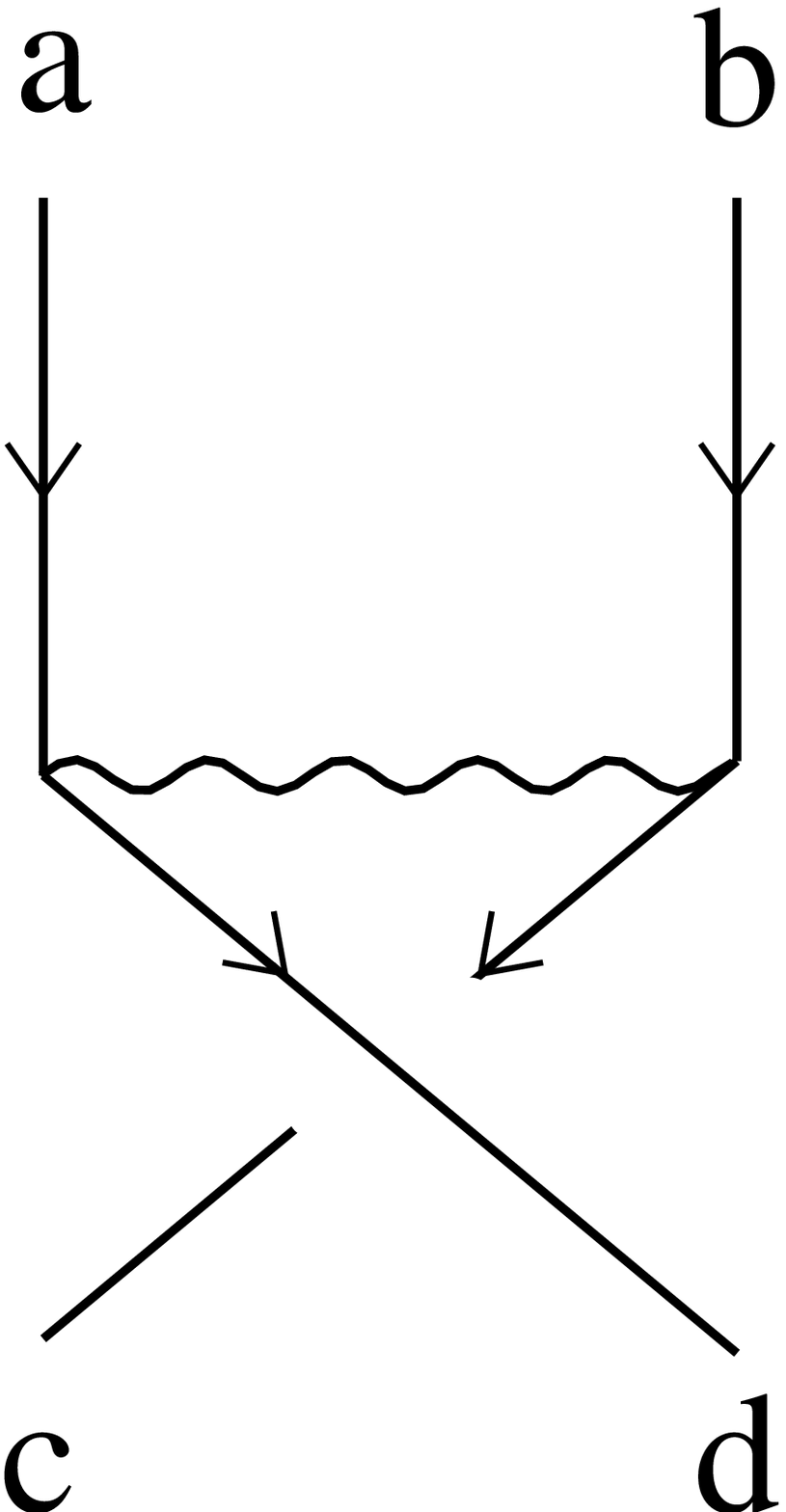}
\end{minipage}
&=\frac{1}{2}\times2\times
\left(
 \frac{f_{pa_1b_1}}{\gym^2}\frac{f_{pa_2b_2}}{\gym^2}
+\frac{f_{pa_1b_2}}{\gym^2}\frac{f_{pa_2b_1}}{\gym^2}
\right)
\left(\frac{\gym^2}{4\pi^2}\right)^5\\[-.5cm]
&\quad\times\int d^4ud^4v
\frac{1}{(u-v)^2}
\frac{1}{(x-u)^2}\lrp_{u^{\mu}}\frac{1}{(y-u)^2}
\frac{1}{(x-v)^2}\lrp_{v_{\mu}}\frac{1}{(y-v)^2}\notag
\end{align}
Performing partial integral and using the formula (\ref{delta}), 
the integrand can be transformed as
\begin{align}
\frac{1}{2}(\Box_{x}+\Box_{y})H_{xyxy}
+4\pi^2\left(
\frac{2}{(x-y)^2}(Y_{xxy}+Y_{xyy})
-X_{xxyy}
\right),\label{afterPD}
\end{align}
where we have defined $X_{abcd}$ and $Y_{abc}$ as
\begin{align}
H_{abcd}&=\int d^4u d^4v
\frac{1}{(a-u)^2}\frac{1}{(b-u)^2}\frac{1}{(u-v)^2}
\frac{1}{(c-v)^2}\frac{1}{(d-v)^2},\\
X_{abcd}&=\int d^4u
\frac{1}{(a-u)^2}\frac{1}{(b-u)^2}\frac{1}{(c-u)^2}\frac{1}{(d-u)^2},\\
Y_{abc}&=\int d^4u
\frac{1}{(a-u)^2}\frac{1}{(b-u)^2}\frac{1}{(c-u)^2}.
\end{align}
As long as we are concerned with the logarithmic divergence,
the first term in (\ref{afterPD}) can be neglected and 
the other terms can be evaluated with the use of the formulae 
(\ref{x4y4}) and (\ref{x4y2}). The result is
\begin{align}
\frac{1}{2}\,\left(\frac{g^2_{YM}}{4\pi^2}\right)^3
(f_{pa_1b_1}f_{pa_2b_2}+f_{pa_1b_2}f_{pa_2b_1})
\frac{\ln(x-y)^2\Lambda^2}{(x-y)^4}.
\end{align}

On the other hand, the scalar 4-point interaction is easily calculated as
\begin{align}
\begin{minipage}{2cm}
\renewcommand{\h}{2cm}
\psfrag{a}{$\hs{-.1cm}Z^{a_1}$}\psfrag{b}{$Z^{a_2}$}
\psfrag{c}{\begin{minipage}{1cm}\vs{.2cm}\hs{-.2cm}$\bar Z^{b_1}$\end{minipage}}
\psfrag{d}{\begin{minipage}{1cm}\vs{.2cm}\hs{-.1cm}$\bar Z^{b_2}$\end{minipage}}
\includegraphics[height=\h]{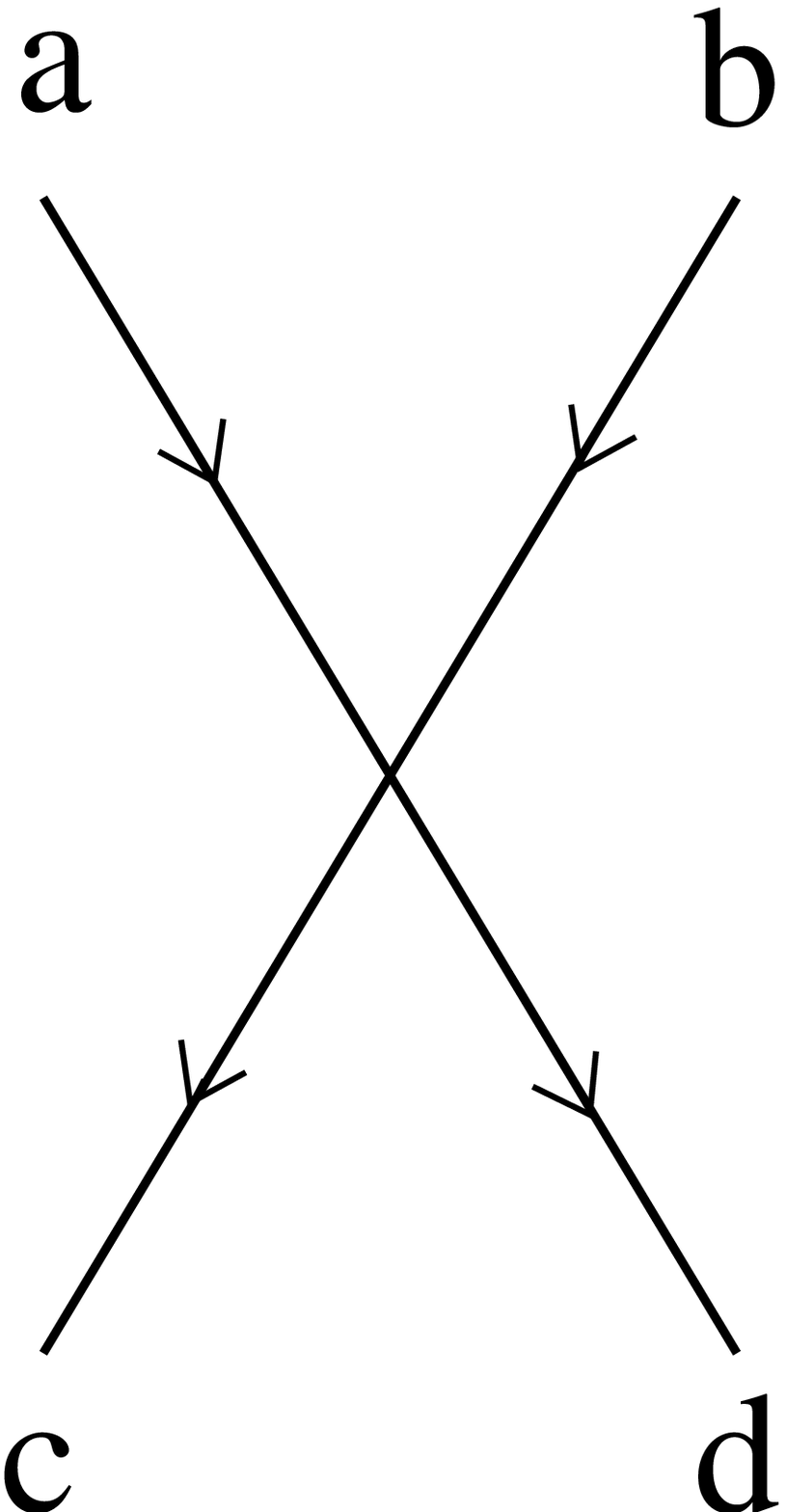}
\end{minipage}
&=\frac{1}{\gym^2}
(f_{pa_1b_1}f_{pa_2b_2}+f_{pa_1b_2}f_{pa_2b_1})
\left(\frac{\gym^2}{4\pi^2}\right)^4
\int d^4 u \frac{1}{(x-u)^4}\frac{1}{(y-u)^4}\notag\\[-.5cm]
&=\frac{1}{2}\left(\frac{\gym^2}{4\pi^2}\right)^3
(f_{pa_1b_1}f_{pa_2b_2}+f_{pa_1b_2}f_{pa_2b_1})
\frac{\ln(x-y)^2\Lambda^2}{(x-y)^4}.
\end{align}
Then, we obtain 
\begin{align}
\begin{minipage}{1.8cm}
\renewcommand{\h}{2.5cm}
\psfrag{a}{$\hs{-.1cm}Z^{a_1}$}\psfrag{b}{$Z^{a_2}$}
\psfrag{c}{\begin{minipage}{1cm}\vs{.1cm}\hs{-.1cm}$\bar Z^{b_1}$\end{minipage}}
\psfrag{d}{\begin{minipage}{1cm}\vs{.1cm}\hs{-.1cm}$\bar Z^{b_2}$\end{minipage}}
\includegraphics[height=\h]{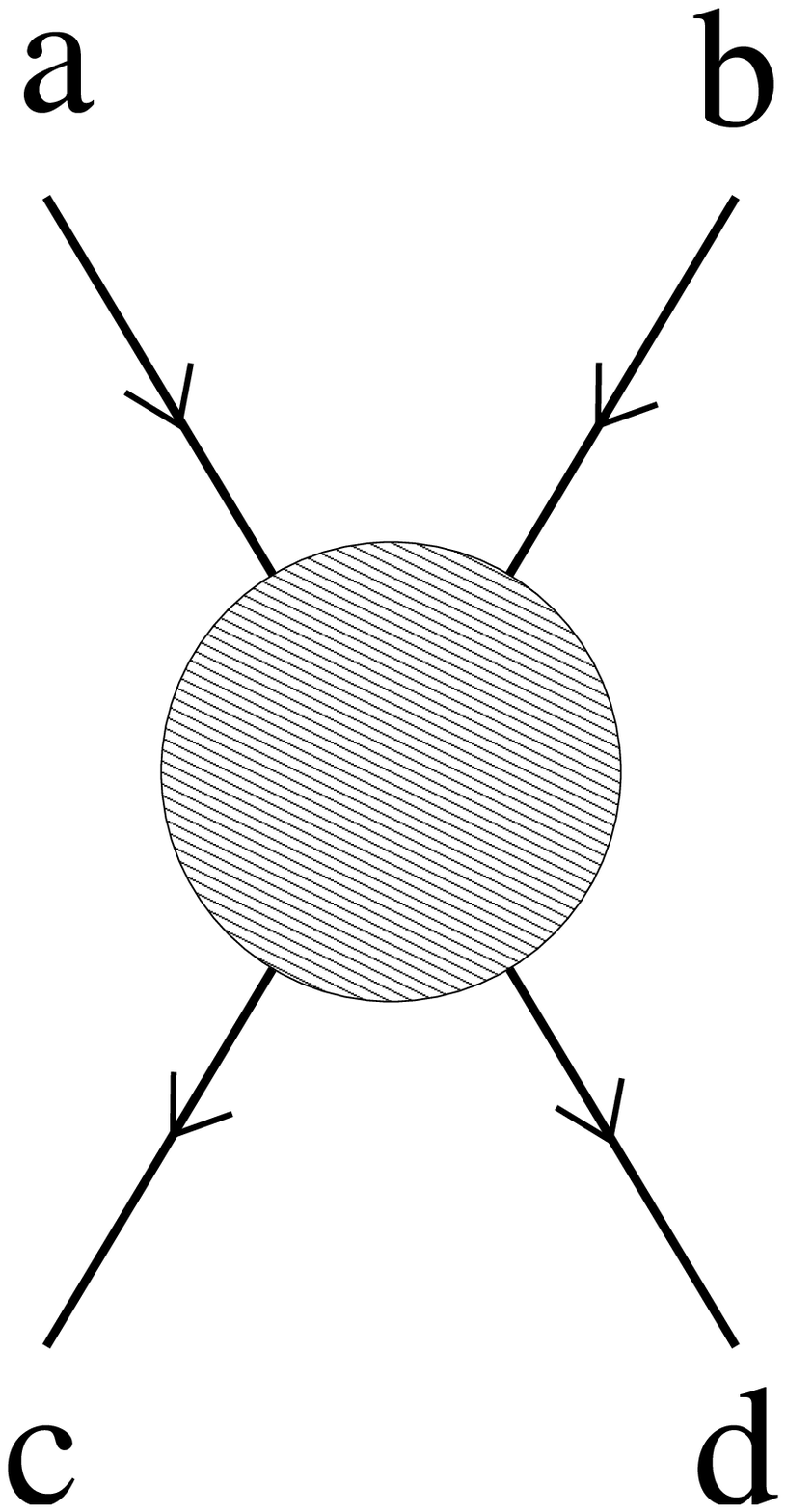}
\end{minipage}
\hs{-.3cm}=
\psfrag{a}{}\psfrag{b}{}\psfrag{c}{}\psfrag{d}{}
\begin{minipage}{1.5cm}
\renewcommand{\h}{2.5cm}
\includegraphics[height=\h]{ZZ-G-exch.eps}
\end{minipage}
+
\begin{minipage}{1.5cm}
\renewcommand{\h}{2.5cm}
\includegraphics[height=\h]{ZZ-G-exch2.eps}
\end{minipage}
+\hs{-.2cm}
\begin{minipage}{1.5cm}
\renewcommand{\h}{2.5cm}
\includegraphics[height=\h]{4Z.eps}
\end{minipage}
\hs{-.3cm}=
\left(\frac{\gym^2}{4\pi^2}\right)^3
(f_{pa_1b_1}f_{pa_2b_2}+f_{pa_1b_2}f_{pa_2b_1})
\frac{\ln(x-y)^2\Lambda^2}{(x-y)^4}
\end{align}
\subsubsection{$
\langle\lambda_{r\a}^{a_1}(x)Z^{a_2}(x)
\bar \lambda_{s\b}^{b_1}(y)\bar Z^{b_2}(y)
\rangle$}

The log divergence which comes through the Yukawa coupling interaction is
\begin{align}
\begin{minipage}{2cm}
\renewcommand{\h}{3cm}
\psfrag{a}{\hs{-.05cm}$\lambda_{r\a}^{a_1}$}\psfrag{b}{$Z^{a_2}$}
\psfrag{c}{\begin{minipage}{1cm}\vs{.1cm}\hs{-.1cm}$\bar\lambda_{s\dot\a}^{b_1}$\end{minipage}}
\psfrag{d}{\begin{minipage}{1cm}\vs{.1cm}\hs{-.1cm}$\bar Z^{b_2}$\end{minipage}}
\psfrag{e}{\hs{-.05cm}$\lambda$}
\includegraphics[height=\h]{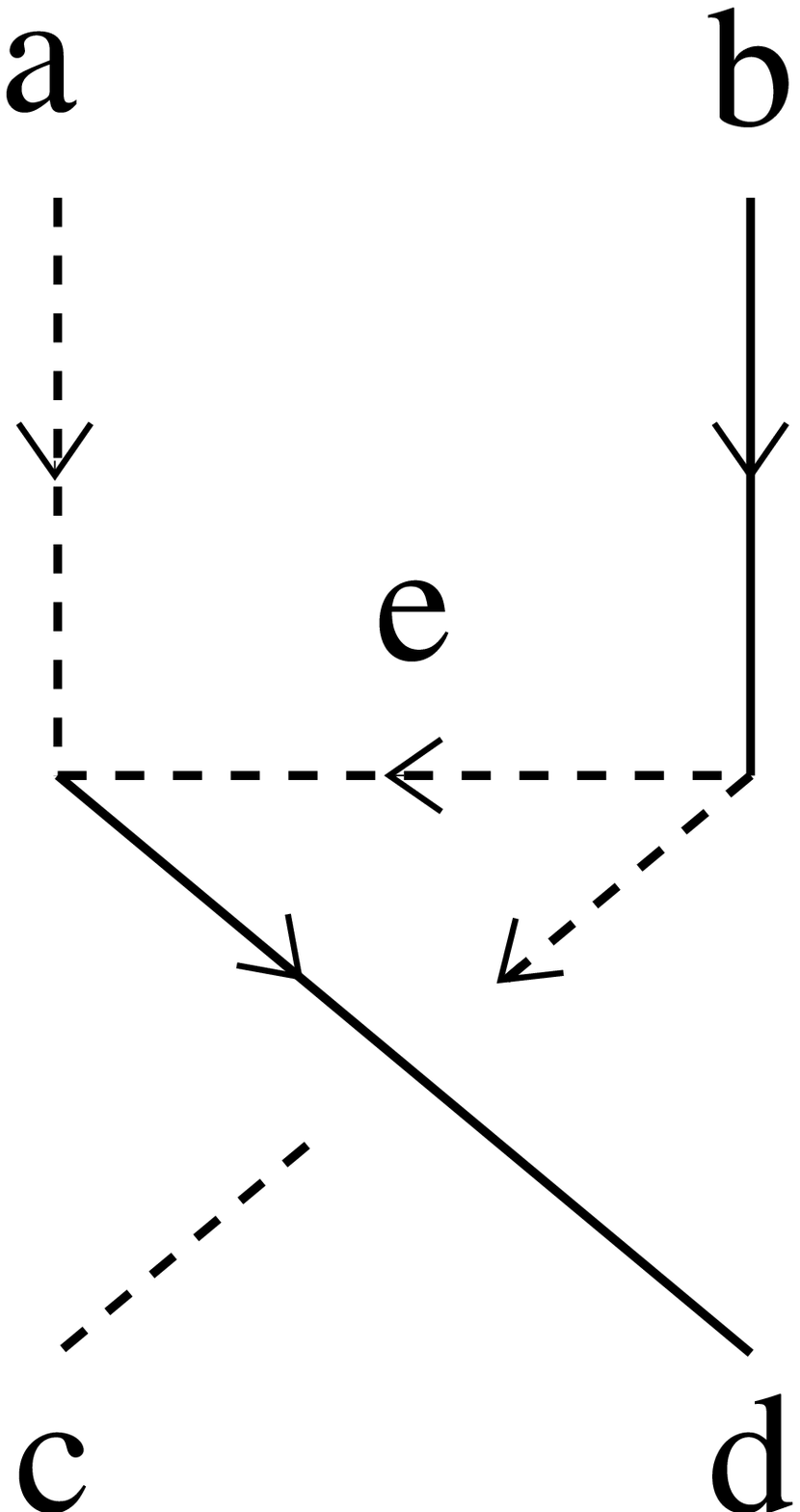}
\end{minipage}
&=
\frac{1}{2}\times8 \times
\frac{if_{pa_1b_2}}{\sqrt{2}\gym^2}\frac{-if_{pa_2b_1}}{\sqrt{2}\gym^2}\epsilon_{rs}
\left(\frac{\gym^2}{4\pi^2}\right)^5\\[-.8cm]
&\quad\times(\sigma^\mu\bar\sigma^\nu\sigma^\rho)_{\a\dot\a}\int d^4u d^4v
\partial_{x^\mu}\frac{1}{(x-u)^2}\frac{1}{(y-u)^2}
\partial_{v^{\nu}}\frac{1}{(u-v)^2}
\partial_{v^\rho}\frac{1}{(y-v)^2}\frac{1}{(x-v)^2}.\notag
\end{align}
The integral in the second line can be evaluated as
\begin{align}
\sigma^{\mu}_{\a\dot\a}
\left(\frac{1}{2}\Box_{x}H'_{\mu}+
2\pi^2\partial_{x^\mu}X_{xxyy}\right),\label{transformed}
\end{align}
where $H'_{\mu}$ is defined as
\begin{align}
 \int d^4ud^4v
\frac{1}{(x-u)^2}
\frac{1}{(y-u)^2}
\partial_{u^\mu}\frac{1}{(u-v)^2}\cdot
\frac{1}{(x-v)^2}
\frac{1}{(y-v)^2}.
\end{align}
The first term in (\ref{transformed}) does not give the log divergence,
and the result is
\begin{align}
\left(\frac{\gym^2}{4\pi^2}\right)^3\epsilon_{rs}
f_{pa_1b_2}f_{pa_2b_1}\frac{1}{(x-y)^2}(\sigma^\mu)_{\a\dot\a}
\partial_{x^\mu}\left(\frac{1}{(x-y)^2}\right)\ln(x-y)^2\Lambda^2.
\end{align}

The gluon exchange gives the contribution
\begin{align}
\label{LZ-G-exch}
\begin{minipage}{2cm}
\renewcommand{\h}{3cm}
\psfrag{a}{\hs{-.1cm}$\lambda_{r\a}^{a_1}$}\psfrag{b}{\hs{-.1cm}$Z^{a_2}$}
\psfrag{c}{\begin{minipage}{1cm}\vs{.1cm}\hs{-.1cm}$\bar\lambda_{s\dot\a}^{b_1}$\end{minipage}}
\psfrag{d}{\begin{minipage}{1cm}\vs{.1cm}\hs{-.1cm}$\bar Z^{b_2}$\end{minipage}}
\includegraphics[height=\h]{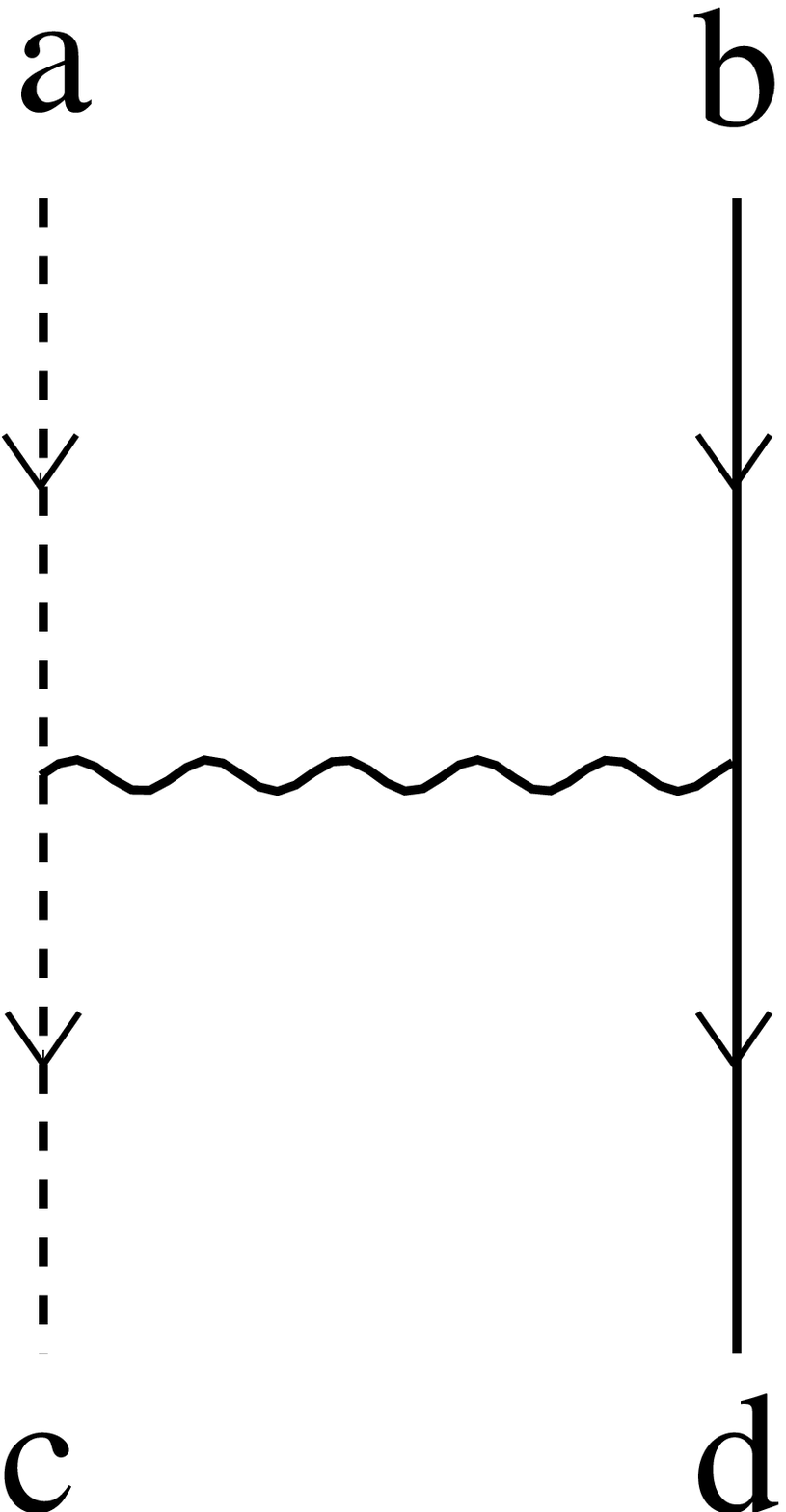}
\end{minipage}
&=\frac{1}{2}\times2\times 
\frac{f_{pa_1b_1}}{\gym^2}\frac{f_{pa_2b_2}}{\gym^2}\epsilon_{rs}
(\sigma^\mu\bar\sigma^\nu\sigma^\rho)_{\a\dot\a}
\\[-1cm]
&\quad\times\left(\frac{\gym^2}{4\pi^2}\right)^5\int d^4u d^4v
\partial_{x^\mu}\frac{1}{(x-u)^2}
\partial_{u^\rho}\frac{1}{(y-u)^2}
\frac{1}{(u-v)^2}
\frac{1}{(x-v)^2}\lrp\frac{1}{(x-v)^2}\notag\\
&=\left(\frac{\gym^2}{4\pi^2}\right)^5
\frac{f_{pa_1b_1}}{\gym^2}\frac{f_{pa_2b_2}}{\gym^2}
\epsilon_{rs}
\sigma^\mu_{\a\dot\a}
\left(
\frac{1}{4}\Box_x \partial_{x^\mu}H_{xxyy}
+8\pi^2\frac{1}{(x-y)^2}\partial_{x^\mu}Y_{xxy}
\right)\notag\\
&=\frac{1}{2}
\left(\frac{\gym^2}{4\pi^2}\right)^3
\epsilon_{rs}f_{pa_1b_1}f_{pa_2b_2}
\frac{1}{(x-y)^2}(\sigma^\mu)_{\a\dot\a}
\partial_{x^\mu}\left(\frac{1}{(x-y)^2}\right)\ln(x-y)^2\Lambda.\notag
\end{align}
Then the net result is
\begin{align}
\begin{minipage}{1.5cm}
\renewcommand{\h}{2.5cm}
\psfrag{a}{\hs{-.1cm}$\lambda_{r\a}^{a_1}$}
\psfrag{b}{\hs{-.1cm}$Z^{a_2}$}
\psfrag{c}{\begin{minipage}{1cm}\vs{.1cm}\hs{-.1cm}$\bar\lambda_{s\dot\a}^{b_1}$\end{minipage}}
\psfrag{d}{\begin{minipage}{1cm}\vs{.1cm}\hs{-.1cm}$\bar Z^{b_2}$\end{minipage}}
\includegraphics[height=\h]{4LZLZ.eps}
\end{minipage}
=
\psfrag{a}{}\psfrag{b}{}\psfrag{c}{}\psfrag{d}{}\psfrag{e}{}
\begin{minipage}{1.5cm}
\renewcommand{\h}{2.5cm}
\includegraphics[height=\h]{LZ-L-exch.eps}
\end{minipage}
+
\begin{minipage}{1.5cm}
\renewcommand{\h}{2.5cm}
\includegraphics[height=\h]{LZ-G-exch.eps}
\end{minipage}
&=
\left(\frac{\gym^2}{4\pi^2}\right)^3
\epsilon_{rs}
\left(
f_{pa_1b_2}f_{pa_2b_1}+
\frac{1}{2}f_{pa_1b_1}f_{pa_2b_2}
\right)\notag\\[-.5cm]
&\quad\times\frac{1}{(x-y)^2}\sigma^\mu_{\a\dot\a}
\partial_{x^\mu}
\left(\frac{1}{(x-y)^2}\right)\ln(x-y)^2\Lambda^2.
\end{align}
\subsubsection{
$\langle\phi^{ia_1}(x)Z^{a_2}(x)\phi^{jb_1}(y)\bar Z^{b_2}(y)\rangle$
}

The gluon exchange gives
\begin{align}
\begin{minipage}{1.8cm}
\renewcommand{\h}{3cm}
\psfrag{a}{$\hs{-.1cm}Z^{a_1}$}\psfrag{b}{$\phi^{ia_2}$}
\psfrag{c}{$\hs{-.1cm}\bar Z^{b_1}$}\psfrag{d}{$\phi^{jb_2}$}
\includegraphics[height=\h]{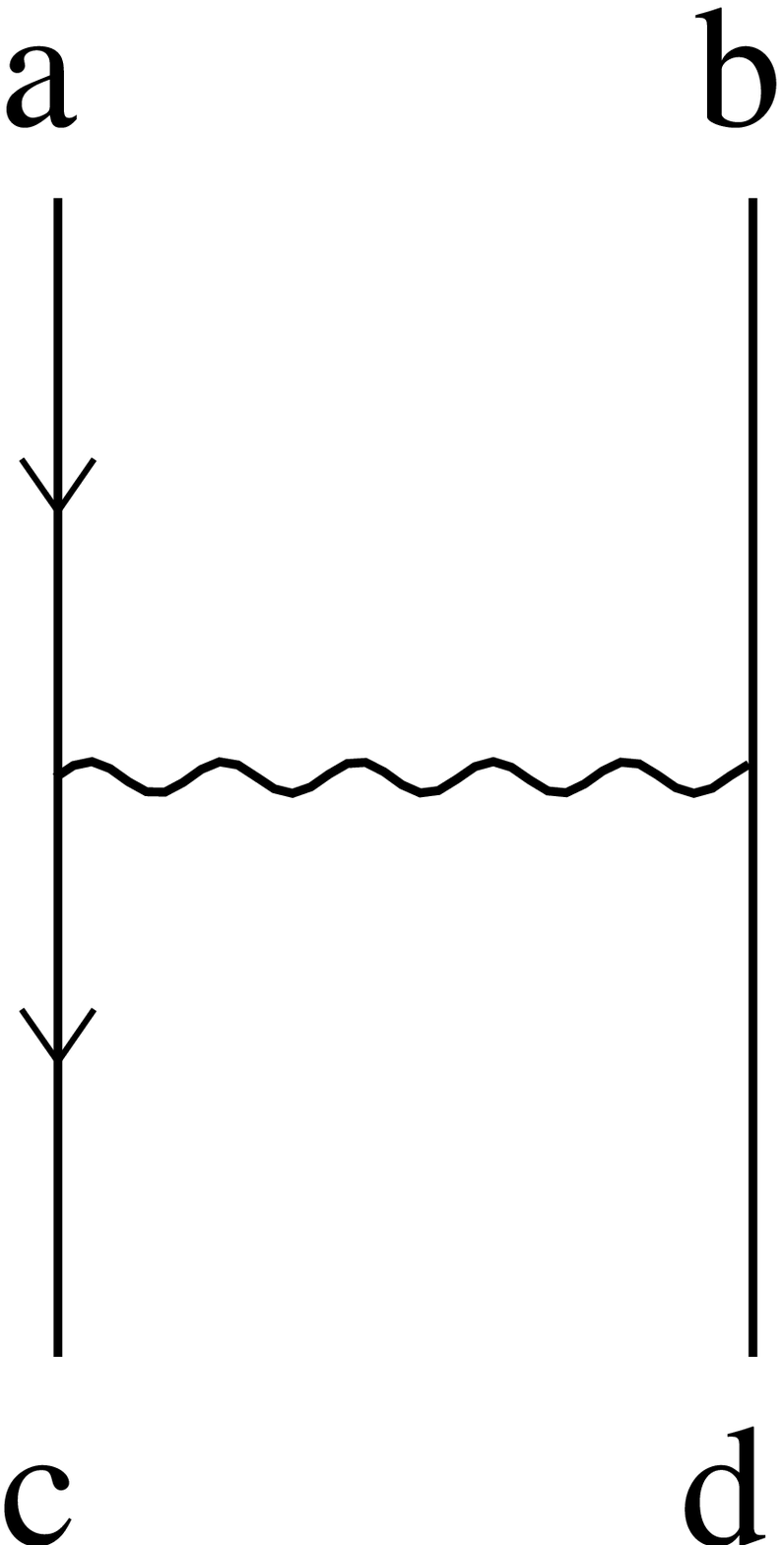}
\end{minipage}
&=\frac{1}{2}\times4\times
\frac{f_{pa_1b_1}}{\gym^2}\frac{f_{pa_2b_2}}{2\gym^2}
\delta_{ij}\notag\\[-1cm]
&\quad\times\left(\frac{\gym^2}{4\pi^2}\right)^5\int d^4u d^4v
\frac{1}{(u-v)^2}\frac{1}{(x-u)^2}
\lrp_{u_\mu}\frac{1}{(y-u)^2}
\frac{1}{(x-v)^2}\lrp_{v^\mu}\frac{1}{(y-v)^2}\notag\\
&=\frac{1}{2}
\left(\frac{\gym^2}{4\pi^2}\right)^3
f_{pa_1b_1}f_{pa_2b_2}\delta_{ij}\frac{\ln(x-y)^2\Lambda^2}{(x-y)^4}
\end{align}
while the scalar 4-point function is
\begin{align}
\begin{minipage}{2cm}
\renewcommand{\h}{2.5cm}
\psfrag{a}{$\hs{-.1cm}Z^{a_1}$}\psfrag{b}{$\phi^{ia_2}$}
\psfrag{c}{\begin{minipage}{1cm}\vs{.1cm}\hs{-0cm}$\hs{-.1cm}\bar Z^{b_1}$\end{minipage}}
\psfrag{d}{\begin{minipage}{1cm}\vs{.1cm}\hs{-.1cm}$\phi^{jb_2}$\end{minipage}}
\includegraphics[height=\h]{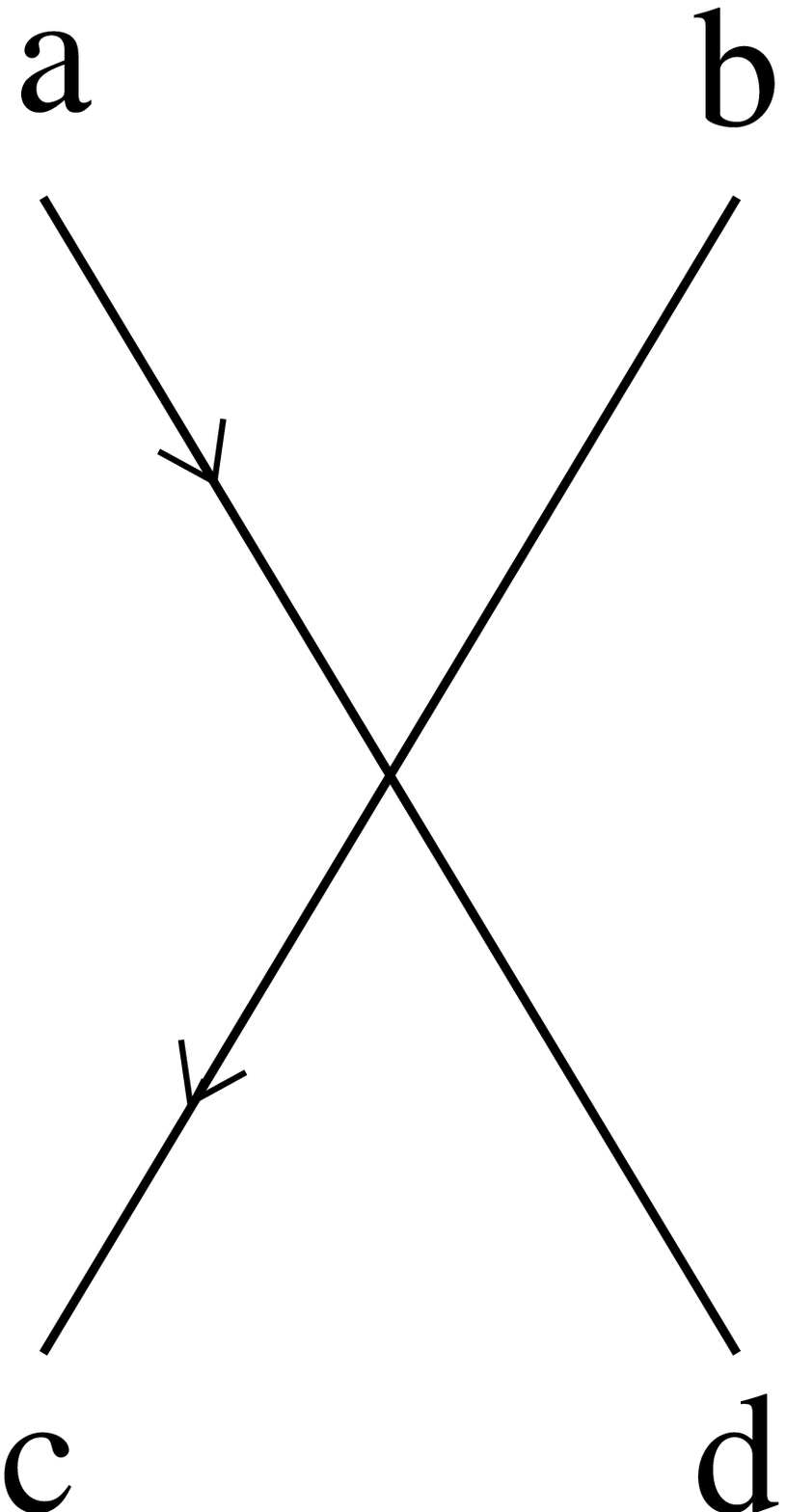}
\end{minipage}
= \frac{1}{2}\left(\frac{\gym^2}{4\pi^2}\right)^3
(f_{pa_1b_2}f_{pa_2b_1}-f_{pa_1a_2}f_{pb_1b_2})\delta_{ij}
\frac{\ln(x-y)^2\Lambda}{(x-y)^4}
\end{align}
Then, using the Jacobi identity, we obtain
\begin{align}
\begin{minipage}{1.8cm}
\renewcommand{\h}{2.5cm}
\psfrag{a}{$\hs{-.1cm}Z^{a_1}$}\psfrag{b}{$\phi^{ia_2}$}
\psfrag{c}{\begin{minipage}{1cm}\vs{.1cm}\hs{-.1cm}$\bar Z^{b_1}$\end{minipage}}
\psfrag{d}{\begin{minipage}{1cm}\vs{.1cm}\hs{-.1cm}$\phi^{jb_2}$\end{minipage}}
\includegraphics[height=\h]{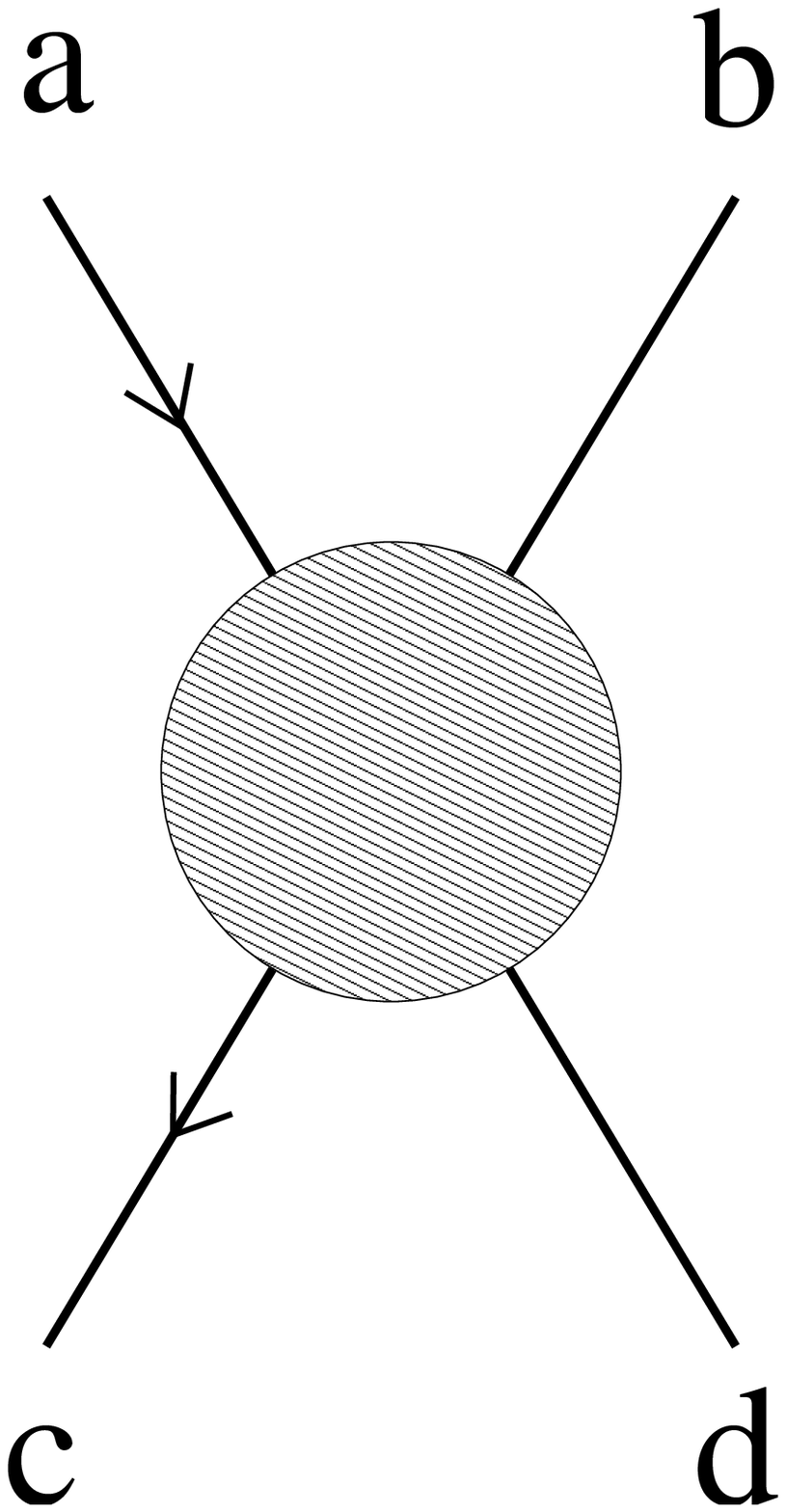}
\end{minipage}
=\begin{minipage}{1.8cm}
\psfrag{a}{}\psfrag{b}{}\psfrag{c}{}\psfrag{d}{}
\renewcommand{\h}{2.5cm}
\includegraphics[height=\h]{ZP-G-exch.eps}
\end{minipage}
+
\begin{minipage}{1.5cm}
\renewcommand{\h}{2.5cm}
\psfrag{a}{}\psfrag{b}{}\psfrag{c}{}\psfrag{d}{}
\includegraphics[height=\h]{2ZP.eps}
\end{minipage}
=
\left(\frac{\gym^2}{4\pi^2}\right)^3
f_{pa_1b_2}f_{pa_2b_1}\delta_{ij}
\frac{\ln(x-y)^2\Lambda^2}{(x-y)^4}.
\end{align}

\subsubsection{
$\langle
\lambda_{r\a}^{a_1}(x)
\phi^{ia_2}(x)
\bar\lambda^{b_1}_{s\dot\a}(y)
\phi^{ib_2}(y)
\rangle$}

There are two possible contraction through Yukawa interactions:
\begin{align}
\begin{minipage}{2cm}
\renewcommand{\h}{2.5cm}
\psfrag{a}{\hs{-.1cm}$\lambda_{r\a}^{a_1}$}\psfrag{b}{$\phi^{ia_2}$}
\psfrag{c}{\begin{minipage}{1cm}\vs{.1cm}\hs{-.1cm}$\bar\lambda_{s\dot\a}^{b_1}$\end{minipage}}
\psfrag{d}{\begin{minipage}{1cm}\vs{.1cm}\hs{-.1cm}$\phi^{jb_2}$\end{minipage}}
\psfrag{e}{$\theta$}
\includegraphics[height=\h]{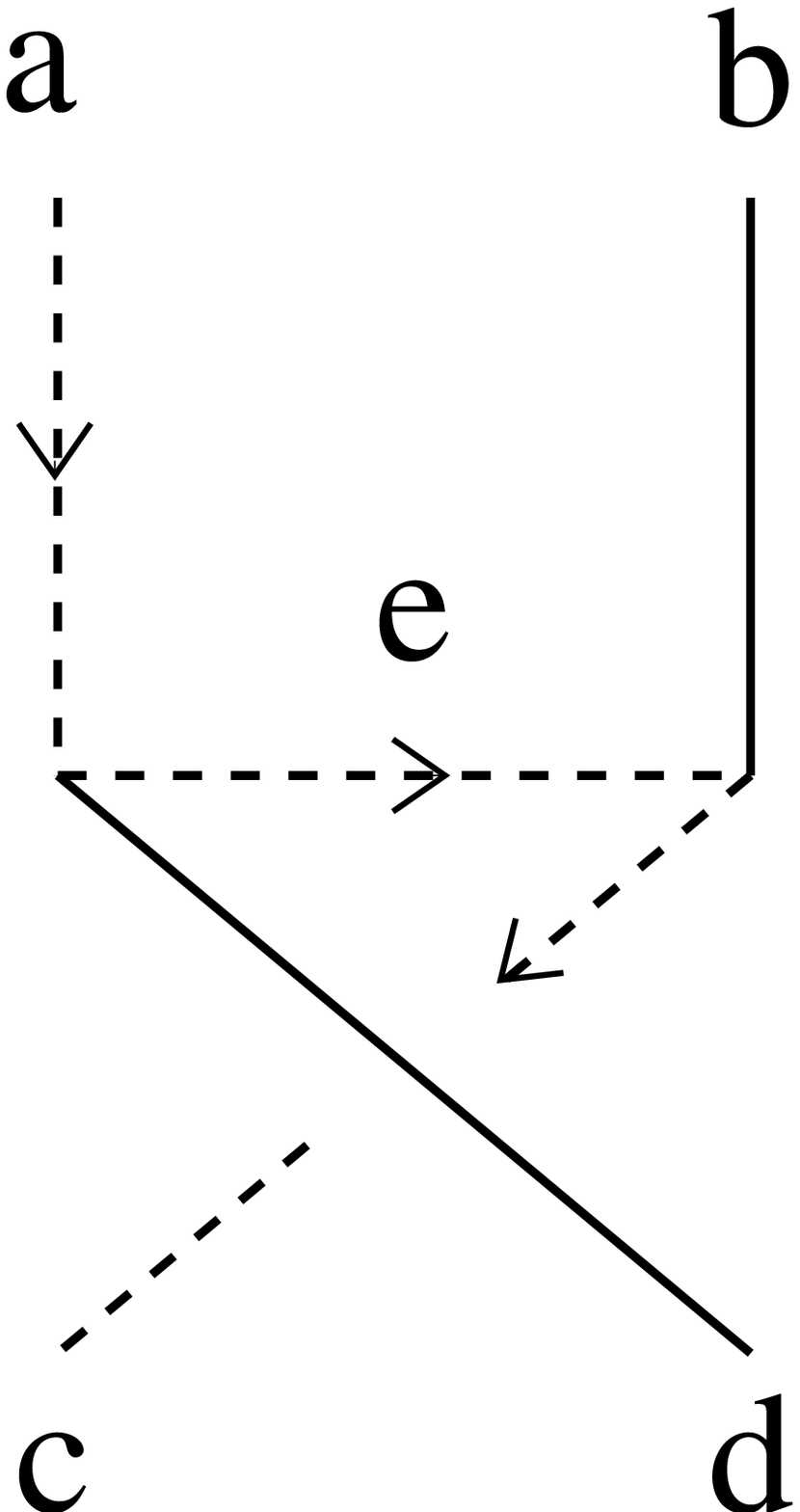}
\end{minipage}
&=\frac{1}{2}\times2\times\frac{if_{pa_1b_2}}{\gym^2}\frac{if_{pb_1a_2}}{\gym^2}
(\bar \tau^i)^{\dot t}_{~s}(\tau^{j})_{r\dot t}
(\sigma^\mu\bar\sigma^\nu\sigma^\rho)_{\a\dot\a}\\[-.7cm]
&\quad\times
\left(\frac{\gym^2}{4\pi^2}\right)^5
\int d^4u d^4v\frac{1}{(x-v)^2}\frac{1}{(y-u)^2}
\partial_{x^\mu}\frac{1}{(x-u)^2}\partial_{u^\nu}\frac{1}{(u-v)^2}
\partial_{v^\rho}{\frac{1}{(y-v)^2}}\notag\\
&=
-\left(\frac{\gym^2}{4\pi^2}\right)^5
\frac{f_{pa_1b_2}}{\gym^2}\frac{f_{pa_2b_1}}{\gym^2}
(\tau^j\bar\tau^i)_{rs}
(\sigma^\mu)_{\a\dot\a}
\left(
2\pi^2\partial_{x^\mu}X_{xxyy}+\frac{1}{2}\Box_x H'_{\mu}
\right)
\notag\\
&=
-\frac{1}{2}
\left(\frac{\gym^2}{4\pi^2}\right)^3
(\tau^j\bar\tau^i)_{rs}f_{pa_1b_2}f_{pa_2b_1}
\frac{1}{(x-y)^2}(\sigma^{\mu})_{\a\dot\a}
\partial_{x^\mu}\left(\frac{1}{(x-y)^2}\right)\ln(x-y)^2\Lambda^2,\notag
\end{align}
\begin{align}
\begin{minipage}{2cm}
\renewcommand{\h}{2.5cm}
\psfrag{a}{\hs{-.1cm}$\lambda_{r\a}^{a_1}$}\psfrag{b}{$\phi^{ia_2}$}
\psfrag{c}{\begin{minipage}{1cm}\vs{.1cm}\hs{-.1cm}$\bar\lambda_{s\dot\a}^{b_1}$\end{minipage}}
\psfrag{d}{\begin{minipage}{1cm}\vs{.1cm}\hs{-.1cm}$\phi^{jb_2}$\end{minipage}}
\psfrag{e}{$\theta$}
\includegraphics[height=\h]{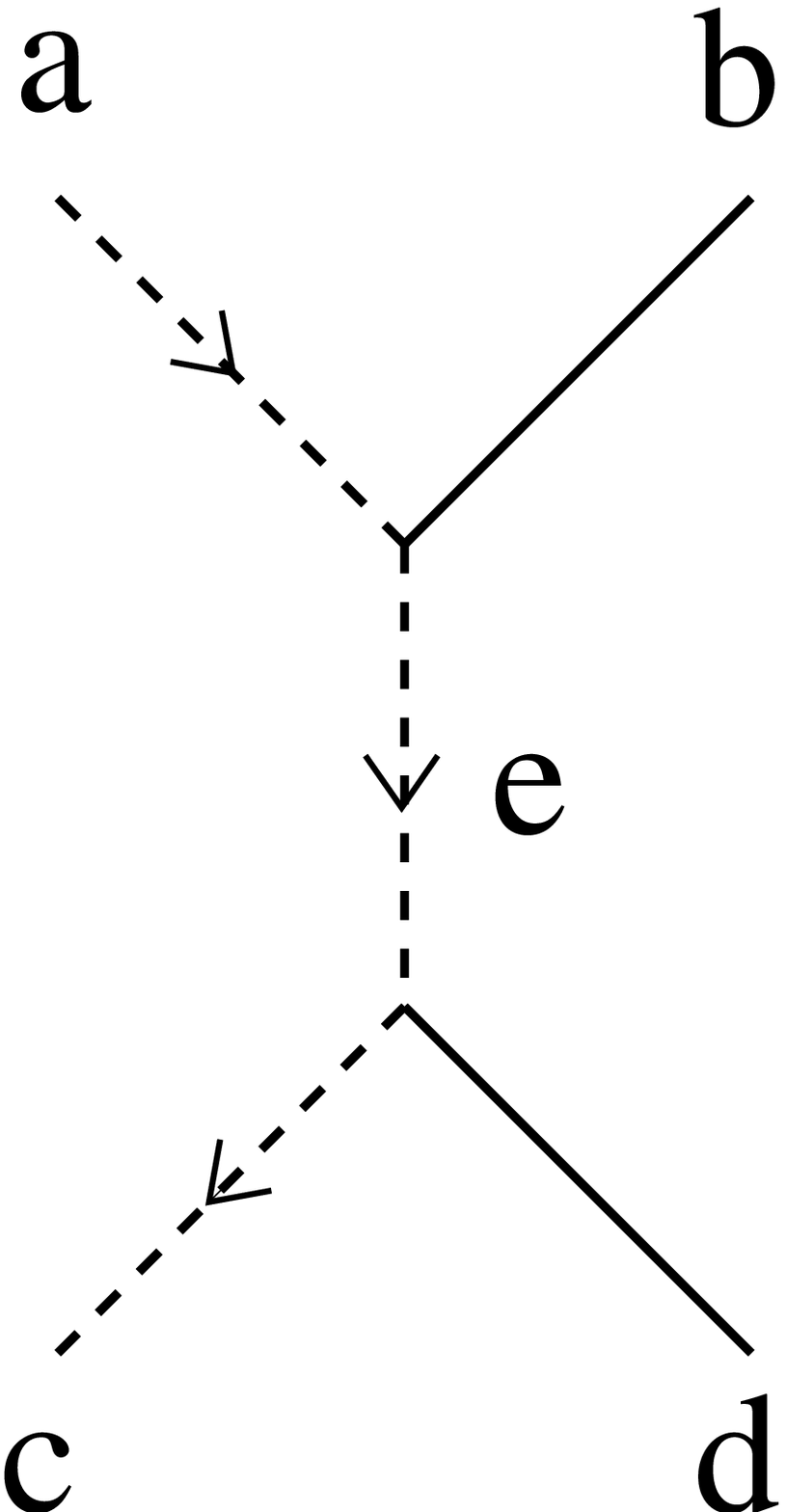}
\end{minipage}
&=
\frac{if_{pa_1a_2}}{\gym^2}\frac{if_{pb_1b_2}}{\gym^2}
(\bar \tau^j)^{\dot t}_{~s}(\tau^{i})_{r\dot t}
(\sigma^\mu\bar\sigma^\nu\sigma^\rho)_{\a\dot\a}\\[-.8cm]
&\quad \times
\left(\frac{\gym^2}{4\pi^2}\right)^5
\int d^4ud^4v\frac{1}{(x-u)^2}\frac{1}{(y-v)^2}
\partial_{x^\mu}\frac{1}{(x-u)^2}\partial_{u^\nu}\frac{1}{(u-v)^2}
\partial_{v^\rho}{\frac{1}{(y-v)^2}}\notag\\
&=
\left(\frac{\gym}{4\pi^2}\right)^5
\frac{f_{pa_1a_2}}{\gym^2}\frac{f_{pb_1b_2}}{\gym^2}
(\tau^i\bar\tau^j)_{rs}
(\sigma^\mu)_{\a\dot\a}\pi^2\partial_{x^\mu}X_{xxyy}\notag\\
&=
\frac{1}{4}\left(\frac{\gym^2}{4\pi^2}\right)^3f_{pa_1a_2}f_{pb_1b_2}
(\tau^i\bar\tau^j)_{rs}
\frac{1}{(x-y)^2}
(\sigma^{\mu})_{\a\dot\a}
\partial_{x^\mu}\left(\frac{1}{(x-y)^2}\right)\ln(x-y)^2\Lambda^2\notag
\end{align}
The gluon exchange is the same as (\ref{LZ-G-exch}),
\begin{align}
\begin{minipage}{2cm}
\renewcommand{\h}{2.5cm}
\psfrag{a}{\hs{-.1cm}$\lambda_{r\a}^{a_1}$}\psfrag{b}{$\phi^{ia_2}$}
\psfrag{c}{\begin{minipage}{1cm}\vs{.1cm}\hs{-.1cm}$\bar\lambda_{s\dot\a}^{b_1}$\end{minipage}}
\psfrag{d}{\begin{minipage}{1cm}\vs{.1cm}\hs{-.1cm}$\phi^{jb_2}$\end{minipage}}
\includegraphics[height=\h]{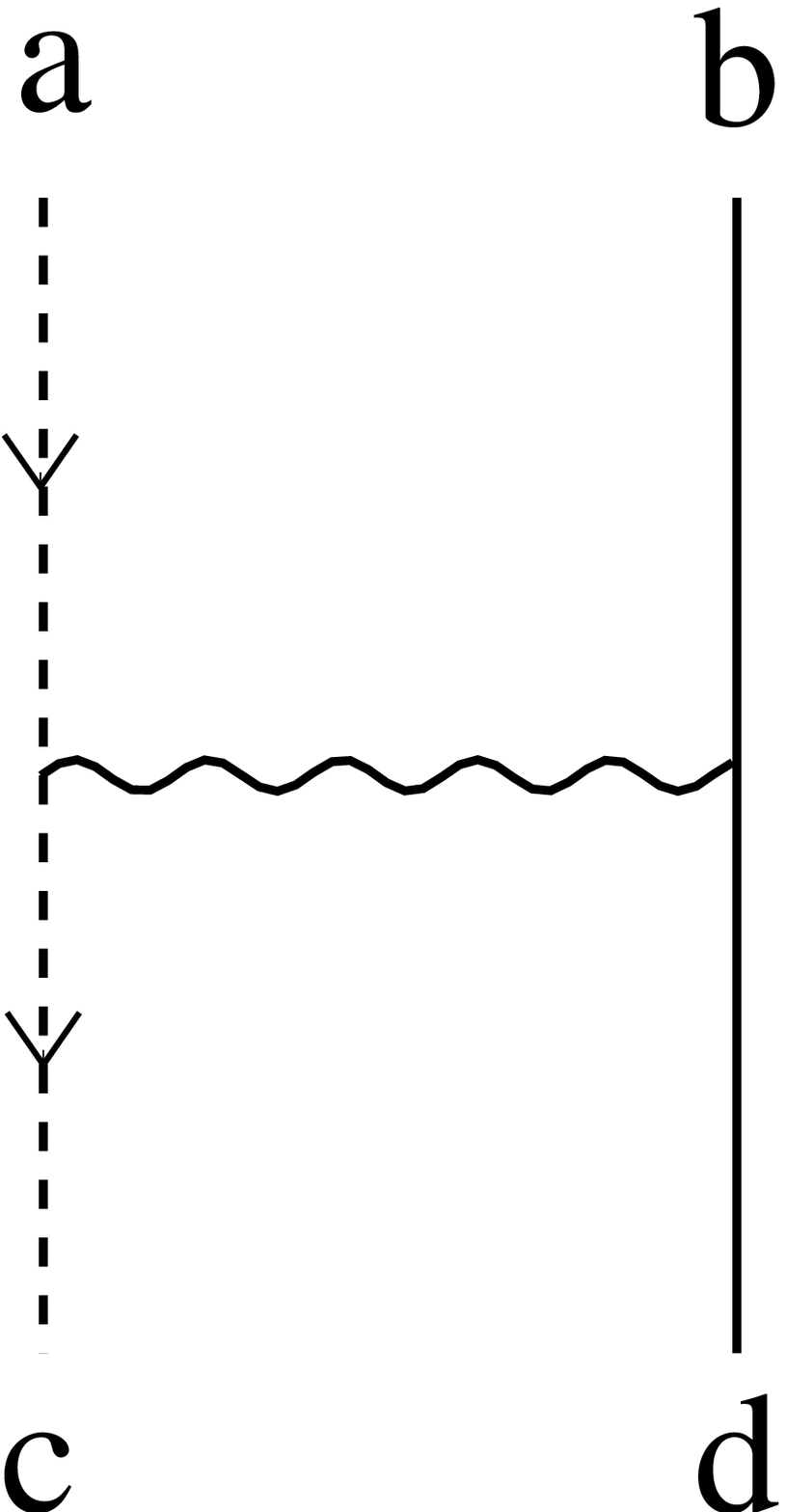}
\end{minipage}
=\frac{1}{2}\left(\frac{\gym^2}{4\pi^2}\right)^3
f_{pa_1b_1}f_{pa_2b_2}\epsilon_{rs}\frac{1}{(x-y)^2}(\sigma^\mu)_{\a\dot\a}
\partial_{x^\mu}\left(\frac{1}{(x-y)^2}\right)\ln(x-u)^2\Lambda^2.
\end{align}
Then we obtain
\begin{align}
\begin{minipage}{1.5cm}
\renewcommand{\h}{2.5cm}
\psfrag{a}{\hs{-.1cm}$\lambda_{r\a}^{a_1}$}\psfrag{b}{$\phi^{ia_2}$}
\psfrag{c}{\begin{minipage}{1cm}\vs{.1cm}\hs{-.1cm}$\bar\lambda_{s\dot\a}^{b_1}$\end{minipage}}
\psfrag{d}{\begin{minipage}{1cm}\vs{.1cm}\hs{-.1cm}$\phi^{jb_2}$\end{minipage}}
\psfrag{e}{$\theta$}
\includegraphics[height=\h]{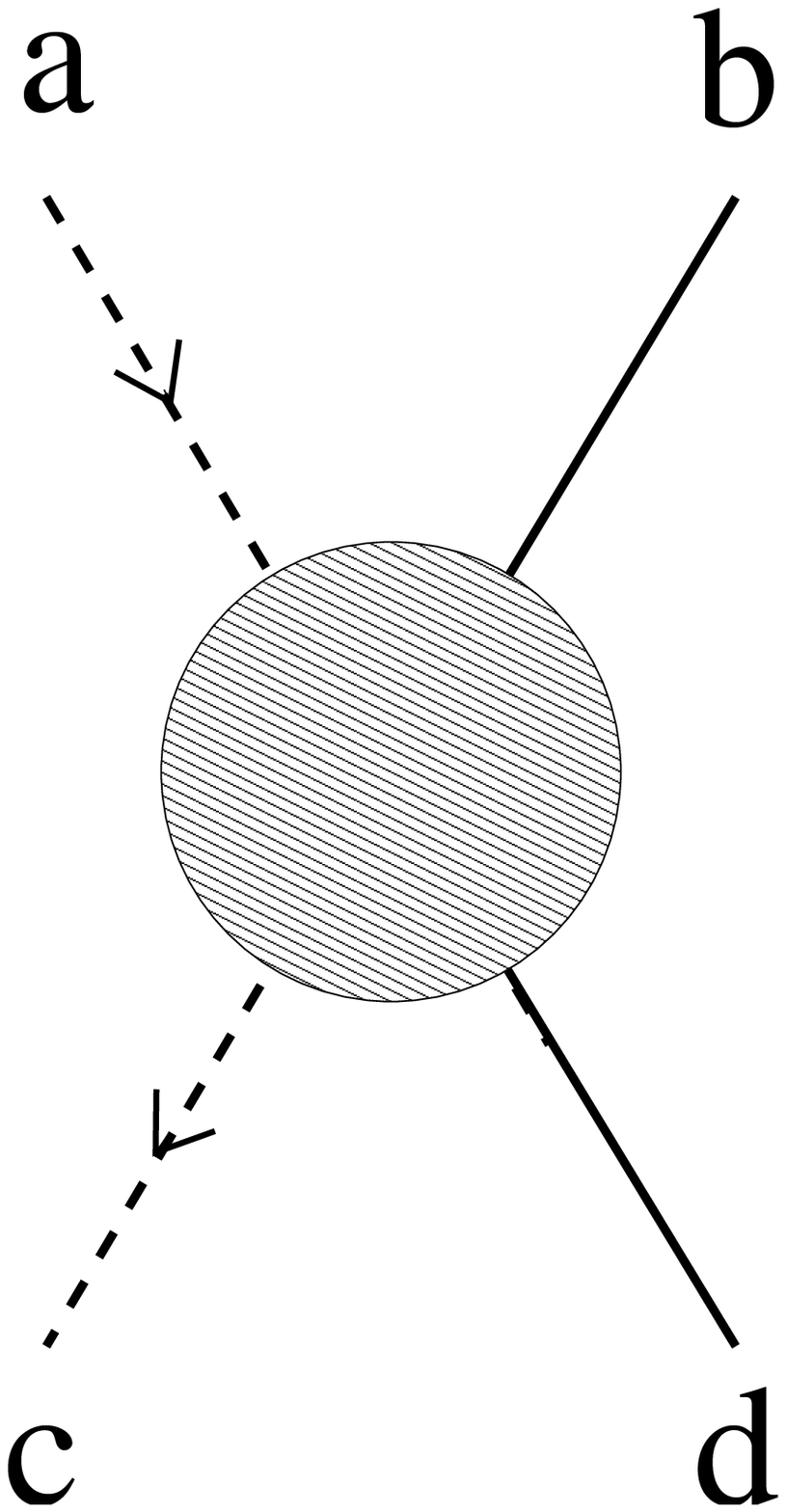}
\end{minipage}
&=
\psfrag{a}{}\psfrag{b}{}\psfrag{c}{}\psfrag{d}{}\psfrag{e}{$\theta$}
\begin{minipage}{1.5cm}
\renewcommand{\h}{2.5cm}
\psfrag{e}{\hs{-.05cm}$\theta$}
\includegraphics[height=\h]{LP-T-exch.eps}
\end{minipage}
+
\begin{minipage}{1.5cm}
\renewcommand{\h}{2.5cm}
\psfrag{e}{$\theta$}
\includegraphics[height=\h]{LP-T-exch2.eps}
\end{minipage}
+\begin{minipage}{2cm}
\renewcommand{\h}{2.5cm}
\psfrag{a}{}\psfrag{b}{}\psfrag{c}{}\psfrag{d}{}
\includegraphics[height=\h]{LP-G-exch.eps}
\end{minipage}\\
&=
\frac{1}{2}
\left(\frac{\gym^2}{4\pi^2}\right)^3
\left(
-(\tau^j\bar\tau^i)_{rs}f_{pa_1b_2}f_{pa_2b_1}
+\frac{1}{2}(\tau^i\bar\tau^j)_{rs}f_{pa_1a_2}f_{pb_1b_2}
+\delta_{ij}\epsilon_{rs}f_{pa_1b_1}f_{pa_2b_2}
\right)\notag\\
&\hspace{1cm}\times\frac{1}{(x-y)^2}\sigma^\mu_{\a\dot\a}
\partial_{x^\mu}
\left(\frac{1}{(x-y)^2}\right)\ln(x-y)^2\Lambda^2\notag.
\end{align}

\subsubsection{
$\langle\lambda^{a_1}_{r\a}(x)\phi^{ia_2}(x)
\bar\theta^{b_1}_{s\dot\a}(y)\bar Z^{b_2}(y)\rangle$
}
There are two kinds of the Yukawa interaction such as
\begin{align}
\begin{minipage}{2cm}
\renewcommand{\h}{2.5cm}
\psfrag{a}{\hs{-.05cm}$\lambda_{r\a}^{a_1}$}\psfrag{b}{$\phi^{ia_2}$}
\psfrag{c}{\begin{minipage}{1cm}\vs{.1cm}\hs{-.1cm}$\bar\theta_{\dot r\dot\a}^{b_1}$\end{minipage}}
\psfrag{d}{\begin{minipage}{1cm}\vs{.1cm}\hs{-.1cm}$\bar Z^{b_2}$\end{minipage}}
\psfrag{e}{\hs{-.05cm}$\lambda$}
\includegraphics[height=\h]{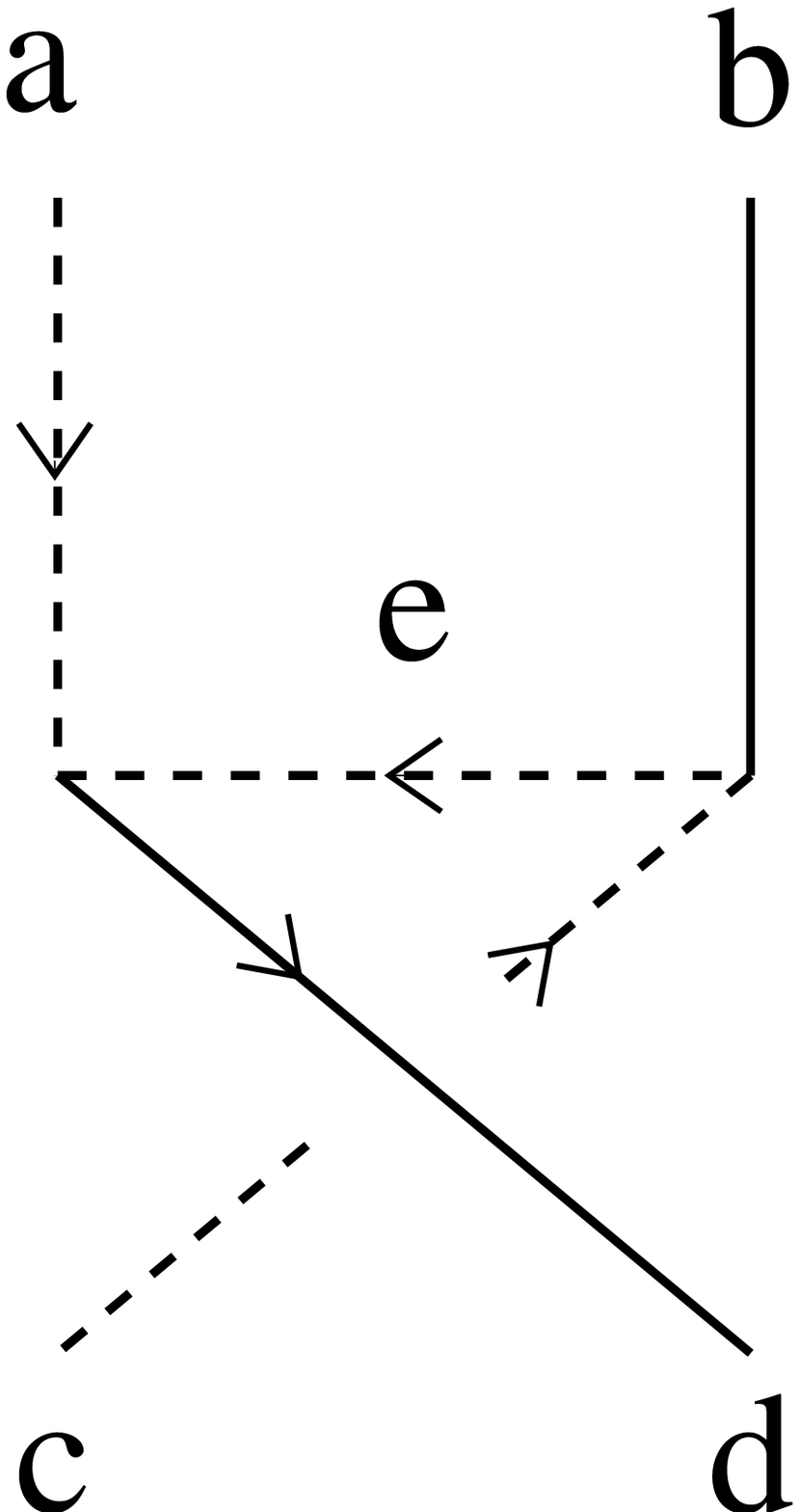}
\end{minipage}
&=
\frac{1}{2}\times4\times
\frac{-i f_{pa_1b_2}}{\sqrt{2}\gym^2}\frac{-if_{pa_2b_1}}{\gym^2}
(\tau^i)_{r\dot r}
(\sigma^\mu\bar\sigma^\nu\sigma^\rho)_{\a\dot\a}\\[-.8cm]
&\quad\times
\left(\frac{\gym^2}{4\pi^2}\right)^5
\int d^4ud^4v\frac{1}{(x-v)^2}\frac{1}{(y-u)^2}
\partial_{x^\mu}\frac{1}{(x-u)^2}
\partial_{v^\nu}\frac{1}{(u-v)^2}
\partial_{y^\rho}\frac{1}{(y-v)^2}\notag\\
&=\frac{1}{\sqrt{2}}
\left(\frac{\gym^2}{4\pi^2}\right)^3
(\tau^i)_{r\dot r}f_{pa_1b_2}f_{pa_2b_1}
\frac{1}{(x-y)^2}(\sigma^\mu)_{\a\dot\a}
\partial_{x^\mu}\left(\frac{1}{(x-y)^2}\right)\ln(x-y)^2\Lambda^2\notag
\end{align}
\begin{align}
\begin{minipage}{2cm}
\renewcommand{\h}{2.5cm}
\psfrag{a}{$\lambda_{r\a}^{a_1}$}\psfrag{b}{$\phi^{ia_2}$}
\psfrag{c}{\begin{minipage}{1cm}\vs{.1cm}\hs{-.1cm}$\bar\theta_{\dot r\dot\a}^{b_1}$\end{minipage}}
\psfrag{d}{\begin{minipage}{1cm}\vs{.1cm}\hs{-.1cm}$\bar Z^{b_2}$\end{minipage}}
\psfrag{e}{$\theta$}
\includegraphics[height=\h]{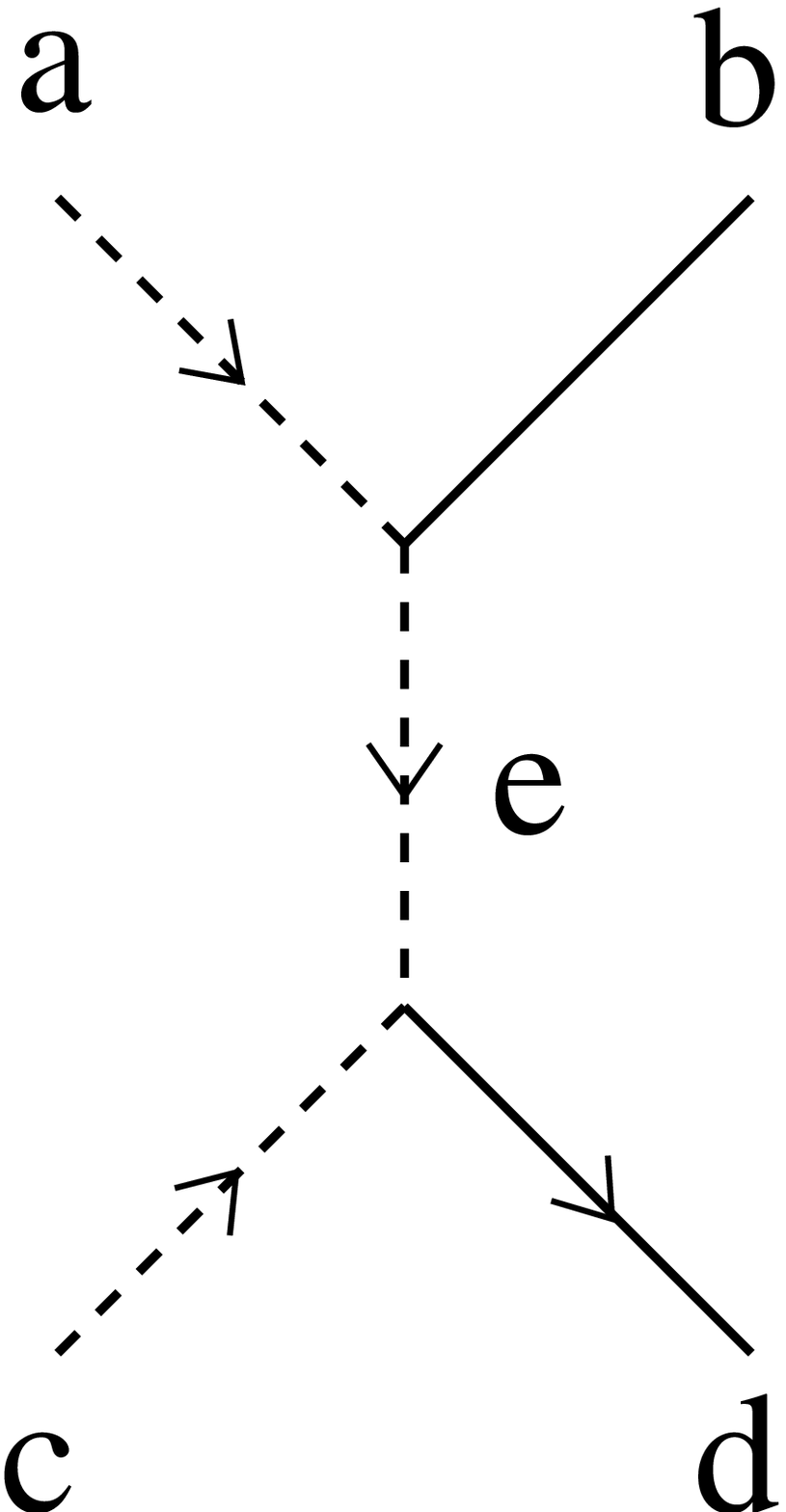}
\end{minipage}
&=\frac{1}{2}\times4\times
(\tau^i)_{r\dot r}
\frac{-if_{pa_1a_2}}{\gym^2}
\frac{if_{pb_1b_2}}{\sqrt{2}\gym^2}
(\sigma^\mu\bar\sigma^\nu\sigma^\rho)_{\a\dot\a}\\[-.8cm]
&\quad\times
\left(\frac{\gym^2}{4\pi^2}\right)^5
\int d^4ud^4v
\frac{1}{(x-u)^2}\frac{1}{(y-v)^2}
\partial_{x^\mu}\frac{1}{(x-u)^2}
\partial_{u^\nu}\frac{1}{(u-v)^2}
\partial_{y^\rho}\frac{1}{(y-v)^2}\notag\\
&=\frac{1}{2\sqrt{2}}
\left(\frac{\gym^2}{4\pi^2}\right)^3
(\tau^i)_{r\dot r}f_{pa_1a_2}f_{pb_1b_2}
\frac{1}{(x-y)^2}(\sigma^\mu)_{\a\dot\a}
\partial_{x^\mu}\left(\frac{1}{(x-y)^2}\right)
\ln(x-y)^2\Lambda^2\notag
\end{align}
Then the sum of the above contribution gives the result (\ref{LPTZ}):
\begin{align}
&\begin{minipage}{1.5cm}
\renewcommand{\h}{2.5cm}
\psfrag{a}{$\lambda_{r\a}^{a_1}$}\psfrag{b}{$\phi^{ia_2}$}
\psfrag{c}{\begin{minipage}{1cm}\vs{.1cm}\hs{-.1cm}$\bar\theta_{\dot r\dot\a}^{b_1}$\end{minipage}}
\psfrag{d}{\begin{minipage}{1cm}\vs{.1cm}\hs{-.1cm}$\bar Z^{b_2}$\end{minipage}}
\psfrag{e}{$\lambda$}
\includegraphics[height=\h]{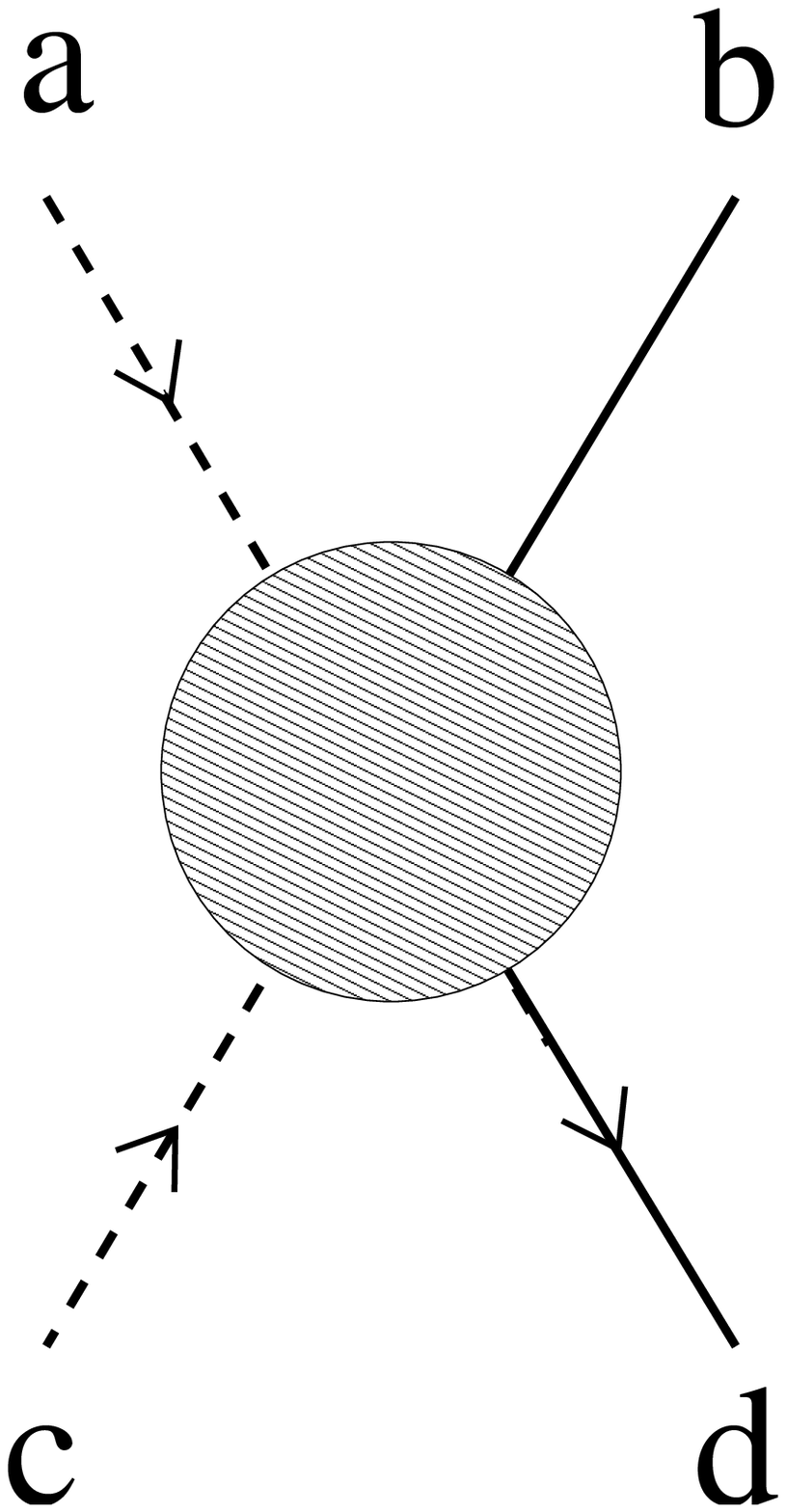}
\end{minipage}
=
\psfrag{a}{}\psfrag{b}{}\psfrag{c}{}\psfrag{d}{}
\begin{minipage}{1.5cm}
\renewcommand{\h}{2.5cm}
\psfrag{e}{$\lambda$}
\includegraphics[height=\h]{LPTZ-L-exch.eps}
\end{minipage}
+
\begin{minipage}{1.5cm}
\renewcommand{\h}{2.5cm}
\psfrag{e}{$\theta$}
\includegraphics[height=\h]{LPTZ-T-exch.eps}
\end{minipage}\\
&=\frac{1}{\sqrt{2}}
\left(\frac{\gym^2}{4\pi^2}\right)^{3}
(\tau^{i})_{r\dot r}
(f_{pa_1b_2}f_{pa_2b_1}+\frac{1}{2}f_{pa_1a_2}f_{pb_1b_2})
\frac{1}{(x-y)^2}
(\sigma^{\mu})_{\a\dot\a}\partial_{x^\mu}
\left(\frac{1}{(x-y)^2}\right)\\[-0cm]
&\hspace{12cm}\times\ln(x-y^2)\Lambda^2.\notag
\end{align}

\subsubsection{
$\langle
\theta^{a_1}_{\dot r\a}(x)Z^{a_2}(x)
\bar\theta^{b_1}_{\dot s\dot\a}(y)\bar Z^{b_2}(y)\rangle$
}

The contribution from the Yukawa coupling is
\begin{align}
\begin{minipage}{2cm}
\renewcommand{\h}{2.5cm}
\psfrag{a}{\hs{-.1cm}$\theta_{\dot r\a}^{a_1}$}\psfrag{b}{$Z^{a_2}$}
\psfrag{c}{\begin{minipage}{1cm}\vs{.1cm}\hs{-.1cm}$\bar\theta_{\dot s\dot\a}^{b_1}$\end{minipage}}
\psfrag{d}{\begin{minipage}{1cm}\vs{.1cm}\hs{-.1cm}$\bar Z^{b_2}$\end{minipage}}
\psfrag{e}{$\theta$}
\includegraphics[height=\h]{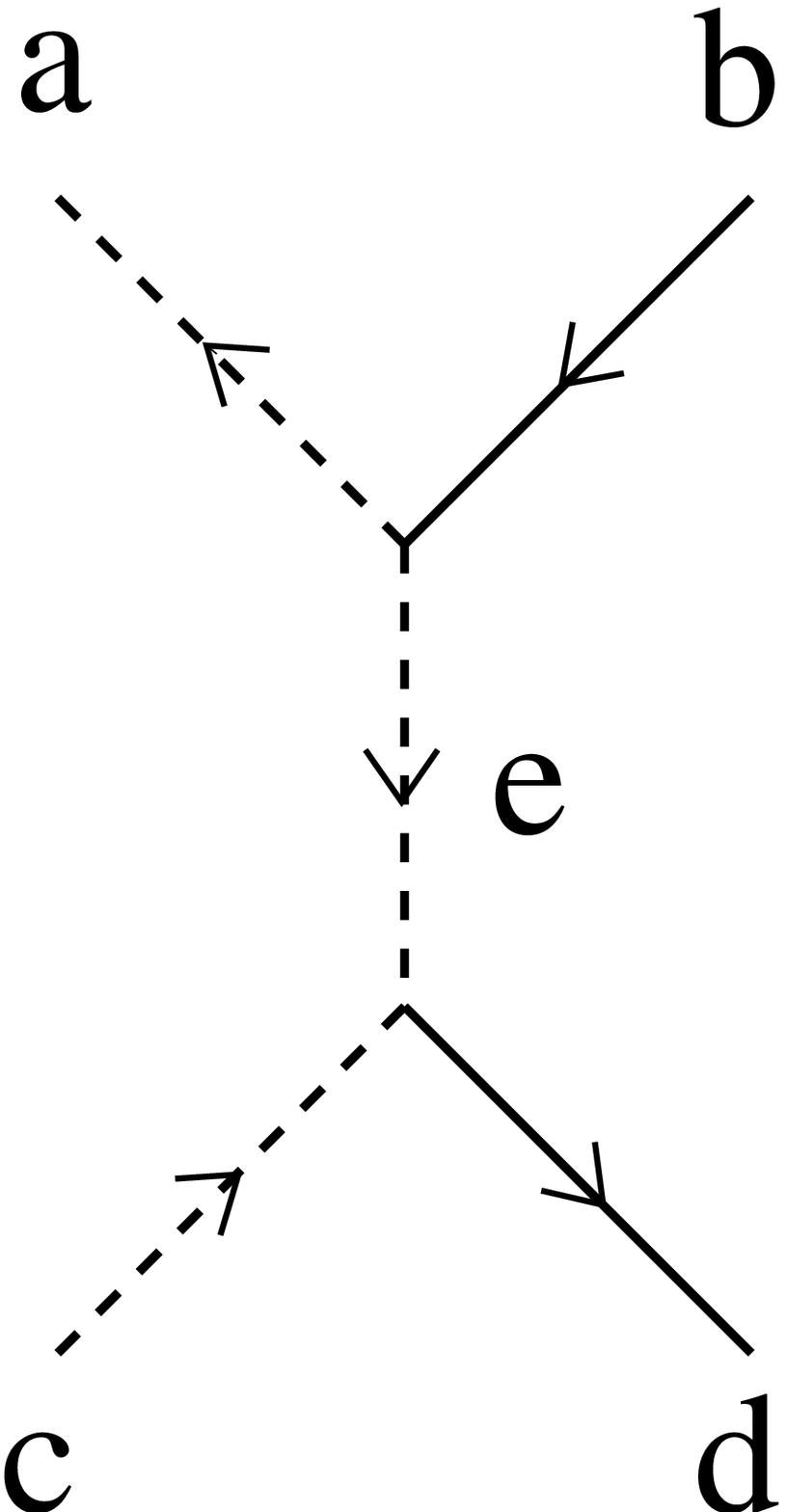}
\end{minipage}
&= \frac{1}{2}\times 8\times 
\frac{f_{pa_1a_2}}{\gym^2}\frac{f_{pb_2b_1}}{\gym^2}
\epsilon_{\dot r \dot s}
(\sigma^\mu\sigma^\nu\sigma^\rho)_{\a\dot\a}\\[-.7cm]
&\quad\times
\left(\frac{\gym^2}{4\pi^2}\right)^5
\int d^4u d^4v\frac{1}{(x-u)^2}\frac{1}{(y-v)^2}
\partial_{u^{\mu}}\frac{1}{(x-u)^2}
\partial_{v^{\nu}}\frac{1}{(u-v)^2}
\partial_{y^{\rho}}\frac{1}{(y-v)^2}\notag\\
&=-\frac{1}{2}\left(\frac{\gym^2}{4\pi^2}\right)^3
f_{pa_1a_2}f_{pb_1b_2}
\epsilon_{\dot r \dot s}
\frac{1}{(x-y)^2}
(\sigma^{\mu})_{\a\dot\a}\partial
\left(\frac{1}{(x-y)^2}\right)\ln(x-y)^2\Lambda^2\notag
\end{align}
The gluon exchange is the same as (\ref{LZ-G-exch}),
\begin{align}
\begin{minipage}{2cm}
\renewcommand{\h}{2.5cm}
\psfrag{a}{\hs{-.1cm}$\theta_{\dot r\a}^{a_1}$}\psfrag{b}{\hs{-.1cm}$Z^{a_2}$}
\psfrag{c}{\begin{minipage}{1cm}\vs{.1cm}\hs{-.1cm}$\bar\theta_{\dot s\dot\a}^{b_1}$\end{minipage}}
\psfrag{d}{\begin{minipage}{1cm}\vs{.1cm}\hs{-.1cm}$\bar Z^{b_2}$\end{minipage}}
\includegraphics[height=\h]{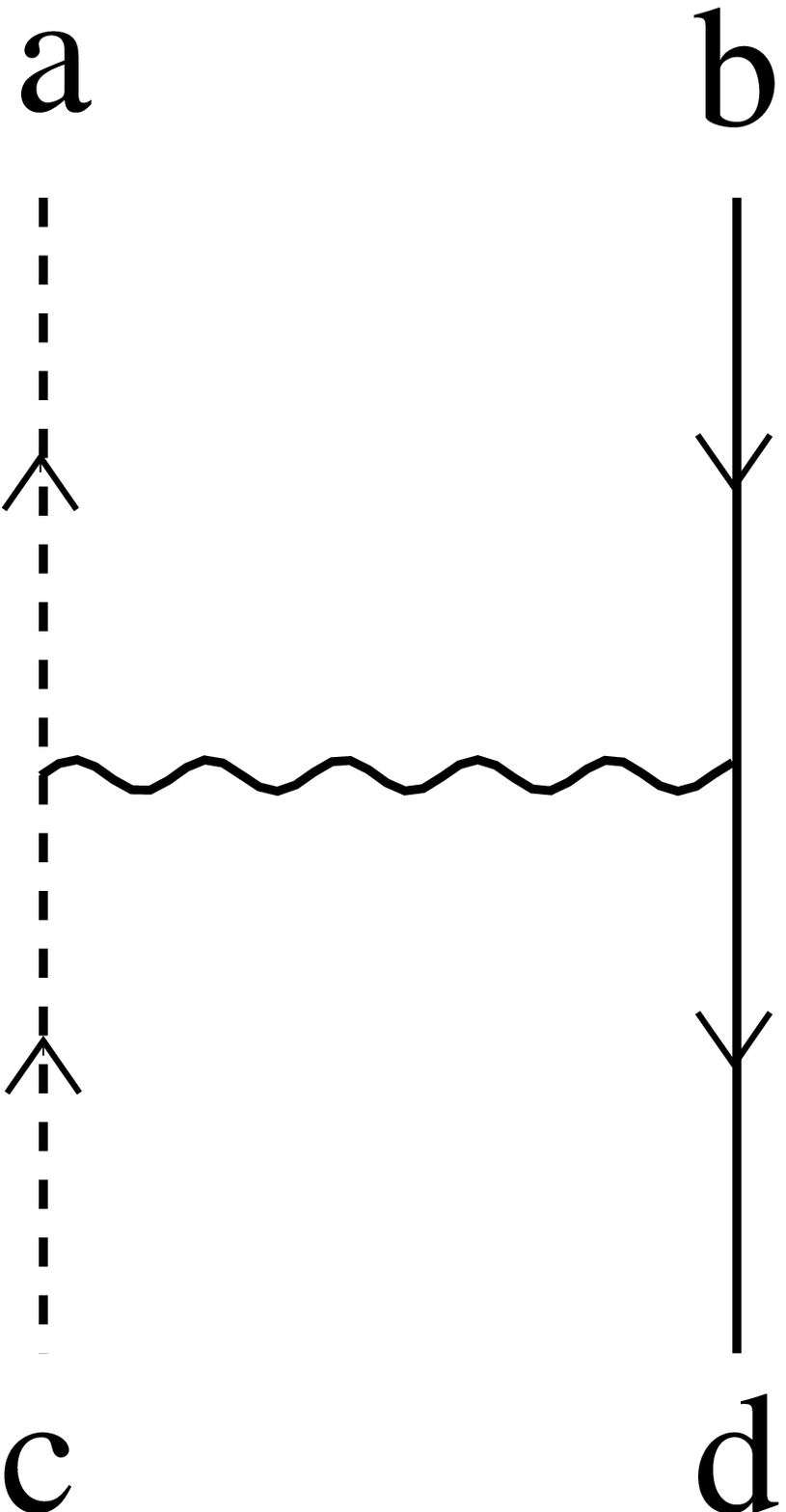}
\end{minipage}
=\frac{1}{2}\left(\frac{\gym^2}{4\pi^2}\right)^3
f_{pa_1b_1}f_{pa_2b_2}\epsilon_{\dot r\dot s}
\frac{1}{(x-y)^2}(\sigma^\mu)_{\a\dot\a}
\partial_{x^\mu}\left(\frac{\ln(x-u)^2\Lambda^2}{(x-y)^2}\right).
\end{align}
Therefore we obtain
\begin{align}
\begin{minipage}{1.5cm}
\renewcommand{\h}{2.5cm}
\psfrag{a}{\hs{-.1cm}$\theta_{\dot r\a}^{a_1}$}\psfrag{b}{\hs{-.1cm}$Z^{a_2}$}
\psfrag{c}{\begin{minipage}{1cm}\vs{.1cm}\hs{-.1cm}$\bar\theta_{\dot s\dot\a}^{b_1}$\end{minipage}}
\psfrag{d}{\begin{minipage}{1cm}\vs{.1cm}\hs{-.1cm}$\bar Z^{b_2}$\end{minipage}}
\includegraphics[height=\h]{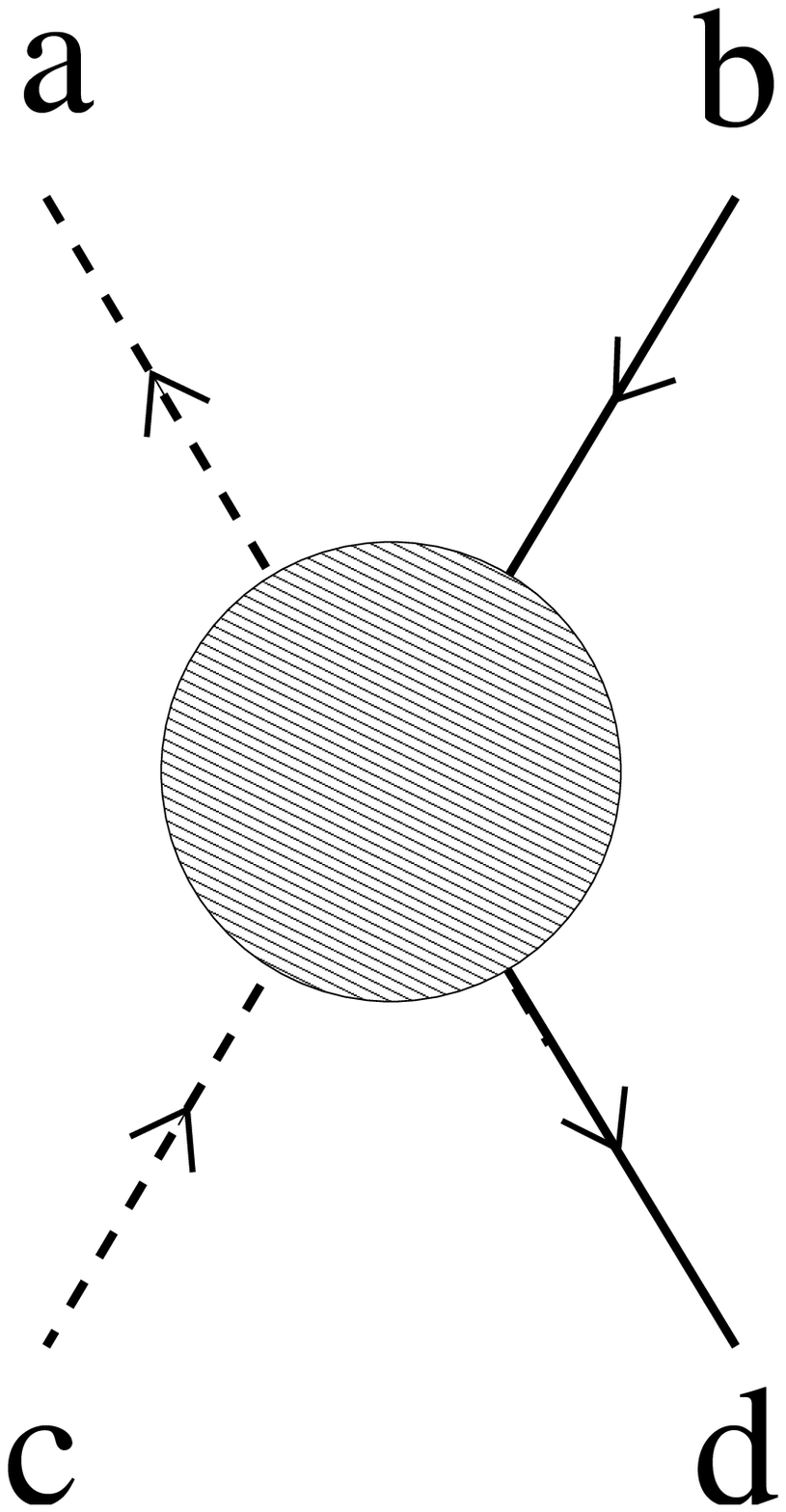}
\end{minipage}
=
\psfrag{a}{}\psfrag{b}{}\psfrag{c}{}\psfrag{d}{}
\begin{minipage}{1.5cm}
\renewcommand{\h}{2.5cm}
\psfrag{e}{$\theta$}
\includegraphics[height=\h]{TZ-T-exch.eps}
\end{minipage}
+
\begin{minipage}{1.5cm}
\renewcommand{\h}{2.5cm}
\includegraphics[height=\h]{TZ-G-exch.eps}
\end{minipage}
&=
-\frac{1}{2}\left(\frac{\gym^2}{4\pi^2}\right)^3
\left(
f_{pa_1b_2}f_{pa_2b_1}
-f_{pa_1b_1}f_{pa_2b_2}
\right)\epsilon_{\dot r\dot s}\notag\\[-.5cm]
&\quad\times\frac{1}{(x-y)^2}\sigma^\mu_{\a\dot\a}
\partial_{x^\mu}
\left(\frac{1}{(x-y)^2}\right)\ln(x-y)^2\Lambda^2.
\end{align}

\section{Prefactor}\label{AppPre}
The form of the prefactor of the interaction vertex $\HSV$ is\footnote{
This definition differs form the one
in (3.28) of \cite{Pan2} by a factor $-{2\a}/{\a'}$.
Note also 
the difference of the total factor of $K^i$ between here and there.}
\begin{eqnarray}
&&\hspace{-2cm}\HSV=P_{SV}\ket{\E},\\
&&\hspace{-2cm}P_{SV}=\frac{1}{2}\left[
 \left(K^{i}\widetilde K^{j}+\mu
\delta^{ij}\right)V_{ij}
-\left(K^{\mu}\widetilde K^{\nu}+\mu
\delta^{\mu\nu}\right)V_{\mu\nu}
\right.\nonumber\\
&&\left.-K^{\dot\a_1\a_1}\widetilde K^{\dot\a_2\a_2}
 S_{\a_1\a_2}(\YI)S^{*}_{\dot\a_1\dot\a_2}(\YII)
-\widetilde K^{\dot\a_1\a_1} K^{\dot\a_2\a_2}
 S^{*}_{\a_1\a_2}(\YI)S_{\dot\a_1\dot\a_2}(\YII)
\right]
\end{eqnarray}
where $K^I$, $\widetilde K^I$ 
($I=i/\mu$ denotes a scalar/vector direction) and
$Y^{\a_1\a_2}$, $Z^{\dot\a_1\dot\a_2}$ are 
bosonic and fermionic constituents of the prefactor defined as
\begin{align}
&K^{J}=\XI^{J}+\XII^{J},\qquad \widetilde K^J=\XI^J-\XII^J,\\
&\XI= \sum_{r=1}^{3}\sum_{n=0}^{\infty}\bar F_n^{(r)}a_{n}^{(r)\dagger},
\quad \XII=\sum_{r=1}^{3}\sum_{n=1}^{\infty}
 i U_{n(r)}\bar F_n^{(r)}a_{-n}^{(r)\dagger},\\
&Y^{\a_1\a_2}=
\sum_{r=1}^3\sum_{n=0}^{\infty}\bar G_n^{(r)}b_n^{(r)\a_1\a_2\dagger}, 
\quad Z^{\dot\a_1\dot\a_2}=
\sum_{r=1}^3\sum_{n=0}^{\infty}
\bar G_n^{(r)}b_{-n}^{(r)\dot\a_1\dot\a_2\dagger},
\end{align}
with
\begin{align}
&\bar F_n^{(r)}=\sqrt{-\frac{\a'}{\a}}(1-4\mu\a K)^{1/2}F_n^{(r)},
\quad
\bar G_n^{(r)}=\sqrt{-\frac{\a'}{\a}}(1-4\mu\a K)^{1/2}G_n^{(r)},\\
&F^{(\inI)}=-\sqrt{\frac{2}{\a'}}\sqrt{\mu\ainI}\ainII,\quad
F^{(\inII)}=\sqrt{\frac{2}{\a'}}\sqrt{\mu\ainII}\ainI,\quad
F^{(\out)}=0,\\
&F^{(r)}_n=-\frac{\a}{\sqrt{\a'}}\,\frac{1}{1-4\mu\a K}\,
\frac{1}{\ar}
(U_{(r)}^{-1}C_{(r)}^{1/2}CN^r)_{n} \quad(n>0),\\
&
G^{(\inI)}=-\sqrt{\frac{1}{\a'}}\sqrt{\ainI}\ainII,\quad
G^{(\inII)}=\sqrt{\frac{1}{\a'}}\sqrt{\ainII}\ainI,\quad
G^{(\out)}=0,\\
&
G^{(r)}_n=-\frac{\a}{\sqrt{\a'}}\,\frac{1}{1-4\mu\a K}\,
\frac{e(\ar)}{\sqrt{|\ar|}}
(U_{(r)}^{-1/2}C_{(r)}^{1/2}C^{1/2}N^r)_{n} \quad(n>0).
\end{align}
The other quantities in the prefactor is defined as 
\begin{eqnarray}
&&\hspace{-1cm}V_{ij}\equiv \delta_{ij}
 \left[
 1+\frac{1}{12}(Y^4+Z^4)+\frac{1}{144}Y^4 Z^4
 \right] \nonumber\\
&&\hspace{3cm}-\frac{i}{2}\left[
 \YI{}_{ij}^2(1+\frac{1}{12}Z^4)-Z^2_{ij}(1+\frac{1}{12}Y^4)
 \right]
+\frac{1}{4}(Y^2Z^2)_{ij},\\
&&\hspace{-1cm}V_{\mu\nu}\equiv\delta_{\mu\nu}
 \left[
 1-\frac{1}{12}(Y^4+Z^4)+\frac{1}{144}Y^4Z^4
 \right] \nonumber\\
&&\hspace{3cm}-\frac{i}{2}\left[
 \YI{}_{\mu\nu}^2(1-\frac{1}{12}Z^4)-Z^2_{\mu\nu}(1-\frac{1}{12}Y^4)
 \right]
+\frac{1}{4}(Y^2Z^2)_{\mu\nu},\\
&&\hspace{-1cm}S(Y)\equiv Y+\frac{i}{3}Y^3,
\end{eqnarray}
with
\begin{eqnarray}
&& K^{\dot\a_r\b_r}\equiv K^i\sigma_i^{\dot\a_r\b_r},\quad
\widetilde K^{\dot\a_r\b_r}\equiv K^i\sigma_i^{\dot\a_r\b_r},\qquad
(r=1,2)\\
&&Y^2_{\a_1\b_1}\equiv Y_{\a_1\a_2}Y^{~~\a_2}_{\b_1},\quad
Y^2_{\a_2\b_2}\equiv Y_{\a_1\a_2}Y^{\a_1}_{~~~\b_2},\\
&&Y^3_{\a_1\b_2}\equiv Y^2_{\a_1\b_1}Y^{\b_1}_{~~~\b_2},\quad
Y^4\equiv Y^2_{\a_1\b_1}Y^{2\a_1\b_1},\\
&&Y^{2ij}\equiv Y^{2\a_1\b_1}\sigma^{ij}_{\a_1\b_1},\quad
Z^{2ij}\equiv Z^{2\dot\a_1\dot\b_1}\sigma^{ij}_{\dot\a_1\dot\b_1},\quad
(Y^2Z^2)^{ij}\equiv Y^{2k(i}Z^{2j)k},
\end{eqnarray}
We refer the reader to the ref.\cite{Pan2} for details.

On the other hand,
the interaction vertex presented in \cite{DPPRT}, which is of the form
\begin{eqnarray}
\HD= \sum_{r=1}^{3}\sum_{m=-\infty}^{\infty}
\frac{\omega_m^{(r)}}{\ar}
\left(\sum_{I=1}^{8}
a_m^{(r)I\dagger}a_{m}^{(r)I}
+\sum_{a=1}^{8}
b_m^{(r)a\dagger}b_{m}^{(r)a}
\right)\ket{\E}
\end{eqnarray} can be written as
\begin{eqnarray} 
&&\hspace{-1cm}\HD=P_{D}\ket{\E},\\
&&\hspace{-1cm}P_{D}=\frac{1}{4}\left(K^2+\widetilde K^2\right)  
-Y^{\a_1\a_2}\widetilde Y_{\a_1\a_2}
-Z^{\dot\a_1\dot\a_2} \widetilde Z_{\dot\a_1\dot\a_2},
\end{eqnarray}
where
\begin{eqnarray}
\widetilde Y^{\a_1\a_2}
=\sum_{r=1}^3\frac{n}{\ar}G_n^{(r)}b_n^{(r)\a_1\a_2\dagger},\qquad
\widetilde Z^{\dot\a_1\dot\a_2}
=\sum_{r=1}^3\frac{n}{\ar}G_n^{(r)}b_{-n}^{(r)\a_1\a_2\dagger}.
\end{eqnarray}

\section{Large $\mu$ behavior }\label{AppLaM}
The large $\mu$ behavior of $\widetilde N_{mn}^{rs}$
is given \cite{He et. al.},
for $(m,n)\neq(0,0)$, by
\begin{eqnarray}\label{Neumann for large mu}
&&\hspace{-1cm}\widetilde N_{mn}^{\inI\inI}=\frac{(-1)^{m+n}}{4\pi\mu |\aout|y },
\quad
 \widetilde N_{mn}^{\inI\inII}=\frac{(-1)^{m+1}}
{4\pi\mu|\aout| \sqrt{y(1-y)}}\\
&&\hspace{-1cm}\widetilde N_{mn}^{\inII\inII}=\frac{1}{4\pi\mu |\aout|(1-y)},
\quad
 \widetilde N_{mn}^{\out\out}=\frac{(-1)^{m+n+1}\sin(\pi m y)\sin(\pi n y)}
{\pi \mu|\aout|}\\
&&\hspace{-1cm} \widetilde N_{mn}^{\inI\out}
=\frac{(-1)^{m+n+1}\sin(\pi n y)}
{\pi \sqrt{y}(n-m/y)},\quad
\widetilde N_{mn}^{\inII\out}=\frac{(-1)^{n}\sin(\pi n y)}
{\pi \sqrt{1-y}(n-m/(1-y))}
\end{eqnarray}
and, for $m=n=0$, by
\begin{eqnarray}
&&\hspace{-1cm}\widetilde N_{00}^{\out\out}=0,\quad
 \widetilde N_{00}^{\out\inI}=-\sqrt{y},\quad
 \widetilde N_{00}^{\out\inII}=-\sqrt{1-y},\\
&&\hspace{-1cm}\widetilde N_{00}^{\inI\inII}
=-\frac{1}{4\pi\mu|\aout| \sqrt{y(1-y)}},
\quad
\widetilde N_{00}^{\inI\inI}=\frac{1}{4\pi\mu|\aout| y},
\quad
 \widetilde N_{00}^{\inII\inII}=\frac{1}{4\pi\mu|\aout| (1-y)}
\end{eqnarray}

When we compute string amplitudes for fermions, the large $\mu$ behavior 
of $\bar F_n^{(r)}$, $\bar G_n^{(r)}$, $U_m^{(r)}$ and $1-4\mu\a K$ are also
useful\footnote{
Note that the definition of the Neumann vector $N^{r}_{n}$ here,
with which $F_n^{(r)}$ is defined, differs by 
$C_{(r)}^{-1/2}U_{r}$ from that of the ref.\cite{He et. al.}.
}:
\begin{eqnarray}
 &&\hspace{-1.5cm}
\bar F_{n}^{(\inI)}=
\frac{(-1)^{n+1}\sqrt{\mu}}{\sqrt{\pi\mu|\a_{(1)}|y}},\quad
\bar F_{n}^{(\inII)}=
\frac{\sqrt{\mu}}{\sqrt{\pi\mu|\a_{(1)}|(1-y)}},\quad
\bar F_{n}^{(\out)}=\frac{(-1)^{n+1}\sqrt{\mu}n\sin(\pi ny)}
{\sqrt{\pi(\mu|\a_{(1)}|)^3}}\\
 &&\hspace{-1.5cm}
\bar F_{0}^{(\inI)}=
-\frac{\sqrt{\mu}}{\sqrt{2\pi\mu|\a_{(1)}|y}},\quad
\bar F_{0}^{(\inII)}=
\frac{\sqrt{\mu}}{\sqrt{2\pi\mu|\a_{(1)}|(1-y)}},\quad
\bar F_{0}^{(\out)}=0\\
&&\hspace{-1.5cm}
\bar G^{(\inI)}_n=\frac{(-1)^{n+1}}{\sqrt{2\pi \mu |\aout| y}},\quad
\bar G^{(\inII)}_n=\frac{1}{\sqrt{2\pi \mu |\aout| (1-y)}},\quad
\bar G^{(\out)}_n=\frac{(-1)^{n+1}\sqrt{2} \sin(\pi n y)}{\sqrt{\pi \mu |\aout|}}\label{asymptotic forms}\\
&&\hspace{-1.5cm}
\bar G^{(\inI)}_0=-\frac{1}{\sqrt{4\pi \mu |\aout| y}},\quad
\bar G^{(\inII)}_0=\frac{1}{\sqrt{4\pi \mu |\aout| (1-y)}},\quad
\bar G^{(\out)}_0=0\\
&&\hspace{-1.5cm}U_n^{(\inI)}=\frac{n}{2 \mu |\aout| y},\qquad
U_n^{(\inII)}=\frac{n}{2 \mu |\aout| (1-y)},\qquad
U_n^{(\out)}=\frac{2\mu |\aout| }{n}, \\
&&\hspace{-1.5cm} 1-4\mu \a K = \frac{1}{4\pi \mu |\aout| y(1-y)},
\end{eqnarray}

\end{document}